\documentclass[twocolumn,aps,amssymb,footinbib]{revtex4-1}
\usepackage{amssymb}
\usepackage{graphicx}
\usepackage{amsmath}
\usepackage{times}
\usepackage{color}
\usepackage{subfigure}
\usepackage{feynmp}
\usepackage{bm}
\usepackage{mathrsfs}

\newcommand{\brr}{\bar{\mathbf{r}}}
\newcommand{\bpp}{\bar{\mathbf{p}}}
\newcommand{\rr}{\mathbf{r}}
\newcommand{\qq}{\mathbf{q}}
\newcommand{\pp}{\mathbf{p}}

\newcommand{\RR}{\mathbf{r}}

\newcommand{\dd}[1]{\frac{\mathrm{d}^2 #1}{(2\pi)^2}}
\newcommand{\LE}{0}

\newcommand{\MM}{\mathcal{M}}

\newcommand{\drr}{\nabla_\RR}
\newcommand{\dpp}{\nabla_\pp}
\newcommand{\Sig}{\Sigma}

\newcommand{\DD}{\mathscr{D}}
\newcommand{\II}{\mathscr{I}}
\newcommand{\Langle}{\langle\hspace{-2.5pt}\langle}
\newcommand{\Rangle}{\rangle\hspace{-2.5pt}\rangle}

\newcommand{\rb}{\bar{r}}
\newcommand{\pb}{\bar{p}}

\begin{document}

\begin{abstract}
We study the collective excitations of polarized single-component quasi-two-dimensional dipolar fermions in an isotropic harmonic trap by solving the collisional Boltzmann-Vlasov (CBV) equation. We study the response to both monopole and quadrupole perturbations of the trap potential and investigate the character of excitations in each case. Simple analytic formulas are provided based on the linearized scaling ansatz and accurate numerical results are obtained by satisfying the first eight moments of the CBV equation. Except for the lowest lying monopole mode that exhibits a negligible damping in all of the studied cases, the quadrupole and the higher order monopole modes undergo a transition from the collisionless regime to a highly dissipative crossover regime and finally to the hydrodynamic regime upon increasing the dipolar interaction strength. For strong vertical confinements (2D limit), we predict the existence of a temperature window within which the characteristics of the collective modes become temperature independent. This behavior, which is a unique feature of the universal near-threshold dipole-dipole scatterings, persists as long as the scattering energies remain in the near-threshold regime. The predictions of this work are expected to be in the reach of current experiments.
\end{abstract}

\title{Collective excitations of quasi-two-dimensional trapped dipolar fermions:\\transition from collisionless to hydrodynamic regime}

\author{Mehrtash Babadi$^1$, Eugene Demler$^1$}
\affiliation{
$^1$ Physics Department, Harvard University, Cambridge, Massachusetts 02138, USA
}

\maketitle

\section{Introduction}
Dipolar quantum gases have been the subject of much interest and significant experimental and theoretical investigations in the recent years. The long-range anisotropic dipole-dipole interactions gives rise to novel phenomena and applications in these systems~(see Ref.~\cite{Baranov2008} and the references therein). In particular, dipolar Bose-Einstein condensates (BECs) with magnetic dipole-dipole interactions have been exhaustively studied both theoretically and experimentally~\cite{Lahaye2009}. The most recent experimental achievement along this line is the BEC of ${}^{168}$Er with a large magnetic dipole moment of $7~\mu_B$~\cite{Aikawa2012}. From an experimental point of view, the dipolar effects are much easier to observe in dipolar BECs compared to dipolar Fermi gases where the large energy scale set by Pauli exclusion requires larger dipoles for the interaction effects to become appreciable. Since electric dipole-dipole interactions are typically stronger than magnetic ones, much of the recent experimental efforts have been focused on the realization of ultracold heternucleus bi-alkali molecules which have large permanent electric dipole moments.

One of the most important experimental achievements in this direction is the realization of a nearly quantum degenerate gas of KRb molecules at JILA~\cite{JILA}. The experiments with other bi-alkali fermionic polar molecules such as LiCs~\cite{Sage2005,Deiglmayr2011} are also making significant progress. At the same time, realization of degenerate fermionic atoms with strong magnetic dipolar interaction, such as ${}^{161}$Dy~\cite{Burdick2012}, as well as microwave dressed Rydberg atoms~\cite{Tanasittikosol2011} are other avenues toward realization of strongly interacting ultracold dipolar fermionic gases.

An important experimental probe for the many-body physics of ultracold gases is the measurement of collective oscillations of trapped gases in response to perturbations of the trap potential. These oscillations constitute the low-lying collective excitations of these systems. The measurement of the frequency and damping of these oscillations can be utilized to understand the properties of the ground state and to extract important information such as the role self-energy corrections, the equilibrium equation of state, collisional relaxation rates and kinetic coefficients. Moreover, the possibility of carrying out extremely precise measurements of these quantities allows us to put our theoretical understanding of the system to the test. For instance, by measuring the frequency of the radial breathing mode for a two-component Fermi gas near the BEC-BCS crossover with a $10^{-3}$ accuracy level, the Innsbruck group could clearly verify the Quantum Monte-Carlo result for the unitary gas and invalidate the predictions of the BCS theory~\cite{Altemeyer2007}. Another good example is the recent measurement of the universal quantum viscosity of the unitary gas~\cite{Cao2010} that confirmed the theoretical $T^{3/2}$ scaling and also provided evidence for a conjecture on the lower bound for the viscosity over entropy ratio obtained using string theory methods~\cite{Kovtun2005}. At the moment, the collective oscillations of trapped BECs~\cite{Dalfovo1999} and two-component atomic gases with $s$-wave interactions in three dimensions~\cite{Grimm2007} are both understood fairly well. Recently, the experimental and theoretical study of the 2D Fermi gas in the strongly interacting regime has also shown a significant progress~\cite{Vogt2012,Schaefer2012,Taylor2012,Enss2012,Wu2012}.

Generally speaking, the low-lying collective excitations of an interacting system may be either described as collisionless (CL), hydrodynamical (HD), or in the crossover between these two limits. The CL limit is achieved when the the gas is either rarefied, or the interactions are negligibly weak or a certain dynamical symmetry forbids collisions. In this case, no dissipation occurs and the collective modes are undamped. The HD limit, on the other hand, is achieved either when the gas is in a superfluid state or in case of strongly interacting Fermi liquids, when the collision rate is much faster than the frequency of the collective modes so that a local equilibrium can be maintained~\cite{KB}. In either case, the dynamics can be described well using simple HD equations in this limit, which are simply statements of conservation of mass, momentum and energy~\cite{LL,KB}. The collective modes are again dissipationless in this limit. A realistic system, however, lies somewhere between these two ideal limits, i.e. either the collision rate is not fast enough to maintain the local equilibrium or in the case of superfluid systems, the non-condensed component leads to collisions. An important aspect of understanding a many-body system is to determine where it lies within this spectrum, both qualitatively and quantitatively.

\begin{figure}
\center
\includegraphics[scale=0.9]{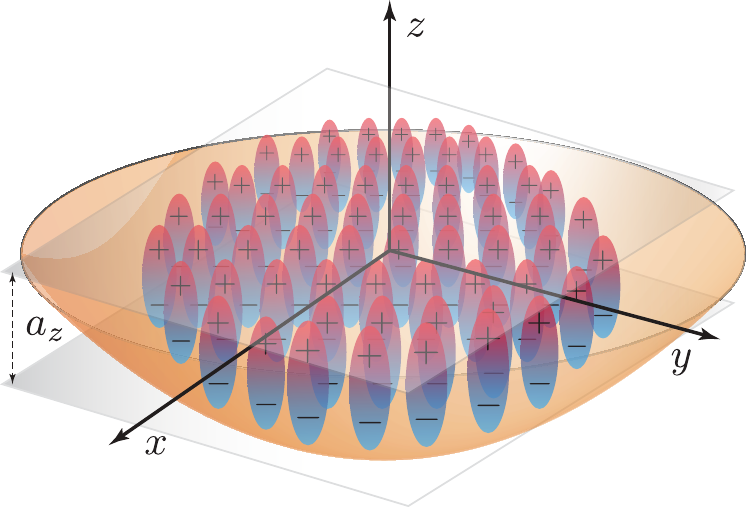}
\caption{(Color online) A schematic picture of quasi-two-dimensional dipolar fermions in a shallow and isotropic in-plane (transverse) trap. Application of a strong dc field aligns the dipoles along the vertical axis ($z$). The quasi-two-dimensional limit corresponds to the situation where $a_z\equiv [\hbar/(m\omega_z)]^{1/2}$ is much smaller than all of $a_0\equiv [\hbar/(m\omega_0)]^{1/2}$, interparticle separation $n_\mathrm{2D}^{-1/2}$ and the thermal de Broglie wavelength $\lambda_T \equiv h/(2\pi m k_B T)^{1/2}$.}
\label{fig:schem}
\end{figure}

In this paper, we address this question for polarized single-component quasi-two-dimensional (quasi-2D) dipolar fermionic gases~(see Fig.~\ref{fig:schem}) which has been the subject of much interest recently. In this setting, dipole-dipole interactions have a repulsive character and can be utilized to produce a strongly correlated Fermi liquid. Moreover, this particular configuration is found to be necessary in experiments with a wide range of bi-alkali polar molecules as a mechanism to suppress ultracold chemical reactions~\cite{JILA}.\\

Several authors have already investigated certain aspects of this problem. In particular, Lima {\it et al.} have studied the collective oscillations in traps with various degrees of anistropy by assuming the validity of a hydrodynamic description~\cite{Lima2010R,Lima2010} while Sogo~{\it et al.} studied the collisionless limit~\cite{Sogo2009}. More recently, Abad~{\it et al.} have studied both regimes separately and gave a comparison of the predictions of each~\cite{Abad2012}. However, none of the mentioned works have given a quantitatively reliable condition for the validity of their approaches beyond simple order of magnitude analyses. Moreover, the intermediate regime in which one expects to observe the interesting physics of dissipation and collisional damping is not addressed in any of these works.\\

In this paper, we make no prior assumption about where the system lies in the CL to HD spectrum. Instead, we use the framework of quantum kinetic equations, in particular, the collisional Boltzmann-Vlasov limit, which in principle allows one to study the dynamics in both limits in a unified way, including the crossover regime. The CL and HD limits naturally emergence when the right conditions are met. We evaluate the linear response of gas to monopole-like and quadrupole-like perturbations in the trap potential ($x^2 + y^2$ and $x^2 - y^2$ respectively) and study the oscillation frequency and damping of the generated excitations. We restrict our analysis to situations where the scattering energies lie well within the near-threshold regime so that Born approximation is applicable~\cite{Ticknor2009,Arnecke2008}. The collision integrals are treated without resorting to the usual relaxation time approximation.

We carry out the calculations in two stages. First, we neglect the self-energy corrections to quasiparticle dispersions and utilize the widely used linearized scaling ansatz approximation~\cite{Scaling} to obtain a simple semi-analytic picture. In the next stage, we add mean-field corrections to quasiparticle dispersions and also extend the scaling ansatz approximation by satisfying all moments of the kinetic equation up to the eight order in order to obtain accurate numerical results. We find that both of these improvements result in significant quantitative corrections. Also, inclusion of higher moments also allows us to go beyond the study of lowest lying modes and to look at higher order modes as well.\\

Before delving into the formalism and details, we find it useful to briefly summarize the main results of this work, some of which are unique and novel features of dipolar fermions in 2D. Without self-energy corrections, the scaling ansatz analysis makes the well-known prediction of undamped monopole oscillations at a fixed frequency of $2\omega_0$, independent of the interaction strength and temperature~\cite{Vichi2000,Boltzmann1909}. Here, $\omega_0$ is the in-plane (transverse) trap frequency. Taking mean-field corrections into account, we find that while the lowest lying monopole mode assumes only a negligibly small damping, its oscillation frequency will significantly increase from $2\omega_0$ due to the repulsive dipole-dipole interactions (see Fig.~\ref{fig:monOmGam}). The higher order monopole modes, however, are significantly affected by collisions: they go through a strongly dissipative crossover regime upon increasing the interaction strengths and finally approach the dissipationless HD regime.

The quadrupole modes, including the lowest lying one, exhibit the same CL to HD transition as well. The oscillation frequency of the lowest lying quadrupole mode approaches $\sqrt{2}\omega_0$ in the HD limit, which is the universal, system-independent, frequency of quadrupole ``surface'' mode~\cite{Griffin1997} (see Fig.~\ref{fig:quadevol}). The emergence of surface mode is a clear indicator of hydrodynamics.

\begin{figure}
\includegraphics[scale=0.65]{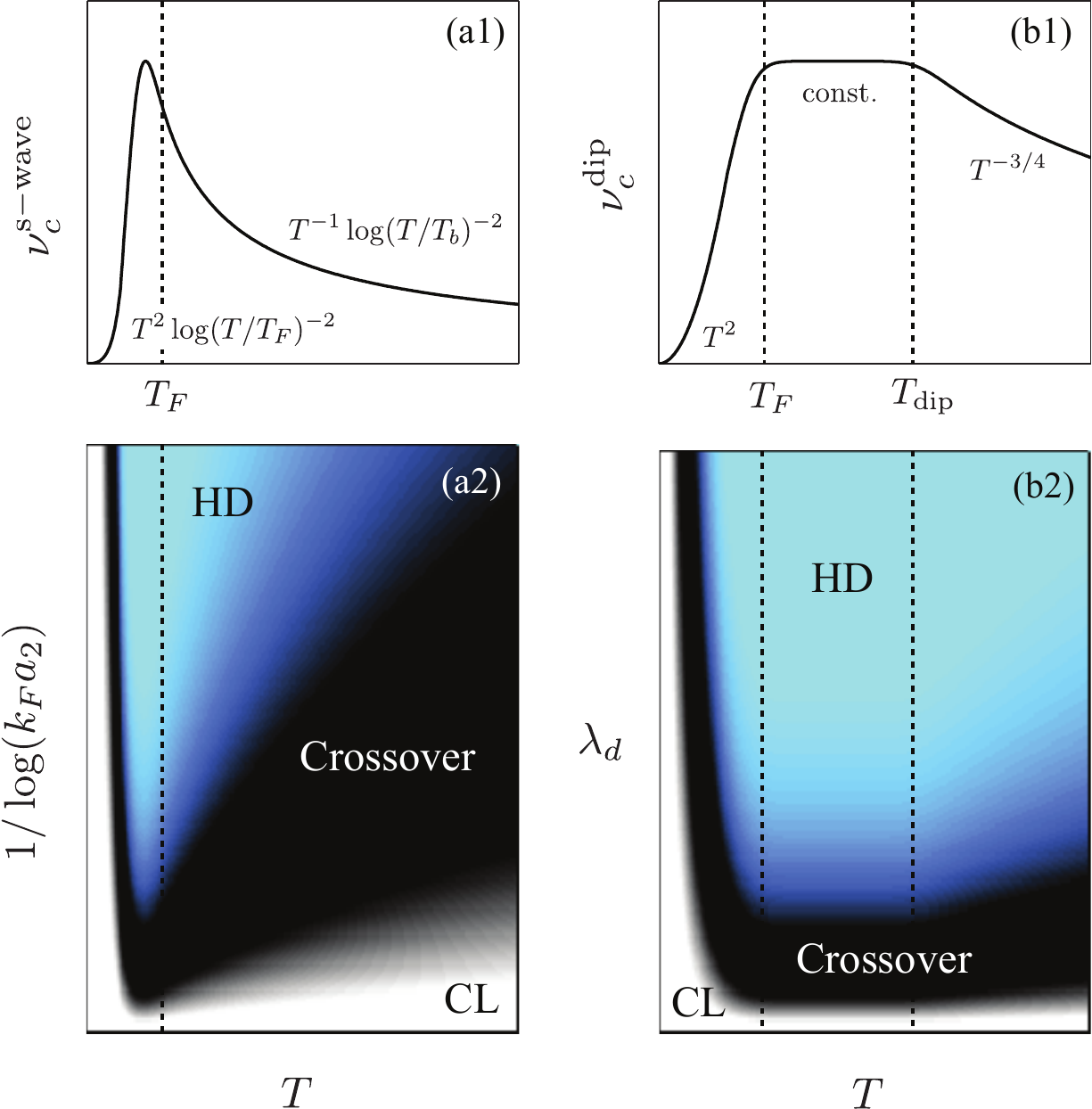}
\caption{(Color online) Qualitative comparison of the dynamical regimes of quadrupole collective modes of 2D $s$-wave and dipolar fermions in harmonic traps. (a1) and (b1) show the temperature dependence of collisional relaxation rates of $s$-wave and dipolar fermions respectively. (a2) and (b2) show the dynamical regimes of quadrupole collective modes as a function of interaction strength and temperature. The asymptotics of $\nu_c^\mathrm{s-wave}$ is due to~\cite{Schaefer2012}. $k_F$ is the trap Fermi momentum, $a_2$ is the 2D scattering length and $T_b = \hbar^2/(m k_B a_2^2)$. See Eqs.~\ref{eq:defs} and~\ref{eq:Tdip} for the definitions of the parameters appearing in (b1) and (b2). Please refer to the main text for details.}
\label{fig:comp}
\end{figure}

We find simple semi-analytic results for quadrupole oscillations using the linearized scaling ansatz and by dropping self-energy corrections. In this approximate picture, the frequency and damping of the quadrupole oscillations are controlled by a single dimensionless parameter, the collisional relaxation rate $\nu_c$ (Ref. to Sec.~\ref{sec:quadscl}). Small and large values of $\nu_c$ correspond to collisionless and hydrodynamic regimes respectively. In the collision dominated regime, the viscosity sum rule yields $\nu_c \sim \langle P/\eta_\mathrm{s} \rangle_\mathrm{trap}$, where $P$ and $\eta_\mathrm{s}$ denote the local pressure and shear viscosity respectively~\cite{Schaefer2012}. By $\langle \ldots\rangle_\mathrm{trap}$, we imply averaging over the trap. Also, we identify $\nu_c \sim \omega_0\tau_c^{-1}$ in the thermal regime ($T \gg T_F \equiv \sqrt{2N}\hbar\omega_0$), where $\tau_c$ is the singleparticle collision time.

For small $T/T_F$, we obtain $\nu_c \sim T^2$ which is due to Pauli blocking. For large $T/T_F$, the behavior of $\nu_c$ depends on the degree of quasi-two-dimensinality (quantified by $\eta$, see Eq.~\ref{eq:defs}). In the strictly 2D limit ($\eta=0$), we find that $\nu_c$ reaches a plateau for $T \gtrsim T_F$. The existence of this plateau, which is a unique feature of 2D dipolar fermions, results from the balance between rarefaction of the gas at higher temperatures on one hand, and the growth of the dipolar scattering cross section on the other hand. The high temperature cut-off for this plateau is $T_\mathrm{dip} \equiv \hbar/(m a_d^2 k_B)$, where $a_d \equiv m D^2/\hbar^2$ is the ``dipolar length''. Here, $m$ and $D$ denote the mass and the dipole moment of a single particle. For $T \gtrsim T_\mathrm{dip}$, we find $\nu_c \sim T^{-3/4}$.

Fig.~\ref{fig:comp} shows a qualitative comparison between the behavior of quadrupole oscillations in 2D two-component fermions interacting via a $s$-wave Feshbach resonance (simply, $s$-wave fermions) and 2D dipolar fermions. The top and bottom panels show the temperature dependence of $\nu_c$ and the resulting dynamical regimes for the collective excitations as a function of interaction parameters and temperature. The discussed regimes of $\nu_c$ for 2D dipolar fermions can be seen in panel (b1). It is worthy of mention that the temperature window in which $\nu_c$ is appreciably large is ``universal'' for 2D $s$-wave fermions. For 2D dipolar fermions, however, this window is amenable to experimental tuning (Ref. to Sec.~\ref{sec:quadscl}).

We looked into the effect of mean-field corrections to quasiparticle dispersions and found that their inclusion yields significant correction in the quantum degenerate regime. This is again in contrast to the case of $s$-wave fermions where self-energy corrections are often found to have a negligible effect on the frequency of collective modes~\cite{Chiacchiera2011}.

Finally, going beyond the scaling ansatz and satisfying higher order moments of the CBV equation, we found that the scaling ansatz overestimates the collision rates in agreement to the findings of Ref.~\cite{Chiacchiera2011} in the context of $s$-wave fermions. We also found that the corrections to the energy of low-lying modes become negligible after 4th moments. The higher order modes were also briefly studied and we found that their behavior is qualitatively similar to the other modes. Finally, we discussed the experimental outlook of this work and gave predictions for the experiments with KRb. We found that although the HD regime is not achieved in the current experiments, there is a significant collisional damping rate which can be easily measured.\\

This paper is organized as follows. In Sec.~\ref{sec:form}, we describe the model in detail and define the response functions. We briefly review quantum kinetic equation and the approximations leading to the CBV equation and their validity conditions in Sec.~{sec:kinetic}. We discuss the equilibrium state of the trapped gas in Sec.~\ref{sec:eqb}. The linear response theory of the CBV equation is described in Sec.~\ref{sec:linresBV} and the variational calculation of the response functions using the method of moments is discussed. The linearized scaling ansatz analysis in given in Sec.~\ref{sec:scans}, followed by the its extension to higher order moments and inclusion of mean-field corrections in Sec.~\ref{sec:ext}. Finally, we discuss the results in Sec.~\ref{sec:disc} and the experimental outlook of this work in Sec.~\ref{sec:expt}. Most of the technical details and tedious calculations are left to the Appendices.

\section{The Formalism}\label{sec:form}

\subsection{The Hamiltonian}
The Hamiltonian for trapped dipolar fermions prepared in a single hyperfine state and put in a strong polarizing dc field (electric for polar molecules, magnetic for atoms with permanent magnetic dipoles) can be written as:
\begin{multline}
H_{\mathrm{3D}} = \int\mathrm{d}^3\rr\,\psi^\dagger(\rr)\left(-\frac{\nabla^2}{2m}+U^\mathrm{3D}_\mathrm{trap}(\RR)\right)\,\psi(\rr)\\
+ \int\mathrm{d}^3\rr\,\mathrm{d}^3\rr'\,\mathcal{V}^\mathrm{3D}_\mathrm{dip}(\rr-\rr')\,\psi^\dagger(\rr)\,\psi^\dagger(\rr')\,\psi(\rr')\,\psi(\rr),
\end{multline}
where:
\begin{equation}
U^\mathrm{3D}_\mathrm{trap}(\rr) = \frac{1}{2}\,m\omega_z^2 z^2 + \frac{1}{2}m\omega_0^2(x^2 + y^2),
\end{equation}
is the axially symmetric trap potential and:
\begin{equation}
\mathcal{V}^\mathrm{3D}_{\mathrm{dip}}(\rr) = \frac{D^2}{|\rr|^5}\,\left(|\rr|^2 - 3z^2\right).
\end{equation}
We set $\hbar=1$ throughout this paper unless it appears explicitly. A schematic picture of the system is shown in Fig.~\ref{fig:schem}. We have assumed that the electric dipoles are polarized along the $z$-axis. Here, $\psi^{(\dagger)}(\rr)$ is the fermion annihilation (creation) operator is 3D space. In the limit $\omega_z \gg \omega_0, \epsilon_F, k_B T$ (where $\epsilon_F$ and $T$ denote the Fermi energy and the temperature), the particles will be confined to the lowest transverse subband and we can reduce the above Hamiltonian to an effective two-dimensional model:
\begin{multline}
H_\mathrm{2D} = \int\mathrm{d}^2\rr\,\psi^\dagger_0(\rr)\left(-\frac{\nabla^2}{2m}+U^\mathrm{2D}_\mathrm{trap}(\RR) + \frac{\omega_z}{2}\right)\,\psi^{\phantom{\dagger}}_0(\rr)\\
+ \int\mathrm{d}^3\rr\,\mathrm{d}^3\rr'\,\mathcal{V}^\mathrm{2D}_\mathrm{dip}(\rr-\rr')\,\psi^\dagger_0(\rr)\,\psi^\dagger_0(\rr')\,\psi^{\phantom{\dagger}}_0(\rr')\,\psi^{\phantom{\dagger}}_0(\rr).
\end{multline}
Here, $\rr=(x,y)$ denote the 2D transverse coordinates and $\psi^{(\dagger)}_0(\rr)$ denotes the fermion annihilation (creation) in the lowest subband. We have neglected the constant zero point energy $\hbar\omega_z/2$ of the lowest subband. $U^\mathrm{2D}_\mathrm{trap}(\RR) = m\omega_0^2(x^2 + y^2)/2$ is the transverse part of the original trap potential and $\mathcal{V}^\mathrm{2D}_\mathrm{dip}(\rr)$ is the effective dipole-dipole interaction in the lowest subband:
\begin{align}
\mathcal{V}^\mathrm{2D}_\mathrm{dip}(\rr) &= \int\mathrm{d}z\,\mathrm{d}z'\,|\phi_0(z)|^2\,|\phi_0(z')|^2\,\mathcal{V}_\mathrm{dip}^\mathrm{3D}(\rr,z-z'),
\end{align}
where $\phi_0(z) = e^{-z^2/(2 a_z^2)}/(\sqrt{\pi}\,a_z)^{\frac{1}{2}}$ is the vertical wavefunction of particles in the lowest subband and $a_z \equiv (m\omega_z)^{-1/2}$ is the transverse oscillator length. The above integration can be done analytically and we find:
\begin{multline}
\mathcal{V}^\mathrm{2D}_\mathrm{dip}(r) = \frac{1}{\sqrt{2\pi}}\frac{D^2}{2 a_z^3}\,e^{r^2/(4 a_z^2)}\bigg[\left(2 + \frac{r^2}{a_z^2}\right)K_0\left(\frac{r^2}{4 a_z^2}\right)\\
- \frac{r^2}{a_z^2} K_1\left(\frac{r^2}{4 a_z^2}\right)\bigg],
\end{multline}
where $K_n(x)$ is the modified Bessel function of the second kind. In the momentum space, we get:
\begin{equation}\label{eq:Vq}
\tilde{\mathcal{V}}^\mathrm{2D}_\mathrm{dip}(q) = \frac{2\pi D^2}{a_z}\bigg[\sqrt{\frac{2}{\pi}} - q a_z e^{q^2 a_z^2/2} \mathrm{Erfc}\left(\frac{q a_z}{\sqrt{2}}\right)\bigg].
\end{equation}
We remark that in the long wavelength regime ($q a_z \ll 1$), the dominant contribution to $\mathcal{V}^\mathrm{2D}_\mathrm{dip}$ results from the repulsive side-by-side part of dipole-dipole interactions. We denote $\mathcal{V}^\mathrm{2D}_\mathrm{dip} \equiv \mathcal{V}$, $\tilde{\mathcal{V}}^\mathrm{2D}_\mathrm{dip} \equiv \tilde{\mathcal{V}}$ and $U^\mathrm{2D}_\mathrm{dip} \equiv U$ in the remainder of the paper for brevity.\\

It is worthwhile to study the behavior of the effective 2D interaction in various limits. For $qa_z \ll 1$, one finds:
\begin{equation}\label{eq:Vq1}
\tilde{\mathcal{V}}(q) \simeq \frac{4\sqrt{2\pi}D^2}{3 a_z} - 2\pi D^2 q + \mathcal{O}(q^2),
\end{equation}
whereas for $q a_z \gg 1$, we get:
\begin{equation}\label{eq:Vq2}
\tilde{\mathcal{V}}(q) \simeq -\frac{2 D^2 \sqrt{2\pi}}{3 a_z} \left(1-\frac{3}{q^2 a_z^2} + \mathcal{O}(q^{-4} a_z^{-4})\right).
\end{equation}
Apart from the constant term in Eq.~(\ref{eq:Vq1}), which is immaterial as long as we are concerned with a single hyperfine state, we notice an initial linear growth with respect to $q$ which eventually saturates to a constant constant for $q \sim 1/a_z$. We shall see later that this linear growth results in an interesting behavior for the low-lying collective excitations.

In real space, for small $r/a_z$, one finds a behavior similar to the 2D Coulomb gas:
\begin{equation}\label{eq:Vrsmall}
\mathcal{V}(r) \approx \frac{D^2}{\sqrt{2\pi}a_z^3}\left\{-2 -\gamma - \ln[r^2/(8 a_z^2)] + \mathcal{O}(r^2 \ln r)\right\},
\end{equation}
and for large $r/a_z$, the $r^{-3}$ dipole-dipole interaction is recovered:
\begin{equation}\label{eq:Vrlarge}
\mathcal{V}(r) \approx D^2/r^3 + \mathcal{O}(a_z/r^4).
\end{equation}
It is useful to define a ``dipolar length'':
\begin{equation}\label{eq:ad}
a_d \equiv \frac{m D^2}{\hbar^2}, 
\end{equation}
which is the length scale associated to dipolar interactions, as well as the following dimensionless parameters:
\begin{align}\label{eq:defs}
\lambda_d &\equiv \frac{m D^2}{\hbar^2}\left(\frac{m\omega_0}{\hbar}\right)^{\frac{1}{2}}(2N)^{\frac{1}{4}} \equiv \left(\frac{a_d}{a_0}\right)(2N)^\frac{1}{4},\nonumber\\
\eta &\equiv (2N)^\frac{1}{4}\left(\frac{\omega_0}{\omega_z}\right)^\frac{1}{2},
\end{align}
where $a_0 \equiv [\hbar/(m\omega_0)]^\frac{1}{2}$ is the transverse oscillator length and $N$ is the number of trapped particles. $\lambda_d$ is a measure of dipolar interaction strength and is of the order of the typical value of interaction energy over the kinetic energy in the quantum degenerate regime. $\eta$ is a measure of ``quasi-two-dimensionality'' and is of the order of the vertical oscillator length $a_z$ divided by the zero temperature Thomas-Fermi radius of the trapped gas. The strict 2D limit $\omega_z \rightarrow \infty$ corresponds to $\eta=0$.

\subsection{Linear response theory}
A typical experiment for measuring the collective excitations of trapped particles is the following: the gas prepared in a thermal equilibrium state at $t<0^-$. At $t>0^-$, the system will be subject to a local perturbation, such as a kick or modulation of the trap potential. A certain observable will be then monitored either with an in situ or absorption imaging technique. If the frequency and amplitude of the perturbing potential is weak compared to the corresponding microscopic scales, such an experiment can be theoretically investigated within the linear response theory. Let us denote the perturbing potential and the observable as $\delta U(\rr,t)$ and $O(\rr)$ respectively, and their corresponding second quantized operators are $\delta \hat{U} \equiv \int\mathrm{d}^2\rr\,\psi^\dagger_0(\rr)\,\delta U(\rr,t)\,\psi^{\phantom{\dagger}}_0(\rr)$ and $\hat{O} \equiv \mathrm{d}^2\rr\,\psi^\dagger_0(\rr)\,O(\rr)\,\psi^{\phantom{\dagger}}_0(\rr)$. The usual linear response theory then yields:
\begin{equation}
\langle \hat{O}\rangle_t = \int_0^t\,\mathrm{d}t'\int\mathrm{d}^2\rr\,\mathrm{d}^2\rr'\,\chi^R_{nn}(\rr,\rr';t-t')\,O(\rr)\,\delta U(\rr',t'),
\end{equation}
where $\chi^R_{dd}(\rr,\rr';t-t')$ is the retarded density-density response function:
\begin{equation}
\chi^R_{nn}(\rr,\rr';t-t') \equiv -i\theta(t-t')\mathrm{Tr}\{\hat{\rho}_0[\hat{n}(\rr,t),\hat{n}(\rr',t)]\},
\end{equation}
where $\hat{n}(\rr,t) = \psi^\dagger_0(\rr,t)\psi^{\phantom{\dagger}}_0(\rr,t)$ is the density operator and $\hat{\rho}_0$ is the initial density matrix. At this stage, one may choose a proper many-body approximation scheme and attempt to evaluate the response function using the diagram technique. However, the lack of translational symmetry due to the presence of the trap potential makes this method complicated. In practice, one will have to make assumptions about separation of microscopic and macroscopic time and length scales in order to proceed. It is, however, much more transparent to acknowledge the existence of such a separation of scales from the outset and reduce the complicated evolution equations of the non-equilibrium Green's functions to quantum kinetic equations. One may then formulate and evaluate the linear response functions in the language of quantum kinetic equations. We describe this method in the next section, where we also briefly review the quantum kinetic equations approach, and conclude this section by defining the response functions relevant to monopole and quadrupole oscillations.\\

The monopole oscillations can be excited by choosing $\delta U(\rr,t) \equiv \delta U_m(\rr,t) \equiv \mathcal{A}(t)\,m\omega_0^2 r^2$, where $\mathcal{A}(t)$ is the temporal shape of the perturbation (e.g. a $\delta$-function, a finite pulse or a periodic modulation). We choose $\mathcal{A}(t) \equiv \mathcal{A}_0\,\omega_0^{-1}\delta(t^-)$ for concreteness. Also, the linear response to any other pulse shape can be determined from the impulse response. Note that we have {\em defined} the monopole oscillations as the response of the trapped gas to a $\sim r^2$ perturbation. One may choose any other isotropic trap perturbation (such as $r^4$, etc). However, such choices are expected to excite higher order modes as well, and not necessarily the lowest lying ones. Here, the observable is the variation in the size of the cloud, $\hat{r}^2 - \langle \hat{r}^2 \rangle_0$. We define the ``monopole response function'' as:
\begin{equation}\label{eq:chimon}
\chi_{r^2}(t) = \mathcal{A}_0^{-1} m\omega_0\,\theta(t)\left(\langle\hat{r}^2\rangle_t - \langle \hat{r}^2 \rangle_0\right).
\end{equation}
Likewise, we define the quadrupole oscillations as the response of the trapped gas to $\delta U(\rr,t) \equiv \delta U_q(\rr,t) \equiv \mathcal{A}(t)\,m\omega_0^2 (x^2 - y^2)$ and define the ``quadrupole response function'' as:
\begin{equation}\label{eq:chiquad}
\chi_{x^2 - y^2}(t) = \mathcal{A}_0^{-1}m\omega_0\,\theta(t)\,\langle\hat{x}^2 - \hat{y}^2\rangle_t.
\end{equation}
Note that $\langle\hat{x}^2 - \hat{y}^2\rangle_0$ due to the isotropy of the trap.

\subsection{From quantum kinetic equations to the collsional Boltzmann-Vlasov equation}\label{sec:kinetic}
Quite generally, the dynamics of confined quantum gases can be formulated and studied using the formalism of non-equilibrium Green's functions, i.e. either by solving Kadanoff-Baym equations using a physically relevant conserving approximation~\cite{KB} or by using the Keldysh-Schwinger diagram technique. Such a formulation, however, is only necessary when the spatial and temporal scales of inhomogeneities (the trap and its perturbation) are comparable to the microscopic scales. In experiments dealing with a large number of particles $N$ in a shallow trap, there is a natural separation of temporal and spatial scales between the microscopic (single particle) and macroscopic (collective) dynamics. Exploiting this fact, one can reduce the complicated Kadanoff-Baym equations to the somewhat simplified picture of quantum kinetic equations using the well-known procedure of gradient expansion~\cite{KB,Danielewics1984}.\\

There exist several decent treatments of the subject of quantum kinetics  in the literature and we refer the reader to the excellent pioneering monograph of Kadanoff and Baym~\cite{KB} and Ref.~\cite{Danielewics1984} for details. However, for the purpose self-containedness and in order to clarify the invoked approximations, we provide a very brief review of the basic elements of the kinetic theory. Our starting point is the general quantum kinetic equation for a system composed of a single species of fermions (i.e. a gas prepared in a single hyperfine state):
\begin{equation}\label{eq:kin}
[\mathfrak{Re}(G^{-1})^+,iG^\gtrless] - [i\Sigma^\gtrless,\mathfrak{Re} G^+] = G^<\Sigma^> - G^>\Sigma^<,
\end{equation}
where $G^+(\pp,\omega;\RR,t) \equiv (\omega - p^2/(2m) - U(\RR,t) - \Sigma^+)^{-1}$ and $G^\gtrless(\pp,\omega;\RR,t)$ are the retarded and greater/lesser non-equilibrium Green's functions in the mixed Wigner coordinates~\cite{Danielewics1984}. $U(\RR,t)$ denotes the external potential (i.e. the trap) and is assumed to vary on a scale much larger than the microscopic scales. $\Sigma^+(\pp,\omega;\RR,t)$ and $\Sigma^\gtrless(\pp,\omega;\RR,t)$ are the retarded and greater/lesser self-energies. In the mixed Wigner coordinates, $(\pp,\omega)$ and $(\RR,t)$ denote to the Fourier transformed microscopic coordinates and the slow macroscopic coordinates, respectively. $[A,B]$ denotes the generalized Poisson's bracket defined as:
\begin{multline}
[A,B] = \partial_\omega A\,\partial_t B - \partial_t A\,\partial_\omega B\\ - \nabla_\pp A \cdot \nabla_\RR B + \nabla_\RR A \cdot \nabla_\pp B.
\end{multline}
It is generally understood that $G^+$ encodes the spectral properties of the system (single particle states) while $G^<$ and $G^>$ contains the information about the statistics of particles and holes, respectively. Likewise, the real and imaginary parts of $\Sigma^+$ describe the renormalization of the singleparticle dispersion and the spectrum broadening while $\Sigma^<$ and $\Sigma^>$ describe the collisional scattering -in and -out rates. In analogy to the equilibrium case, it is fruitful to introduce the local spectral function $A(\pp,\omega;\RR,t)$, Wigner's function $f(\pp,\omega;\RR,t)$ and spectral broadening $\Gamma(\pp,\omega;\RR,t)$ (hereafter, we drop the common arguments of the functions unless it is necessary), such that $G^< \equiv i A f$, $A \equiv i(G^>-G^<) \equiv -2\,\mathfrak{Im}(G^+)$ and $\Gamma \equiv i(\Sigma^> - \Sigma^<) \equiv -2\,\mathfrak{Im}(\Sigma^+)$. The kinetic equations can be partially integrated to yield $(G^+)^{-1} = \omega - p^2/(2m) - \mathfrak{Re}(\Sigma^+) + i \Gamma/2$~\cite{KB}. This observation, along with one's choice of a many-body approximation that gives the self-energies as a functional of $G^<$ and $G^>$, and finally the kinetic equation (Eq.~\ref{eq:kin}) for either of $G^<$ or $G^>$ constitute a closed set of partial integro-differential equations for $f$ and $A$ whose solution describes the slow non-equilibrium dynamics of the system. For the case of $\Phi$-derivable many-body approximations, the kinetic equation obeys differential conservation laws for mass, momentum and energy currents. Such conservation laws are essential for formation and propagation of collective modes~\cite{KB}.\\

Although the formalism of quantum kinetics is much simpler than a full non-equilibrium treatment, it is still extremely difficult to solve them in reality without resorting to further approximations. One useful approximation relevant for weakly interacting systems is the quasiparticle approximation. The idea is that in the quantum degenerate regime, only the particle-hole excitations near the Fermi surface are responsible for the slow dynamics. The lifetime of such excitations, $\Gamma^{-1}(p_F, \epsilon_F)$, is proportional to $T_F^2/T^2$ which can be very large. Thus, one may safely neglect the spectral broadening of the Green's functions appearing in the Poisson brackets and approximate the spectral function as $A \approx 2\pi\delta(\omega - p^2/(2m) - U - \Sigma^+)$. This approximation yields as ansatz for the greater/lesser Green's functions:
\begin{align}
G^<_{\mathrm{qp}}(\pp,\omega;\RR,T) &= 2\pi i\,Z_\pp\,\delta (\omega - E_\pp)\,n(\pp;\RR,t),\nonumber\\
G^>_{\mathrm{qp}}(\pp,\omega;\RR,T) &= -2\pi i\,Z_\pp\,\delta (\omega - E_\pp)\,[1-n(\pp;\RR,t)],
\end{align}
where $E_\pp$ is the (local) quasiparticle dispersion and is obtained by solving $\omega - p^2/(2m) - U(\rr,t) - \Sigma^+(\pp,E_\pp;\RR,t)=0$, and $Z_\pp = [1-\partial_\omega\Sigma^+(\pp,\omega = E_\pp;\RR,t)]^{-1}$ is the (local) quasiparticle residue. $n(\pp;\RR,t) \equiv f(\pp,E_\pp;\RR,t)$ is the quasiparticle occupation number. Plugging this ansatz into the kinetic equation, we obtain the collisional Boltzman-Vlasov (CBV) equation:
\begin{multline}\label{eq:BV}
\bigg(\frac{\partial}{\partial t} + \frac{\pp}{m}\cdot\drr + \dpp\Sig^+[n]\cdot\drr - \drr\Sig^+[n]\cdot\dpp\\
- \drr U(\RR,t)\cdot\dpp\bigg)n(\pp;\RR,t) = I_c[n].
\end{multline}
$I_c[n]$ is called the collision integral operator and is given by:
\begin{equation}
I_c[n] \equiv -i Z_\pp\left[(1-n)\,\Sig^< + n\,\Sig^>\right].
\end{equation}
The CBV equation can be thought as a generalization of the usual Boltzmann transport equation of classical gases by (1) including Pauli exclusion effect in the collision integral, and (2) self-energy corrections of quasiparticle dispersions. A crucial observation made by Kadanoff and Baym is that the one may use different conserving many-body approximations for left hand (known as convective or dynamical) and the right hand (collisional) sides of the kinetic equation, without breaking the conservation laws. Intuitively, the dynamical and collisional contributions describe different physics and as long as each respect the conservation laws, the conserving property of the kinetic equation is preserved as a whole.\\

The main goal of this work is to study the effect of interactions to the leading order in the interaction strength on both collisionless quasiparticle transport and elastic quasiparticle collisions. We use the self-consistent Hartree-Fock (HF) approximation on the dynamical side and the Born approximation (which is the lowest order $\Phi$-derivable approximation that leads to collisions) on the collisional side. The retarded self-energy in the HF approximation is instantaneous and is given by:
\begin{multline}\label{eq:senhf}
\Sig^+[n](\pp;\rr,t)=\int\mathrm{d}^2\rr'\,\frac{\mathrm{d}^2\pp'}{(2\pi)^2}\Big[\mathcal{V}(\rr-\rr')\\
-\delta^2(\rr-\rr')\tilde{\mathcal{V}}(\pp-\pp')\Big]n(\pp';\rr',t),
\end{multline}
where $\mathcal{V}(\rr)$ and $\tilde{\mathcal{V}}(\pp)$ are the two-body interactions in the real and momentum space. Dealing with long-range interactions, we have included non-local contributions in the Hatree term. Such contributions are clearly beyond the first order gradient approximation but their inclusion may be necessary for sufficiently long-range interactions (it is exactly the presence of such non-local contributions in the Boltzmann-Vlasov equation for the plasma that leads to plasmon modes and Landau damping). However, we will show momentarily that non-local contributions are negligible for the case of dipole-dipole interactions. Also, note that since $\Sig^+$ has no $\omega$-dependence, the quasiparticle residue is $1$.

The collision integral in the Born approximation is given by~\cite{Danielewics1984}:
\begin{multline}\label{eq:Ic}
\hspace{-8pt}I_c[n] = \int \dd{\pp_1}\,\dd{\pp'}\,\dd{\pp'_1}\,(2\pi)^2\delta^2(\Delta\mathbf{P})(2\pi)\delta(\Delta E)\\
\times\frac{1}{2}\,|\MM|^2\Big[(1-n)(1-n_1)n'n'_1 -n n_1 (1-n')(1-n'_1)\Big],
\end{multline}
where $\mathcal{M} = \tilde{\mathcal{V}}(\pp-\pp') - \tilde{\mathcal{V}}(\pp-\pp'_1)$ is the Born scattering amplitude, $\Delta\mathbf{P} = \pp + \pp_1 - \pp' - \pp'_1$ and $\Delta E = E_\pp + E_{\pp_1} - E_{\pp'} - E_{\pp'_1}$. Note that $E_\pp = p^2/(2m) + U(\rr,t) + \Sigma^+[n](\pp;\rr,t)$. We have also used the shorthand $n \equiv n(\pp;\rr,t)$, $n_1 \equiv n(\pp_1;\rr,t)$, etc.\\

We conclude this section by discussing the validity of the adopted approximations. Since we have described the interactions using the lowest order diagrams, the predictions are quantitatively reliable only as long as the system is in the weakly interacting regime, i.e. $\lambda_d \ll 1$ (see Eq.~\ref{eq:defs}). For dipolar interactions, this condition is equivalent to diluteness $n_\mathrm{2D} a_d \ll 1$, where $n_\mathrm{2D}$ is the 2D density and $a_d$ is the dipolar length defined earlier (Eq.~\ref{eq:ad}). Dealing with a Fermi liquid with essentially short-range interactions (i.e. $\int\mathrm{d}^2\rr\,\mathcal{V}(r) < \infty$), the major many-body corrections such as the screening of interactions and in-medium T-matrix corrections are expected to change the predictions only quantitatively and the leave qualitative features intact even in the strongly interacting regime ($\lambda_d \gg 1$). Therefore, while we acknowledge the limitations our approach, we allow ourself to extend our analysis to $\lambda_d \sim \mathcal{O}(1)$ as well.

Aside from the many-body physics, the validity of Born approximation in describing two-body scatterings and neglect of multiple scatterings must also be assessed. The Born approximation is valid when $\hbar v \gg \mathcal{V} a$, where $v$ is the typical velocity of the scattering pairs in the center of mass frame and $a$ is range of interactions. Identifying $a$ with $a_d$ and $v \sim [m\max(k_BT, k_B T_F)]^\frac{1}{2}$, this his condition implies:
\begin{equation}\label{eq:Tdip}
\mathrm{max}(k_B T, k_B T_F) \ll T_d \equiv \frac{\hbar^2}{m a_d^2},
\end{equation}
where we have defined the ``dipolar temperature'' $T_\mathrm{dip}$. This is precisely the condition for near-threshold scatterings. Ref.~\cite{Ticknor2009} has studied the 2D dipolar scatterings in detail, both in the near-threshold and semi-classical regimes. The study concludes that Born approximation predicts the correct scaling of the total scattering cross section with respect to the scattering energy provided that $m v a_d /\hbar \lesssim 0.3$. Inclusion of multiple scatterings, however, results in $\mathcal{O}(1)$ quantitative corrections as one approaches the semiclassical regime. In this paper, we confine our analysis to near-threshold scatterings. Therefore, the quantitative validity of our results crucially relies on Eq.~(\ref{eq:Tdip}). Finally, we assume that the following scale separation holds:
\begin{equation}
T_F \ll T_\mathrm{dip}\quad \equiv \quad \frac{a_0}{a_d} \gg N^\frac{1}{4},
\end{equation}
so that we can allow ourselves to investigate both the quantum degenerate regime ($T/T_F \ll 1$) and the thermal regime ($T/T_F \gg 1$) up to $T \sim T_\mathrm{dip}$. We note that this condition is well satisfied in the current experiments with both polar molecules and atoms with permanent magnetic moments. 

\section{The equilibrium state}\label{sec:eqb}
The first step in the linear response analysis using the kinetic equations is to determine the equilibrium distribution about which the perturbation analysis is carried out. Notice the analogy with the linear response analysis using the diagram technique, where the first step is the evaluate of the equilibrium Green's functions.

As mentioned earlier, we assume that the external potential $U(\rr) = m\omega_0^2 r^2/2$ is independent of time for $t<0^-$ and the system is assumed to have reached a thermal equilibrium state. It is easily to show that the CBV equation has a unique equilibrium solution given by:
\begin{multline}\label{eq:fhf}
n_\LE(p;r) =\\
\qquad\left\{\exp\left[\beta\left(\frac{p^2}{2m} + \Sigma_\LE(p;r) + \frac{1}{2}\,m\omega_0^2r^2-\mu\right)\right]+1\right\}^{-1},
\end{multline} 
where we have introduced the shorthand $\Sigma_\LE \equiv \Sigma^+[n_\LE]$. The above equation has to be solved self-consistenty along with the expression for the self-energy, Eq.~(\ref{eq:senhf}). It is easily verified that the above solution satisfies $I_c[n_0]=0$ and at the same time, it solves the left hand side of the CBV equation. The global chemical potential $\mu$ has to be found such that the equilibrium distribution function yields the correct number of trapped particles:
\begin{equation}
\int\mathrm{d}\Gamma\,n_0(\pp;\rr) = N,
\end{equation}
where we have defined the useful shorthand $\mathrm{d}\Gamma = \mathrm{d}^2\rr\,\mathrm{d}^2\pp/(2\pi)^2$. In the case of harmonic traps, it is useful to define the following scales coordinates:
\begin{align}\label{eq:scaled}
&\brr \equiv \frac{\rr}{r_0},\qquad r_0 \equiv [2N/(m\omega_0)^2]^{1/4},\nonumber\\
&\bpp \equiv \frac{\pp}{p_0},\qquad p_0 \equiv [2N(m\omega_0)^2]^{1/4}.
\end{align}
In the scaled coordinates, the equation for the particle number is $\int\mathrm{d}\bar{\Gamma}\,n_0(\bpp;\brr) = 1/2$, where $\mathrm{d}\bar{\Gamma} \equiv \mathrm{d}^2\brr\,\mathrm{d}^2\bpp/(2\pi)^2$. The equilibrium distribution function also reads as:
\begin{multline}\label{eq:fhf2}
n_\LE(\bpp;\brr) =\\
\qquad\left\{\exp\left[\bar{\beta}\left(\frac{\bar{p}^2 + \bar{r}^2}{2} + \bar{\Sigma}_\LE(\bar{r};\bar{p}) -\bar{\mu}\right)\right]+1\right\}^{-1},
\end{multline}
where $\bar{\beta} = \sqrt{2N}\omega_0/(k_B T)$, $\bar{\mu} = \mu/(\sqrt{2N}\omega_0)$, and the dimensionless and scaled self-energy functional is:
\begin{multline}
\bar{\Sig}^+[n](\bpp;\brr,t)=\omega_0^{-1}\int\mathrm{d}\bar{\Gamma}'\Big[\sqrt{2N}\,\mathcal{V}[r_0(\brr-\brr')]\\
-\delta^2(\brr-\brr')\tilde{\mathcal{V}}[p_0(\bpp-\bpp')]\Big]n(\bpp';\brr',t).
\end{multline}
The motivation for the introduced dimensionless coordinates can be understood by investigating the non-interacting equilibrium solution at low temperatures. In this case, one can easily find analytic solutions for the equilibrium density, $\bar{n}_0^{(0)}(\bar{r})$:
\begin{equation}\label{eq:dens0}
\bar{n}^{(0)}_0(\bar{r}) \equiv \int\mathrm{d}^2\bpp\,\bar{n}_0(\bar{p};\bar{r}) = \log\left[1 + e^{\bar{\beta}(\bar{\mu} - \bar{r}^2/2)}\right]/(2\pi\bar{\beta}).
\end{equation}
Integrating over $\brr$, we obtain the following equation for the chemical potential:
\begin{equation}\label{eq:mueq}
\bar{\mu}^2 + \frac{\pi^2}{3}\,\bar{T}^2 + 2\,\bar{T}^2\,\mathrm{Li}_2[-\exp(-\bar{\mu}/\bar{T})]=1,
\end{equation}
where $\bar{T} = \bar{\beta}^{-1}$. At low temperatures, the above equation admits the solution $\bar{\mu} = 1 - \pi^2 \bar{T}^2/6 + \mathcal{O}(e^{-\bar{\beta}})/\bar{\beta}^2$. The zero-temperature Thomas-Fermi radius of the cloud is easily obtained from Eq.~(\ref{eq:dens0}), yielding $R_{\mathrm{TF}}^{(0)} = [2\sqrt{2N}/(m\omega_0)]^{1/2} \equiv \sqrt{2}\,r_0$. Also, the Fermi momentum at the center of the trap is given by $p_F^{(0)} = [2\sqrt{2N}(m\omega_0)]^{1/2} \equiv \sqrt{2}\,p_0$. We note that $N$ does not appear explicitly in the above expression, and the expressions look formally similar. Moreover, the equilibrium distribution function has almost a finite support of radius $\mathcal{O}(1)$ in the scaled coordinates at low temperatures (beyond which it becomes exponentially small).

Once we take the interactions into account, we can no longer obtain analytic solutions and will have to find the equilibrium distribution function numerically. It is useful to investigate the effect of non-local Hartree energy (the first term in Eq.~\ref{eq:senhf2}) before we move on. Carrying out the momentum integration, Hartree contribution of the self-energy can be expressed just as a function of the density:
\begin{equation}
\bar{\Sig}_H^+[n](\brr,t)=\omega_0^{-1}\int\mathrm{d}^2\brr'\,\sqrt{2N}\,\mathcal{V}(r_0\brr')\,n(\brr-\brr',t).
\end{equation}
Observing that the density function is only appreciably large in a region of size $\mathcal{O}(1)$ in the scaled coordinates and the appearance of $r_0 \sim N^{1/4}$ in the argument of interaction potential, the above integral is expected to only depend of the values of the density within a small region of size $\sim N^{-1/4}$ about $\brr$. Assuming that $n(\rr)$ is a smooth function, we may expand $n$ to quadratic order about $\brr$, yielding:
\begin{multline}
\bar{\Sig}_H^+[n](\brr,t)\approx\omega_0^{-1}\int\mathrm{d}^2\brr'\,\sqrt{2N}\,\mathcal{V}(r_0\brr')\,\Big[n(\brr,t)\\
- \brr'\cdot\nabla n(\brr,t) + \bar{r}'_\alpha\bar{r}'_\beta\partial_\alpha\partial_\beta n(\brr,t)/2\Big].
\end{multline}
The first contribution is the usual local density approximation (LDA) expression:
\begin{align}
\bar{\Sig}_H^+[n]^\mathrm{LDA}(\brr,t) &\equiv \sqrt{2N}\omega_0^{-1} n(\brr,t)\int\mathrm{d}^2\brr'\,\,\mathcal{V}(r_0\brr')\nonumber\\
&=\frac{\tilde{\mathcal{V}}(0)}{m\omega_0^2}\,n(\brr,t).
\end{align}
The gradient term vanishes due to the isotropy of $\mathcal{V}(\rr)$. The quadratic term is dominated by the long-range behavior of $\mathcal{V}(\rr)$ assuming that the short-range part of $\mathcal{V}(\rr)$ is integrable (which is the case for dipolar interactions, see Eq.~\ref{eq:Vrsmall}). Observing that the Hessian matrix of the density is also $\mathcal{O}(1)$ in the scaled coordinates, we easily find that the quadratic density variations yield a correction that scales like $N^{1/2-\alpha/4}$ for a potential with power-law tail $\mathcal{V}(\rr) \sim r^{-\alpha}$. For dipolar interactions, $\alpha=3$ and we find that the beyond LDA corrections scale like $N^{-1/4}$ and become irrelevant for large $N$. Note that if we were dealing with an electron gas ($\alpha=1$), such corrections would grow larger with $N$ and had to be retained. This is the reason that one has to treat the Coulomb interactions in its full non-local from when studying the transport in plasmas; on the same note, we remark that the physics of Landau damping is expected to be absent with dipolar fermions in the thermodynamic limit. In the remainder of this paper, we treat the Hartree potential in the LDA approximation and use the following local self-energy functional instead:
\begin{align}\label{eq:senhf2}
\bar{\Sig}^+_{\mathrm{LDA}}[n](\bpp;\brr,t)&=\omega_0^{-1}\int\frac{\mathrm{d}^2\bpp'}{(2\pi)^2}\Big[\tilde{\mathcal{V}}(0)-\tilde{\mathcal{V}}[p_0(\bpp-\bpp')]\Big]\nonumber\\
&\times n(\bpp';\brr',t)\nonumber\\
&=\lambda_d\int\frac{\mathrm{d}^2\bpp'}{(2\pi)^2}\,u(|\bpp-\bpp'|,\eta)\,n(\bpp';\brr',t).
\end{align}
In the last last, we have defined:
\begin{align}\label{eq:udef}
u(x,\eta) &= 2\pi x\,\mathrm{Erfcx}\left(\frac{x\eta}{\sqrt{2}}\right),
\end{align}
where $\mathrm{Erfcx}(x)\equiv e^{x^2}\mathrm{Erfc}(x)$. The dimensionless parameters $\lambda_d$ and $\eta$ were defined earlier (Eq.~\ref{eq:defs}) Note that the dependence on $N$ enters the equations only through these two parameters.\\

We obtain the equilibrium distribution function using a simple iterative numerical method as follows: at the initial step, we set $\bar{\Sigma}_0 = 0$ and define the function $n_0(\bar{\mu}) \equiv n[\bar{\Sigma}_0,\bar{\mu}]$ (i.e. the distribution function obtained using the self-energy $\bar{\Sigma}_0=0$ and chemical potential $\bar{\mu}$). Keeping $\Sigma_0$ fixed, we find $\mu_0$ such that $\int \mathrm{d}\bar{\Gamma}\,n_0(\mu_0) = 1/2$. To proceed from $i$'th step to $(i+1)$'th step, we set $\bar{\Sigma}_{i+1} = \bar{\Sigma}^+[n_i]$, define $n_{i+1}(\bar{\mu}) \equiv n[\bar{\Sigma}_{i+1},\bar{\mu}]$ and for a fixed $\bar{\Sigma}_{i+1}$, we find $\bar{\mu}_{i+1}$ such that $\int \mathrm{d}\Gamma'\,n_{i+1}(\bar{\mu}_{i+1}) = 1/2$. At the end of the step, we set $n_{i+1} \rightarrow (1-\lambda) n_{i} + \lambda\,n_{i+1}$, where $0<\lambda<1$. The last step is to stabilize the iterative procedure and damp possible oscillations that prevent convergence. We found the above iterative procedure to converge to the solution in less than ten steps within a relative error tolerance of $10^{-8}$.\\
\begin{figure}[hb!]
\center
\includegraphics[scale=0.65]{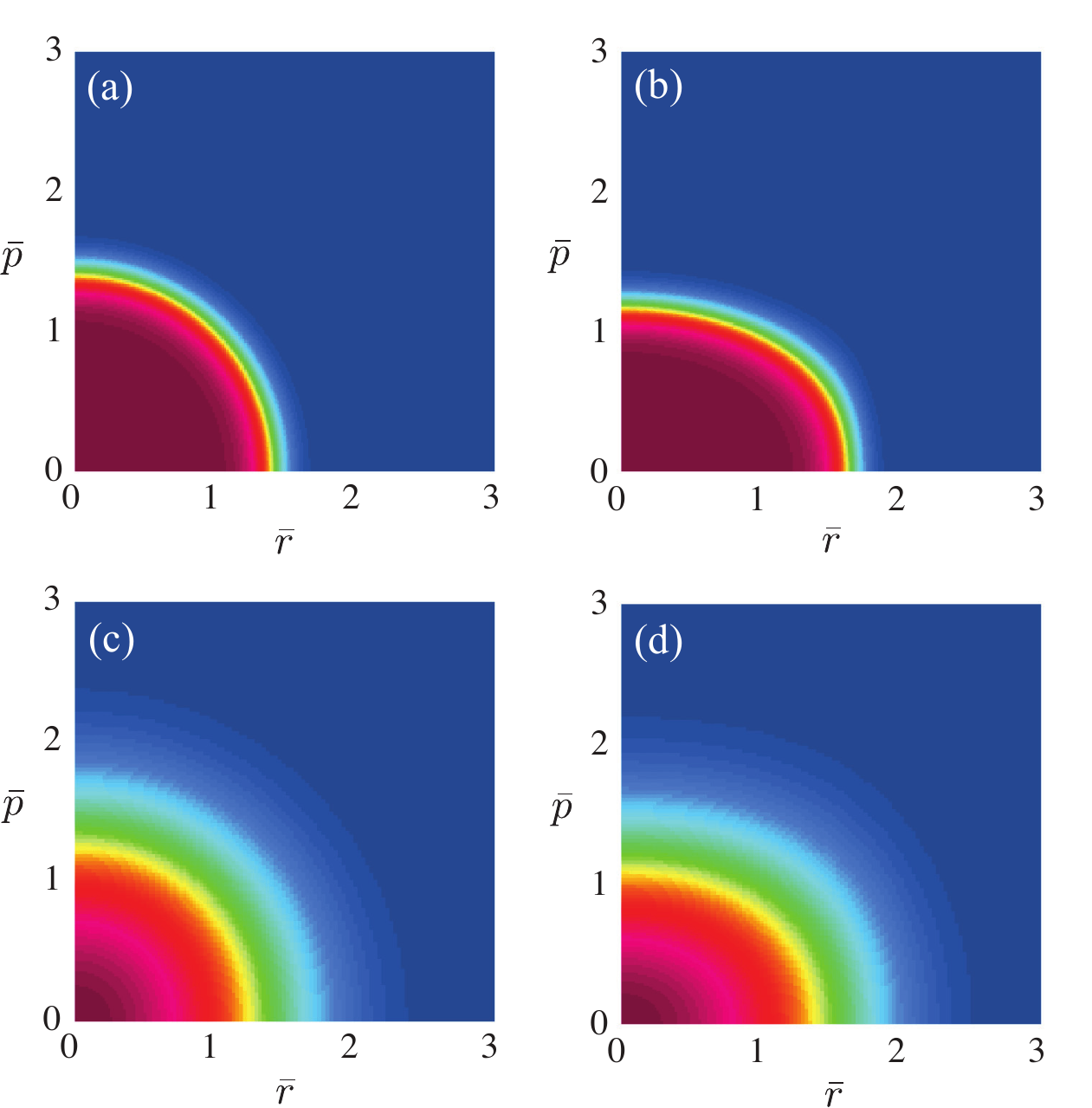}
\caption{(Color online) Equilibrium quasiparticle distribution function of quasi-2D dipolar fermions for various temperatures and interactions strengths ($\omega_z = 2\pi \times 23\,\mathrm{kHz}$, $\omega_0 = 2\pi \times 36\,\mathrm{Hz}$, $N=2200$). (a) $T/T_F = 0.1$, $\lambda_d = 0$, (b) $T/T_F = 0.1$, $\lambda_d = 1$, (c) $T/T_F = 0.5$, $\lambda_d = 0$, (d) $T/T_F = 0.5$, $\lambda_d = 1$. Red and blue regions correspond to occupied and empty states respectively.}
\label{fig:fwig}
\end{figure}

Fig.~\ref{fig:fwig} shows the equilibrium quasiparticle distribution function as a function of $\bar{p}$ and $\bar{r}$ for several temperatures and interaction strengths. As one expects, the presence of interactions, which are effectively repulsive, results in the expansion of the gas in the trap (compare panels a and b) and thermal fluctuations smear the Fermi surface (compare panels a and c).

\begin{figure}
\center
\includegraphics[scale=0.65]{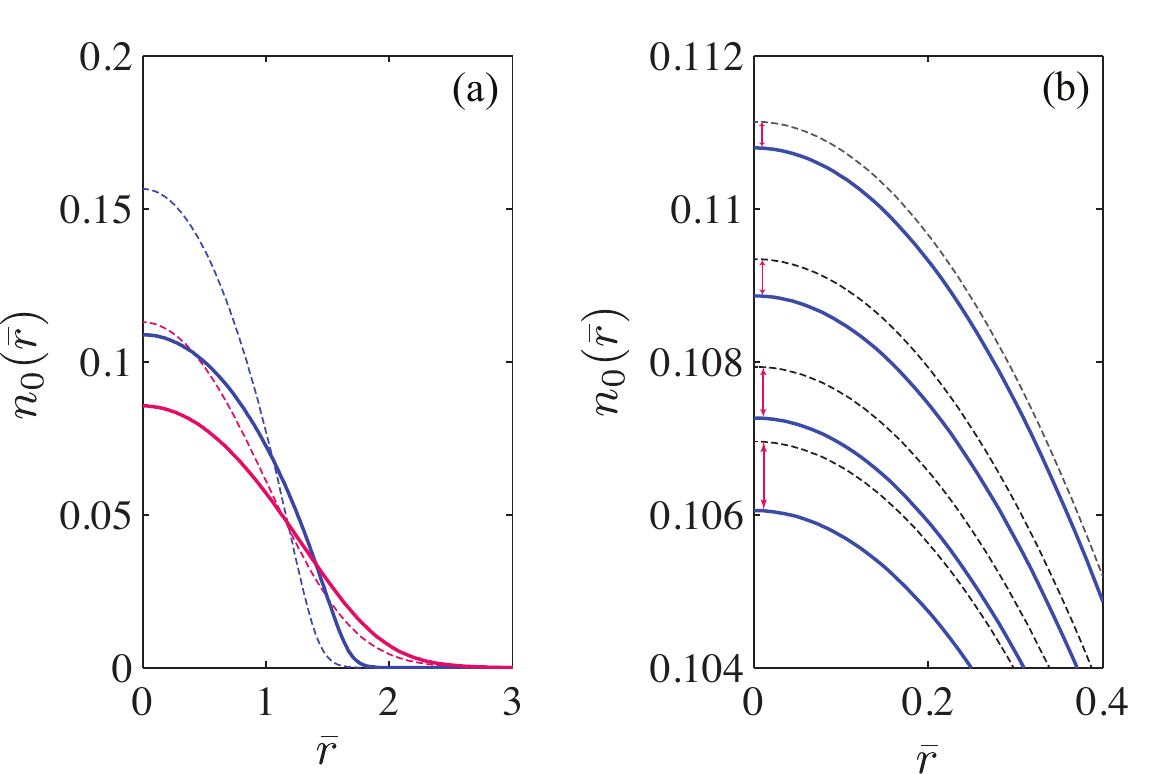}
\caption{(Color online) Equilibrium quasiparticle density of quasi-2D dipolar fermions ($\omega_z = 2\pi \times 23\,\mathrm{kHz}$, $\omega_0 = 2\pi \times 36\,\mathrm{Hz}$). (a) dashed and solid lines correspond to the non-interacting ($\lambda_d = 0$) and interacting ($\lambda_d = 1$), blue (top) and red (bottom) lines correspond to $T/T_F = 0.1$ and $0.5$ respectively. In all cases, $N=2200$. (b) A comparison between the LDA (solid lines) and non-local (dashed lines) Hartree self-energy functionals. From bottom to top, $N=500,\,1000,\,2200,$ and $5000$. $\lambda_d = 1$ and $T/T_F = 0.1$ in all cases. The non-local corrections are clearly negligible and become smaller for larger choices of $N$.}
\label{fig:dens}
\end{figure}

The equilibrium density is shown in Fig.~\ref{fig:dens}a. The exponentially decaying tail of the density at higher temperatures and reduction of the density at the center of the trap at low temperatures due to repulsive interactions can be clearly seen. We also compare the LDA and non-local Hartree self-energy functions in Fig.~\ref{fig:dens}b for various number of particles in the trap. The relative correction to the density is in the order of $10^{-3}$ for realistic number of trapped particles and as argued earlier, becomes smaller for larger system sizes.\\

Knowing the equilibrium state, we can move on to the investigation of the low-lying collective excitations about the equilibrium state. To this end, we discuss the linear response theory of the CBV equation in the next section.

\section{Analysis of the collective modes:\\linear response theory of the collisional Boltzmann-Vlasov equation}\label{sec:linresBV}
The linear response can be evaluated using the CBV equation by introducing a perturbation to the external potential, linearizing the resulting equation about deviations from the global equilibrium state, $\delta n(\bpp;\brr,t) \equiv n(\bpp;\brr,t)-n_\LE(\bar{p};\bar{r})$ and solving the resulting linear integro-differential equation. The merits of this approach compared to the diagram technique is the possibility of obtaining approximate solutions using variational methods.

Since we are mostly concerned with low temperatures here, it is fruitful to introduce the following ansatz for $\delta n$:
\begin{align}\label{eq:fvar}
\delta n(\bpp;\brr,t) &\equiv \theta(t)\,\Delta_\LE(\bar{p};\bar{r})\,\Phi(\bpp;\brr,t),
\end{align}
where $\Delta_\LE \equiv \partial n_\LE/\partial \bar{\mu} = \bar{\beta} n_\LE(1-n_\LE)$. We remark that the above ansatz is not restrictive at the moment and since for $T>0$, $\Delta_\LE$ has a unbounded support and any arbitrary deviation from the equilibrium state can be represented with a proper choice of $\Phi$. The only exception is $T=0$ where $\Delta_0$ restricts the deviations to the local Fermi surface, which is in fact a favorable feature. Since the low-lying collective excitations essentially depend on the particle-hole excitations about the Fermi surface and observing that the pre-factor $\Delta_0$ peaks about the local Fermi surface, we expect the solution of the linearized CBV equation to be representable with a smooth choice of $\Phi$~\cite{LL}. As we shall see, this feature allows us to construct decent approximate solutions by choosing a linear combination of smooth functions as a variational ansatz for $\Phi$. Plugging this ansatz into the CBV equation, expanding to first order in $\Phi$ and taking a Fourier transform in time, we obtain the following linear integral equation for $\Phi(\bpp;\brr,\omega)$:
\begin{multline}\label{eq:phieq}
-i\,\bar{\omega}\,\Delta_\LE \Phi + \DD[\Phi] - \II[\Phi]=\\
-(2N)^{-\frac{1}{2}}\{n_\LE,\delta U(r_0\brr,\omega)\},
\end{multline}
where $\{\phi,\psi\} \equiv \drr \phi \cdot \dpp \psi - \dpp \phi \cdot \drr \psi$ is the ordinary Poisson bracket and $\bar{\omega} \equiv \omega/\omega_0$. $\DD[\Phi]$ describes the collisionless self-consistent mean-field dynamics of quasiparticles:
\begin{align}\label{eq:Ddef}
\DD[\Phi] &= \Delta_\LE\{\Phi,\bar{\mathcal{H}}_0\} + \{n_\LE,\bar{\Sigma}[\Delta_\LE \Phi]\}\nonumber\\
&=\Delta_\LE\{\Phi + \bar{\Sigma}[\Delta_0\Phi],\bar{\mathcal{H}}_0\},
\end{align}
where $\mathcal{\bar{H}}_0 = (\bar{p}^2 + \bar{r}^2)/2 + \bar{\Sigma}_\LE$. To get the second line, we have used the identity $\{n_0,\mathcal{A}\} \equiv - \Delta_0\{\bar{\mathcal{H}}_0,\mathcal{A}\}$ which can be easily proved by direct calculation and is valid for arbitrary $\mathcal{A}$. The first term describes the dynamics in the equilibrium mean-field. The second term describes the the mean-field generated by the deviations and is a consequence of our self-consistent treatment. $\II[\Phi]$ describes the collisional dynamics and reads as:
\begin{multline}\label{eq:Idef}
\II[\Phi] = -\frac{\bar{\beta}(2N)^{\frac{1}{2}}}{2}\int \dd{\bpp_1}\,\dd{\bpp'}\,\dd{\bpp'_1}\,(2\pi)^2\delta^2(\Delta \mathbf{\bar{P}})\\
\times(2\pi)\delta(\Delta \bar{E})\,|\bar{\MM}|^2\,\mathrm{S}\{\Phi\}\,n_\LE n_{\LE,1}(1-n'_\LE)(1-n'_{\LE,1}),
\end{multline}
where $\Delta \bar{E} \equiv \bar{\mathcal{H}}_0(\bpp,\brr) + \bar{\mathcal{H}}_0(\bpp_1,\brr) - \bar{\mathcal{H}}_0(\bpp',\brr) - \bar{\mathcal{H}}_0(\bpp'_1,\brr)$, $\Delta \bar{\mathbf{P}} \equiv \bpp + \bpp_1 - \bpp' - \bpp'_1$, $\bar{\mathcal{M}} = m(\tilde{\mathcal{V}}[p_0(\bpp-\bpp')]-\tilde{\mathcal{V}}[p_0(\bpp-\bpp'_1)])$, and $\mathrm{S}[\Phi] \equiv \Phi(\bpp;\brr,\omega) + \Phi(\bpp_1;\brr,\omega) - \Phi(\bpp';\brr,\omega) - \Phi(\bpp'_1;\brr,\omega)$. Note that we have included Hartree-Fock self-energy corrections in the collisions which is required to be consistent with the inclusion of mean-field effects in the collisionless dynamics. Specializing to the case of dipole-dipole interactions, we get:
\begin{equation}
|\bar{\mathcal{M}}|^2 = \lambda_d^2\left[u(|\bpp-\bpp'|,\eta)-u(|\bpp-\bpp'_1|,\eta)\right]^2.
\end{equation}

Formally, the solution of Eq.~(\ref{eq:phieq}) can be written as:
\begin{equation}\label{eq:Phif}
\Phi = -\left(-i\,\bar{\omega}\Delta_\LE + \DD - \II\right)^{-1}\frac{\{n_0,\delta U(r_0\brr,\omega)\}}{(2N)^{\frac{1}{2}}},
\end{equation}
and the linear response can be determined using Eq.~(\ref{eq:fvar}):
\begin{equation}
\langle O \rangle_t = \int\mathrm{d}\Gamma\int\,\frac{\mathrm{d}\omega}{2\pi}\,e^{-i\omega t}\Delta_\LE(\bar{p};\bar{r})\,\Phi(\bpp;\brr,\omega^+)\,O(\pp;\rr).
\end{equation}
The difficulty is in inverting the operator appearing in the parenthesis in Eq.~(\ref{eq:Phif}). Decent approximate solutions can however be found using a variational technique known as the method of moments. To this end, we restrict the solution space of Eq.~(\ref{eq:phieq}) to a subspace spanned by a set of basis functions (also known as moment functions) $\{\phi_\alpha(\bpp;\brr)\}$ and expand $\Phi$ and $\delta U$ in this basis:
\begin{align}
\Phi(\bpp;\brr,\omega) &= \sum_\alpha \Phi_\alpha(\omega)\,\phi_\alpha(\bpp;\brr),\nonumber\\
(2N)^{-\frac{1}{2}}\delta U(r_0\brr,\omega) &= \sum_\alpha \delta U_\alpha(\omega)\,\phi_\alpha(\bpp;\brr).
\end{align}
Plugging this ansatz into Eq.~(\ref{eq:phieq}) and evaluating the moments of the resulting equation with respect to each of the basis functions, i.e. multiplying the sides of the CBV equation by each of the basis functions and integrating over $\bar{\rr}$ and $\bar{\pp}$, we find a closed set of linear equations for the coefficient set $\{\Phi_\alpha\}$:
\begin{multline}\label{eq:mateq}
-i\bar{\omega}\Langle\phi_\beta\phi_\alpha\Rangle\Phi_\alpha(\omega) + \Langle\phi_\beta\{\phi_\alpha,\mathcal{\bar{H}}_0\}\Rangle\left[\delta U_\alpha(\omega) + \Phi_\alpha(\omega)\right]\\
+ \Langle\phi_\beta\{\bar{\Sigma}[\Delta_\LE\phi_\alpha],\bar{\mathcal{H}}_0\}\Rangle\Phi_\alpha(\omega) - \II_{\beta\alpha}\Phi_\alpha(\omega)=0,
\end{multline}
where we have defined the ``$\Delta_0$-average'' as:
\begin{equation}
\Langle \mathcal{A}(\bpp;\brr) \Rangle \equiv \int\mathrm{d}\bar{\Gamma}\,\Delta_\LE(\bpp;\brr)\mathcal{A}(\bpp;\brr).
\end{equation}
Summation over repeated indices is implied in Eq.~(\ref{eq:mateq}). The matrix elements of the collision integral, $\II_{\alpha\beta} \equiv \int\mathrm{d}\bar{\Gamma}\,\phi_\alpha\II[\phi_\beta]$ can be put in the following symmetric form using the symmetry properties of the collision integral kernel:
\begin{multline}\label{eq:Imat}
\II_{\alpha\beta} = -\frac{\bar{\beta}(2N)^\frac{1}{2}}{8}\int \mathrm{d}^2\brr\int\dd{\bpp}\,\dd{\bpp_1}\,\dd{\bpp'}\,\dd{\bpp'_1}\\
\times(2\pi)\delta(\Delta \bar{E})\,(2\pi)^2\delta^2(\Delta \bar{\mathbf{P}})\,|\bar{\MM}|^2\,\mathrm{S}[\phi_\alpha]\,\mathrm{S}[\phi_\beta]\\
\times \,n_\LE n_{\LE,1}(1-n'_\LE)(1-n'_{\LE,1}).
\end{multline}
The first term on the second line of Eq.~(\ref{eq:mateq}) can be put in a more useful form using the identity $\phi_\beta\{\bar{\Sigma}[\Delta_\LE\phi_\alpha],\bar{\mathcal{H}}_0\} = \{\phi_\beta\Sigma[\Delta_\LE\phi_\alpha],\bar{\mathcal{H}}_0\} - \Sigma[\Delta_\LE\phi_\alpha]\{\phi_\beta,\bar{\mathcal{H}}_0\}$. Taking the $\Delta_0$-average of both sides on this identity, we find that the first term on the left hand side vanishes. To see this, note that $\Langle \{\psi,\bar{\mathcal{H}}_0\} \Rangle = \int \mathrm{d}\bar{\Gamma}\,\Delta_\LE\{\psi,\bar{\mathcal{H}}_0\} = \int \mathrm{d}\bar{\Gamma}\,\{\Delta_\LE\psi,\bar{\mathcal{H}}_0\}$ for arbitrary $\psi$. The last equality holds since $\{\Delta_\LE,\bar{\mathcal{H}}_0\} = 0$. Since $\Delta_\LE \rightarrow 0$ exponentially fast for large $\rr$ or $\pp$, the divergence theorem implies that the last integral vanishes as long as $\psi$ is exponentially bounded. Here, $\psi = \phi_\beta\bar{\Sigma}[\Delta_\LE\phi_\alpha]$ which is in fact exponentially bounded. Finally, Eq.~(\ref{eq:mateq}) can be put in the following matrix form:
\begin{equation}\label{eq:mateq2}
(-i\bar{\omega}\mathsf{M} + \mathsf{H}_0 - \mathsf{\Sigma} - \mathsf{I}_c)\mathbf{\Phi}(\omega) = - \mathsf{H}_0\,\mathbf{\delta U}(\omega),
\end{equation}
where:
\begin{align}\label{eq:mateq3}
(\mathsf{M})_{\alpha\beta} &= \Langle\phi_\alpha\phi_\beta\Rangle,\nonumber\\
(\mathsf{H}_0)_{\alpha\beta} &= \Langle\phi_\alpha\{\phi_\beta,\bar{\mathcal{H}}_0\}\Rangle,\nonumber\\
(\mathsf{\Sigma})_{\alpha\beta} &= \Langle\bar{\Sigma}[\Delta_\LE\phi_\beta]\{\phi_\alpha,\bar{\mathcal{H}}_0\}\Rangle,\nonumber\\
(\mathsf{I}_c)_{\alpha\beta} &= \II_{\alpha\beta},
\end{align}
and $\mathbf{\Phi}(\omega)$ and $\mathbf{\delta U}(\omega)$ are the vectors with entries $\Phi_\alpha(\omega)$ and $\delta v_\alpha(\omega)$ respectively. If the observable $O(\bpp;\brr)$ is also expressible in terms of the basis functions, $O(\bpp;\brr) = \sum_\alpha O_\alpha \phi_\alpha(\bpp;\brr)$, then the linear response can be expressed as:
\begin{align}\label{eq:respphi}
\langle O \rangle_\omega &= \int\mathrm{d}\bar{\Gamma}\,O_\beta\phi_\beta\,\Delta_\LE \Phi_\alpha(\omega^+) \phi_\alpha\nonumber\\
&=\mathbf{O}^T \mathsf{M}\,\mathbf{\Phi}(\omega^+).
\end{align}
Eqs.~(\ref{eq:mateq2})-(\ref{eq:respphi}) are similar to the analysis given in Ref.~\cite{Chiacchiera2011} for the case of $s$-wave fermions. Here, however, we have an additional matrix $\mathsf{\Sigma}$ that accounts for the self-energy corrections.

It is useful to express the ``evolution matrix'', which we define to be $\mathsf{E} \equiv \mathsf{M}^{-1}(\mathsf{H}_0 - \mathsf{\Sigma} - \mathsf{I}_c)$, in its diagonal basis:
\begin{equation}\label{eq:Ediag}
\mathsf{E} \equiv i\,\mathsf{V} \mathsf{\Omega} \mathsf{V}^{-1},
\end{equation}
where $\mathsf{\Omega}$ is a diagonal matrix. Note that in general, $\mathsf{E}$ is a not a Hermitian operator and may have complex eigenvalues. Moreover, it is a non-normal matrix and therefore, its eigenvectors are not orthogonal. As a side note, the same non-normality feature of the linearized BV equation for plasmas lead to Landau damping~\cite{Kampen1955}. However, as we argued earlier, dipole-dipole interactions are not long-ranged enough to give rise to such effects.

Using diagonal form of the evolution matrix, Eq.~(\ref{eq:mateq2}) can be expressed as:
\begin{equation}\label{eq:phidiag}
\mathbf{\Phi}(\omega) = -i \mathsf{V}\,\frac{1}{\bar{\omega} - \mathsf{\Omega}}\,\mathsf{V}^{-1}\mathsf{M}^{-1}\mathsf{H}_0\,\delta\mathbf{U}(\omega).
\end{equation}
The real and imaginary parts of $\mathsf{\Omega}$ determine the oscillation frequency and damping of the corresponding eigenmodes. Clearly, not all of the eigenmodes are expected to get excited in response to a given perturbation. This becomes particularly important when one is dealing with a large variational basis set. In such cases, as we will see later, the evolution matrix will have poles which are very close to each other on the complex frequency plane and it is not a priori clear which one(s) and in what proportion contribute to the response of the system. Using the linear response formalism, however, we don't have to deal with this question separately. Using Eqs.~(\ref{eq:respphi}) and~(\ref{eq:phidiag}), we get:
\begin{align}
\langle O \rangle_\omega &= \sum_\alpha\frac{r_\alpha(\omega)}{\omega - \Omega_\alpha},\nonumber\\
r_\alpha(\omega) &= -i[\mathsf{V}^T\mathsf{M}\mathbf{O}]_\alpha [\mathsf{V}^{-1}\mathsf{M}^{-1}\mathsf{H}_0\,\delta\mathbf{U}(\omega)]_\alpha,
\end{align}
i.e. the residues $r_\alpha$ can be expressed in terms of the known matrices. Note that for a Dirac delta perturbation in time, $\delta \mathbf{U}(\omega)$ in constant and independent of $\omega$ and so is $r_\alpha(\omega)$.\\

Our goal here is to evaluate the linear responses accurately within the approximations made so far. In practice, the reliability of the approximate linear response functions obatined using the method of moments depends on one's choice of the basis functions. This choice can be motivated by the symmetries of the perturbing potential and the equilibrium state. Here, the trap potential is assumed to be isotropic and it is easy to see that $[-i\omega\Delta_\LE + \DD - \II,S_z] = 0$ , where $S_z \equiv S_z^{(r)} + S_z^{(p)}$, and $S_z^{(r)} = i(x \partial_y - y \partial_x)$ and $S_z^{(p)} = i(p_x \partial_{p_y} - p_y \partial_{p_x})$ are the rotation operators in the coordinate and momentum space respectively. Therefore, if $\delta U$ lies in a certain eigenspace of $S_z$, so will the solution of the linearized equation $\Phi$ and one may choose the basis functions within the same eigenspace. Another symmetry which is preserved by the CBV equation is the reflection symmetry. More explicitly, defining the reflection operator as $R_x\phi(p_x,p_y;x,y) = \phi(-p_x,p_y;-x,y)$, it is easy to show that the linearized evolution operator commutes with $R_x$ as well. We will utilize these observations to define appropriate (and extensible) basis sets for monopole and quadrupole oscillations.\\

Before we attempt to present accurate solutions which inevitably requires heavy numerical calculations, we find it useful to make simple analytical predictions using a small basis set as first step. We use the scaling ansatz approach to find such a basis set and neglect self-energy corrections to simplify the calculations. We extend the basis set and include self-energy corrections afterwards and discuss the nature and importance of corrections that arise.

\section{Linearized scaling ansatz analysis}\label{sec:scans}
The scaling ansatz provides a simple and intuitive picture of the collective excitations of confined gases. This method has been applied to various system in both isotropic and anisotropic traps, including Bose gases below and above $T_c$, $s$-wave and dipolar fermions in the collisionless and hydrodynamics regimes~\cite{Scaling,Lima2010R,Lima2010,Sogo2009,Abad2012}. Here, we apply the method to the collisional Boltzman equation which as we shall see, allows us to study both CL and HD limits as well as transition from one regime to the other.

In this method, one assumes that the non-equilibrium quasiparticle distribution function can be approximately described as a scaled copy of the equilibrium distribution:
\begin{align}\label{eq:fscl}
n_\mathrm{sc.}(\bpp;\brr,t) \equiv \frac{1}{\prod_i(b_i \phi_i)}\,n_\LE\big[\phi_i^{-1}(\pb_i - \dot{b}_i \rb_i/b_i); \rb_i/b_i\big],
\end{align}
where $b_i$ and $\phi_i$ ($i=x,y$) are time-dependent scalings of positions and temperature. The pre-factor is to ensure conservation of particle number. The equilibrium solution corresponds to the choice $b_x = b_y = \phi_x = \phi_y = 1$. Introducing the following re-parametrization of the scaling variables:
\begin{align}
b_x(t) &= 1 + \bar{\lambda}(t) + \lambda(t), \quad b_y(t) = 1 + \bar{\lambda}(t) - \lambda(t),\nonumber\\
\phi_x(t) &= 1 + \bar{\nu}(t) + \nu(t), \quad \phi_y(t) = 1 + \bar{\nu}(t) - \nu(t),
\end{align}
and expanding Eq.~(\ref{eq:fscl}) to first order in $\lambda$, $\bar{\lambda}$, $\nu$ and $\bar{\nu}$, we get:
\begin{multline}\label{eq:df2}
\delta n_\mathrm{sc} \approx -2(\bar{\lambda} + \bar{\nu})n_\LE + \Delta_\LE\big[\dot{\bar{\lambda}}\,\brr\cdot\bpp + \bar{\nu}\,\pb^2 + \bar{\lambda}\,\rb^2\big]\\
+ \Delta_\LE\big[\dot{\lambda}\,(\bar{x}\pb_x - \bar{y}\pb_y) + \nu\,(\pb_x^2 - \pb_y^2) + \lambda\,(\bar{x}^2 - \bar{y}^2)\big],
\end{multline}
where in the last equation, we have neglected self-energy corrections for simplicity and used the non-interacting equilibrium solution. Also, $\Delta_0 = \partial n_0 / \partial \bar{\mu} = \bar{\beta} n_0(1-n_0)$ as before. Here, ($\bar{\lambda}, \bar{\nu}$) and ($\lambda, \nu$) correspond to the isotropic (monopole) and anisotropic (quadrupole) scalings. Comparing the last equation with Eq.~(\ref{eq:fvar}), we can recognize the first and second set of terms in the brackets as $\Phi_\mathrm{mon}$ and $\Phi_\mathrm{quad}$, i.e. the variational ansatz that the linearized scaling ansatz provides for monopole and quadrupole modes respectively.

The first term in Eq.~(\ref{eq:df2}), which is a consequence of the normalization prefactor of the scaling ansatz, requires further discussion. First of all, we note that this term in only non-vanishing in the monopole case. Since quadrupole oscillations are purely anisotropic, they do not violate conservation of mass in the linear regime and therefore, no normalization results. The monopole oscillations as described by $\Phi_\mathrm{mon}$, however, violate the conservation of mass and the ansatz must be fixed with a counter term. The scaling ansatz fixes this defect with a uniform scaling of the distribution, leading to the first term in Eq.~(\ref{eq:df2}).

We argue that such an ansatz is not a particularly good choice once collisions are taken into account and must be avoided since it may lead to unphysical conclusions. It is generally understood that the non-equilbrium dynamics of degenerate Fermi gases are governed by excitations near the Fermi surface while the fermions deep inside the Fermi sea remain in place due to their large excitation energy gap. A global rescaling of the quasiparticle distribution, i.e. a uniform rescaling of quasiparticle occupations irrespective of their energy gap implies mobilization of all fermions with the same probably, including those which are deep inside the Fermi sea (the density of states is constant in two dimensions). This is clearly an unphysical assumption and may lead to unphysically large collision rates. We note that it is well-known that the kinetic description of the monopole mode in the absence of self-energy corrections, as we shall also show momentarily, must result in undamped oscillations due to conservation laws.

To fixed this defect, we remove the global normalization factor and address the issue of mass conservation by simply allowing the chemical potential to vary instead. This amounts to adding a term $\sim \delta \bar{\mu}(t)\,\partial n_0 / \partial \bar{\mu} = \Delta_0\,\delta\mu(t)$ to the ansatz, i.e. adding $\phi = 1$ to the monopole basis set. To summarize, we obtain:
\begin{equation}\label{eq:Phimon}
\Phi_\mathrm{mon} = \delta \mu(t) + c_1(t)\,\brr \cdot \bpp + c_2(t)\,\rb^2 + c_3(t)\,\pb^2,
\end{equation}
and:
\begin{equation}\label{eq:Phiquad}
\Phi_\mathrm{quad} = d_1(t)\,(\bar{x}\bar{p}_x - \bar{y}\bar{p}_y) + d_2(t)(\bar{x}^2 - \bar{y}^2) + d_3(t)(\bar{p}_x^2 - \bar{p}_y^2),
\end{equation}
where $\delta\mu(t)$, $c_i(t)$ and $d_i(t)$ are to be determined.

The determination of these unknown functions is usually done by plugging the ansatz into the kinetic equation, multiplying the resulting equation by each of the basis function and integrating over the phase space to obtain a close set of differential equations. This is equivalent to the formalism described in Sec.~\ref{sec:linresBV} and we prefer to do it in the notation of this paper as a warm-up for the later sections where we extend the basis set and include self-energy corrections. We finally note that the various terms appearing in Eqs.~(\ref{eq:Phimon}) and~(\ref{eq:Phiquad}) can be simply understood in physical terms. In particular, $\brr \cdot \bpp$ and $\bar{x}\bar{p}_x - \bar{y}\bar{p}_y$ in $\Phi_\mathrm{mon}$ and $\Phi_\mathrm{quad}$ correspond to isotropic and anisotropic macroscopic velocity fields, $\mathbf{v}_\mathrm{mon} \propto \mathrm{\brr}$ and $\mathbf{v}_\mathrm{quad} \propto \bar{x}\mathbf{e}_{x} - \bar{y}\mathbf{e}_{y}$.

\subsubsection{Monopole oscillations from the scaling ansatz}\label{sec:monscl}
Neglecting self-energy corrections, we have $\mathsf{\Sigma}=0$, $\bar{\mathcal{H}}_0 = (\rb^2 + \pb^2)/2$ and we easily obtain the following simple forms for $\mathsf{M}$ and $\mathsf{H}_0$:
\begin{equation}
\mathsf{M}^\mathrm{mon}_\mathrm{sc.} =
\left(\begin{tabular}{cccc}
$\Langle 1 \Rangle$ & $0$ & $\Langle \rb^2 \Rangle$ & $\Langle \pb^2 \Rangle$\\
$0$ & $\Langle (\brr \cdot \bpp)^2 \Rangle$ & $0$ & $0$\\
$\Langle \rb^2 \Rangle$ & $\Langle 0 \Rangle$ & $\Langle \rb^4 \Rangle$ & $\Langle \rb^2\pb^2 \Rangle$\\
$\Langle \pb^2 \Rangle$ & $0$ & $\rb^2 \pb^2$ & $\Langle \pb^4 \Rangle$
\end{tabular}\right),
\end{equation}
and:
\begin{equation}
\mathsf{H}_{0,\mathrm{sc.}}^\mathrm{mon} =
\left(\begin{tabular}{cccc}
$0$ & $0$ & $0$ & $0$\\
$0$ & $0$ & $2\Langle (\brr \cdot \bpp)^2 \Rangle$ & $-2\Langle (\brr \cdot \bpp)^2 \Rangle$\\
$0$ & $\Langle \rb^2 \pb^2 - \rb^4 \Rangle$ & $0$ & $0$\\
$0$ & $\Langle \pb^4 - \rb^2 \pb^2\Rangle$ & $0$ & $0$
\end{tabular}\right),
\end{equation}
where the basis is chosen in the same order as appears in Eq.~(\ref{eq:Phimon}). The matrix elements of the collision integral vanish due to conservation of energy and momentum (see Eq.~\ref{eq:Imat}, and notice that $\mathrm{S}[1] = \mathrm{S}[\rb^2] = 0$, $\mathrm{S}[\pb^2] = 2\Delta \bar{E}$ and $\mathrm{S}[\brr \cdot \bpp] = \brr \cdot \Delta \bar{\mathbf{P}}$). Therefore, the oscillations will be undamped.\\

It is possible find find analytic expressions for the $\Delta_\LE$-averages appearing in the above matrices. However, using the relations $\Langle \rb^2 \Rangle = \Langle \pb^2 \Rangle $ and $\Langle \rb^4 \Rangle = \Langle \pb^2 \Rangle$ and $\Langle (\brr \cdot \bpp)^2 \Rangle = \Langle \rb^2 \pb^2 \Rangle/2$ (the first two of which are only valid for Harmonic traps), we find that they all factor out from the evolution matrix and we get:
\begin{equation}\label{eq:Emon}
\mathsf{E}^\mathrm{mon}_\mathrm{sc.} =
\left(\begin{tabular}{cccc}
$0$ & $0$ & $0$ & $0$\\
$0$ & $0$ & $2$ & $-2$\\
$0$ & $-1$ & $0$ & $0$\\
$0$ & $1$ & $0$ & $0$
\end{tabular}\right),
\end{equation}
a result which is independent of temperature. The monopole excitation operator can be expressed is $r^2$, which gives the ``excitation vector'' $\delta \mathbf{U} = (0, 0, 1, 0)^T$ in the monopole basis (see the definition of $\delta\mathbf{U}$ after Eq.~\ref{eq:mateq3}). Using Eq.~(\ref{eq:mateq2}), we finally find:
\begin{equation}
\Phi^\mathrm{mon}(\bpp;\brr,\omega) = \left[-2 i \omega (\brr \cdot \bpp) + 2(\rb^2 - \pb^2)\right]/(\bar{\omega}^2 - 4).
\end{equation}
The frequency of oscillations is given by the poles of the denominator, $\bar{\omega}^\mathrm{mon} = \pm 2$, which is a well-known result~\cite{Boltzmann1909}. We state without proof that extending the monopole basis has no effect on this result as long as self-energy corrections are neglected. In fact, it is a well-known fact that the full nonlinear Boltzmann equation (including collisions) admits an exact monopole solution with frequency $2\omega_0$~\cite{Boltzmann1909}. This is deeply related to the fact that the trap potential is harmonic and the particles have quadratic dispersion. Including interaction effects or changing the trap potential will both result in violation of this result.

We remark that besides the $\bar{\omega} = \pm 2$, the evolution matrix above admits two zero eigenvalues that correspond to eigenvector $\Phi \sim 1$ and $\Phi \sim \rb^2 + \pb^2$. Both of these eigenvectors correspond to unphysical excitations as they violate mass conservation. However, it is easy to see that both lie in the null space of $\mathsf{H}_{0,\mathrm{sc.}}^\mathrm{mon}$ at the same time. Therefore, in light of Eq.~(\ref{eq:phidiag}), they will never be excited regardless of one's choice of excitation operator $\delta\mathbf{U}$. 

\subsubsection{Quadrupole oscillations from the scaling ansatz}\label{sec:quadscl}
\begin{figure*}[ht!]
\includegraphics[scale=0.64]{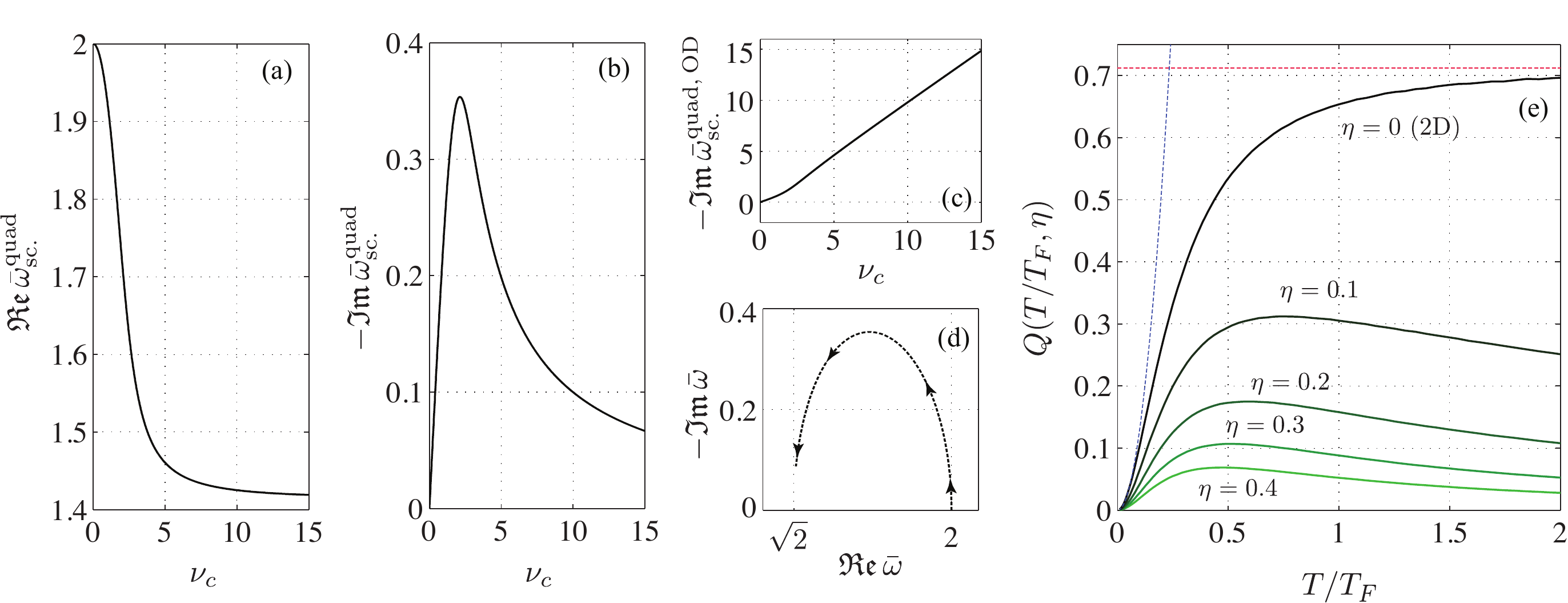}
\caption{(Color online) Frequency and damping of quadrupole oscillations of quasi-two-dimensional dipolar fermions in isotropic harmonic traps. (a) and (b): the frequency and damping of oscillations vs. $\nu_c$ respectively. (c) the damping rate of the overdamped component vs. $\nu_c$. (d) the evolution of the damped oscillatory pole on the complex plane upon increasing $\nu_c$ in the range $[0, 15]$. (e) $Q(T/T_F,\eta)$ as a function of $T/T_F$ for different values of $\eta \equiv (2N)^\frac{1}{4}(\omega_0/\omega_z)^\frac{1}{2}$. $Q$ is related to the dimensionless collisional relaxation rate $\nu_c$ as $v_c = N  (a_d/a_0)^2\,Q(T/T_F,\eta)$. The low temperature and high temperature asymptotes in the 2D limit are shown as blue and red dashed lines respectively.}
\label{fig:quadsc}
\end{figure*}

We find the following forms for $\mathsf{M}$ and $\mathsf{H}_0$ in the quadrupole basis:
\begin{equation}\label{eq:Mquad}
\mathsf{M}^\mathrm{quad}_\mathrm{sc.} = \frac{1}{2}
\left(\begin{tabular}{ccc}
$\Langle \rb^2 \pb^2 \Rangle$ & $0$ & $0$\\
$0$ & $\Langle \rb^4 \Rangle$ & $0$\\
$0$ & $0$ & $\Langle \pb^4 \Rangle$
\end{tabular}\right),
\end{equation}
and:
\begin{equation}\label{eq:Hquad}
\mathsf{H}_{0,\mathrm{sc.}}^\mathrm{quad} = \frac{1}{2}
\left(\begin{tabular}{ccc}
$0$ & $2 \Langle \rb^2 \pb^2 \Rangle$ & $-2\Langle \rb^2 \pb^2 \Rangle$\\
$-\Langle \rb^4 \Rangle$ & $0$ & $0$\\
$-\Langle \pb^4 \Rangle$ & $0$ & $0$
\end{tabular}\right).
\end{equation}
The order of basis functions is the same as it appars in Eq.~(\ref{eq:Phiquad}).
The only non-zero collision matrix element is $\mathscr{I}_{33}$, the rest of which vanish again due to conservation laws (see Eq.~\ref{eq:Imat}, and note that $\mathrm{S}[\bar{x}^2 - \bar{y}^2]=0$ and $\mathrm{S}[\bar{x}\bar{p}_x - \bar{y}\bar{p}_y] = (\bar{x}\mathbf{e}_x - \bar{y}\mathbf{e}_y)\cdot\Delta\bar{\mathbf{P}}$). The collision integral can be expressed as follows using the results of Appendices~\ref{sec:Ic} and~\ref{sec:Icquad} (in particular, see Eq.~\ref{eq:Iinteg}):
\begin{widetext}
\begin{align}\label{eq:I33}
\II_{33}^\mathrm{quad} =& -64\pi(2N)^\frac{1}{2}\lambda_d^2\,\bar{T}^5\int_0^\infty \rho^5\,\mathrm{d}\rho\int_0^{2\pi}\frac{\mathrm{d}\phi}{2\pi}\int_0^{2\pi}\frac{\mathrm{d}\phi'}{2\pi}\int_0^{\frac{\pi}{2}}\mathrm{d}\xi\,\sin^7\xi\,\cos \xi\int_0^{\frac{\pi}{2}}\mathrm{d}\nu\,\sin^5\nu\,\cos\nu\nonumber\\
&\times\sin^2(\phi-\phi')\left[\chi_1\,\mathrm{Erfcx}\left(2\eta\chi_1\sqrt{\bar{T}\rho}\right)-\chi_2\,\mathrm{Erfcx}\left(2\eta\chi_2\sqrt{\bar{T}\rho}\right)\right]^2\nonumber\\
&\times\left[\frac{1}{\cosh(\rho - \bar{\mu}/\bar{T})+\cosh(\rho\,\sin^2\xi\sin 2\nu\cos\phi)}\,\frac{1}{\cosh(\rho - \bar{\mu}/\bar{T})+\cosh(\rho\,\sin^2\xi\sin 2\nu\cos\phi')}\right],
\end{align}
\end{widetext}
where $\chi_1 = \sin\xi\,\sin\nu\,|\sin[(\phi-\phi')/2]|$ and $\chi_2 = \sin\xi\,\sin\nu\,|\cos[(\phi-\phi')/2]|$. The above integration can not be carried out analytically in general and requires a numerical treatment.  The analytical low $T$ and high $T$ asymptotics are given in Appendix~\ref{sec:Qasym}. Note that the (dimensionless) non-interacting chemical potential $\bar{\mu}$ is given implicitly by Eq.~(\ref{eq:mueq}) and depends only on the dimensionless temperature $\bar{T}$. Therefore, except for the prefactor, the above integral is a universal function of $\bar{T}$ and $\eta$. We define the ``collisional relaxation rate'' $\nu_c$ as:
\begin{align}\label{eq:nuc}
\nu_c^\mathrm{\quad} \equiv -\frac{2\II_{33}^\mathrm{quad}}{\Langle \pb^4\Rangle} &\equiv \frac{(2N)^{\frac{1}{2}}\lambda_d^2}{2}\,Q(\bar{T},\eta)\nonumber\\
&\equiv N\left(\frac{a_d}{a_0}\right)^2 Q(\bar{T},\eta).
\end{align}
The last equation also serves as the definition of the universal function $Q(\bar{T},\eta)$. The quadrupole excitation operator is $\delta\mathbf{U}^\mathrm{quad} = (0, 1, 0)^T$ in this basis and finally, a simple calculation similar to the monopole case yields:
\begin{multline}\label{eq:Phiquad2}
\Phi^\mathrm{quad}(\bpp;\brr,\omega) = \big[2\bar{\omega}(\nu_c-i\bar{\omega})(\bar{x}\bar{p}_x - \bar{y}\bar{p}_y) + 2i(\nu_c-i\bar{\omega})\\
\times(\bar{x}^2 - \bar{y}^2) + 2\bar{\omega}(\bar{p}_x^2 - \bar{p}_y^2)\big]/D^\mathrm{quad}(\bar{\omega},\nu_c),
\end{multline}
where $D^\mathrm{quad}(\bar{\omega},\nu_c)$, the ``quadrupole characteristic equation'' is:
\begin{equation}\label{eq:Dquad}
D^\mathrm{quad}(\bar{\omega},\nu_c) = \bar{\omega}(\bar{\omega}^2-4) + i\nu_c(\bar{\omega}^2 - 2).
\end{equation}
The roots of $D^\mathrm{quad}(\bar{\omega},\nu_c$ yield the frequency and damping of the quadrupole mode. We note that Eq.~(\ref{eq:Phiquad2}), along with the characteristic equation given above, are ``generic'' results in the sense that one obtains the same expression for quadrupole oscillations independent of the specific form of interactions. For instance, Refs.~\cite{Vichi2000} and ~\cite{GueryOdelin1999} obtain the same characteristic equation for $s$-wave fermions and a classical gas respectively. The model-specific details are encoded in the collisional relaxation rate $\nu_c$. Therefore, it is worthwhile to study the generic features of the quadrupolar oscillations from Eq.~(\ref{eq:Phiquad2}) in terms of $\nu_c$ as a first step. We return to the analysis of $\nu_c$ afterwards.\\

Two important limits can be recognized for quadrupole oscillations. The collisionless limit corresponds to $\nu_c \rightarrow 0$ and we find:
\begin{multline}
\lim_{\nu_c\rightarrow 0} \Phi^\mathrm{quad}(\bpp;\brr,\omega) \equiv \Phi^\mathrm{quad}_\mathrm{CL}(\bpp;\brr,\omega) =\\ \big[-2i\bar{\omega} (\bar{x}\bar{p}_x - \bar{y}\bar{p}_y) + 2 (\bar{x}^2 - \bar{y}^2) - 2(\bar{p}_x^2 - \bar{p}_y^2)\big]/(\bar{\omega}^2-4).
\end{multline}
Notice the formal similarity to the monopole case. In this limit, we obtain undamped oscillations at $\omega_\mathrm{CL}^\mathrm{quad} = 2 \omega_0$ which correspond to the free motion of particles in the trap. In the limit of very fast collisions, $\nu_c \rightarrow \infty$, we find:
\begin{multline}
\lim_{\nu_c\rightarrow \infty} \Phi^\mathrm{quad}(\bpp;\brr,\omega) \equiv \Phi^\mathrm{quad}_\mathrm{HD}(\bpp;\brr,\omega)\\
= \big[-2i\bar{\omega} (\bar{x}\bar{p}_x - \bar{y}\bar{p}_y) + 2(\bar{x}^2 - \bar{y}^2)\big]/(\bar{\omega}^2-2),
\end{multline}
which describe undamped oscillations at a frequency $\omega^\mathrm{quad}_\mathrm{HD} = \sqrt{2}\omega_0$. This is the well-known quadrupolar ``surface'' mode which may also be obtained by solving hydrodynamics equations for harmonically trapped gases~\cite{Griffin1997}. The absence of damping despite that fact that the collision rate is very high can be understood by noticing that the mean free path of particles becomes much smaller than the system size in this limit and the only role of collisions is to maintain a local equilibrium state for each element of the gas. The entropy is locally extermal in this limit and there is no room for dissipation. This is essentially the physics of first sound. In contrast to the first sound, however, the surface modes only show up in confined gases. Although we have neglected self-energy corrections here, it is known that surface modes have universal frequencies since they correspond to divergence-less flows and are entirely driven by the trap restoring force~\cite{Griffin1997}. We will observe this universality in the later section, where we include self-energy corrections and obtain the same oscillation frequency in the HD limit.\\

Except for the two ideal limits discussed so far, quadrupolar oscillations are generally damped for finite values of $\nu_c$. This is due to the fact that the collisions are not fast enough to maintain the local equilibrium and thus lead to dissipation. The oscillation frequency and damping rates can be found by analyzing the roots $D^\mathrm{quad}(\bar{\omega},\nu_c)$. Fig.~\ref{fig:quadsc}a-c show the real and imaginary parts of the poles as a function of $\nu_c$. In the limit $\nu_c \ll 1$, the poles are approximately at:
\begin{equation}\label{eq:approxpoles1}
\pm\left(2 - \frac{5\nu_c^2}{64}\right) - \frac{i\nu_c}{4} + \mathcal{O}(\nu_c^5),\quad -\frac{i\nu_c}{2} + i\mathcal{O}(\nu_c^3),
\end{equation}
corresponding to a damped oscillatory mode at a frequency slightly less than $2\omega_0$ and a damping rate of $\sim \nu_c\omega_0/2$. Additionally, there is an overdamped component with the same damping frequency to the leading order. In the other limit $\nu_c \gg 1$, we get:
\begin{equation}\label{eq:approxpoles2}
\pm\left(\sqrt{2} + \frac{3}{2\sqrt{2}\nu_c^2}\right) - \frac{i}{\nu_c} + \mathcal{O}(\nu_c^{-3}),\quad -i\nu_c + i\mathcal{O}(\nu_c^{-1}),
\end{equation}
which describe a damped oscillatory mode at a frequency slightly higher than $\sqrt{2}\omega_0$ and a damping rate of $\sim \nu_c^{-1}\omega_0$, accompanied by a (highly) overdamped component with a damping rate of $\omega_0\nu_c$. Studying the residues of the overdamped poles, we find that the contribution of the this component is $\propto \nu_c^2$ and $\propto \nu_c^{-2}$ to leading order in the CL and HD limits respectively. We associate the presence of such an overdamped component to the damping of initial excitations that lie far away from the local equilibrium. Finally, Fig.~\ref{fig:quadsc}d shows the evolution of the pole on the complex frequency plane upon increasing $\nu_c$. It starts off at $2\omega_0$, moves to the lower half plane and finally returns to the real axis at the hydrodynamic frequency $\sqrt{2}\omega$.\\

We conclude this section by studying the behavior of $Q(\bar{T},\eta)$, which is the universal function that yields the collisional relaxation rate $\nu_c$ for dipole-dipole interactions (Eq.~\ref{eq:nuc}). In the collision dominated regime (i.e. $\nu_c \gg 1$) where a viscous hydrodynamic description is admissible, the shear viscosity sum rule yiels $\nu_c$ as $\omega_0 \langle P/\eta_\mathrm{s} \rangle_\mathrm{trap}$, where $P$, $\eta_\mathrm{s}$ and $\omega_0$ are the local pressure, shear viscosity and the trap frequency respectively~\cite{Schaefer2012}. By $\langle \ldots\rangle_\mathrm{trap}$, we imply averaging over the trap. Also, in the classical regime ($T \gg T_F$), one finds $\nu_c \sim \tau_c^{-1}$ where $\tau_c$ is the typical time between two collisions~\cite{Vichi2000}. 

We have calculated $Q$ for several values of $\eta$ as a function of $\bar{T}$ by  evaluating the integral given in Eq.~(\ref{eq:I33}) numerically. The results are shown in Fig.~\ref{fig:quadsc}e. The asymptotic behavior of $Q$ is investigated in Appendix~\ref{sec:Qasym} in the low and high temperature regimes in the 2D limit ($\eta=0$) and are shown on the same figure as red and blue dashed lines. We find that $Q \sim \bar{T}^2$ for small $T$ while it saturates to a constant value for large $\bar{T}$. The low temperature $T^2$ scaling is related to the Pauli blocking effect, however, it is different from the case 2D $s$-wave fermions (and 2D electron gas), where one finds $\nu_c \sim T^2\,\log(T/T_F)^{-2}$~\cite{Schaefer2012,Novikov2006}. This difference can be traced back to the fact that our system is spin polarized and the $s$-wave interaction channel is blocked. The logarithmic enhancement of the shear viscosity (i.e. attenuation of $\nu_c$) originates from the logarithmic divergence of the $s$-wave scattering length in the near-threshold regime in 2D. We remark that the near-threshold cross section of all other scattering channels remains bounded~\cite{Arnecke2008}.

The high temperature plateau is a unique feature of near-threshold dipole-dipole scatterings in the 2D limit and its existence can be understood in terms of the interplay between the temperature dependence of the scattering cross section and rarefaction of the gas. Provided that $T_F \ll T \ll T_\mathrm{dip}$, we can estimate the relaxation rate using a classical analysis by identifying $\nu_c \sim \gamma_c \equiv \tau_c^{-1}$. The Born 2D scattering cross section can be estimated as $\sigma_\mathrm{B} \sim q^{-1}|\tilde{\mathcal{V}}(q)|^2 \sim qa_d^2\,\mathrm{Erfcx}^2(q a_z)$, where $q$ is the typical momentum of scattering particles and is $\sim (m k_B T)^{1/2}$ in the high temperature regime. The collision frequency is $\gamma_\mathrm{c} = \tau_c^{-1} \sim \hbar q l^{-1}_\mathrm{mfp} \equiv \hbar q n\sigma$, where $l_\mathrm{mfp} = (n\sigma)^{-1}$ is the mean free path. The density at the center of the trap is $n_0 = m\omega_0^2 N/(2\pi T)$ and decreases as $\sim T^{-1}$. Combining these results, the collision rate amounts to:
\begin{equation}
\gamma_c \sim N \left(\frac{a_d}{a_0}\right)^2\,\mathrm{Erfcx}^2\bigg[\left(\frac{k_B T}{\hbar\omega_z}\right)^\frac{1}{2}\bigg],\,\,\,\,(T_F \ll T \ll T_\mathrm{dip})
\end{equation}
In the $2D$ limit, $\omega_z \rightarrow \infty$ and we find $\gamma_c = \mathrm{const}$ (note that $\mathrm{Erfcx}(0) = 1$). In other words, the growth of scattering cross section counteracts rarefaction of the gas to yield a constant collision rate. For finite vertical trap frequencies, the effective quasi-two-dimensional dipolar interaction weakens and we find that $\gamma_c$ decays like $\sim 1/T$ (note that $\mathrm{Erfcx}(x) \sim 1/x$ for large $x$). We remark that the single subband picture adopted here no longer holds true in the quasi-2D regime for large $k_B T/(\hbar \omega_z)$ and one must include higher subbands into account. We have shown in a previous work~\cite{Babadi2011} that all inter-subband interaction matrix elements have the same $\mathrm{Erfcx}$ factor and therefore, we expect this qualitative behavior to remain unaffected.

The plateau reached in the 2D limit relies crucially on the applicability of Born approximation. As mentioned earlier, the scatterings enter the semiclassical regime for $T \gtrsim T_\mathrm{dip}$ (see Eq.~\ref{eq:Tdip}) and Born approximation breaks down. In this regime, the total scattering cross section can be estimated using the Eikonal approximation~\cite{Ticknor2009} and one finds $\sigma_\mathrm{SC} \sim (a_d/q)^{1/2}$. Repeating the same analysis with the semiclassical cross section, we find:
\begin{equation}
\gamma_c \sim N \left(\frac{a_d}{a_0}\right)^\frac{1}{2} \left(\frac{\hbar\omega_0}{k_B T}\right)^\frac{3}{4}, \quad\qquad (T \gtrsim T_\mathrm{dip}).
\end{equation}
The qualitative behavior of $\nu_c$ for the full range of temperatures was shown earlier in Fig.~\ref{fig:comp}a1.\\

So far, we have neglected self-energy corrections in the description of the collective modes. We have also restricted our analysis to a variational calculation within a small basis set. In the next section, we extend our analysis to address both of these shortcomings.

\section{Extended basis analysis:\\
the effect of higher order moments and self-energy corrections}\label{sec:ext}
The general formalism described in Sec.~\ref{sec:linresBV} allows one to account for self-energy corrections as well as obtaining a more accurate calculation of the response functions by extending the variational basis set in a controlled way. Using simple symmetry considerations, we introduce extensible polynomial-like variational basis sets relevant for describing monopole and quadrupole dynamics. Finite truncations of these basis sets allows one to satisfy all moments of the CBV equation up to the truncation order, which is an extention of our previous analysis. Since we are dealing with large basis sets and self-energy corrections at finite temperatures, resorting to numerical methods is inevitable and no simple analytic results are expected to be found.

\subsection{Variational basis set for monopole oscillations}
The generator of monopole oscillations, $\delta U_m \sim r^2$, belong to the zero angular momentum representation of $S_z$. An arbitrary function of such type can be expressed as $f(p,r)[(x+iy)(p_x - ip_y)]^n$ for arbitrary $n \in \mathbb{Z}$ and $f(p,r)$. Any smooth function of this type can be written as a power series expansion in $r^2$, $p^2$, $\rr\cdot\pp$ and $\xi \equiv y p_x - x p_y$. Observing that $\xi^2 = r^2p^2 - (\rr \cdot \pp)^2$, the most general basis for such functions can be constructed from the following two classes:
\begin{align}
\phi^+_\alpha &\equiv \phi_{(m_{\alpha},n_{\alpha},k_{\alpha})} = r^{2m_\alpha}\,p^{2n_{\alpha}}(\rr \cdot \pp)^{k_\alpha},\nonumber\\
\phi^-_\alpha &\equiv \phi_{(m_{\alpha},n_{\alpha},k_{\alpha})} = \xi\,r^{2m_\alpha}\,p^{2n_{\alpha}}(\rr \cdot \pp)^{k_\alpha}.
\end{align}
Observing that $R_x \phi^\pm_\alpha = \pm \phi^\pm_\alpha$ and the fact that the equilibrium state and the perturbations are reflection symmetric, we may discard $\{\phi^-_\alpha\}$. We define $\{\phi^+_\alpha\}$ as the ``extended monopole basis'' and drop the $+$ superscript for brevity.
To truncate the basis set, we keep all basis functions satisfying $m + n + k \leq M$, where $M$ is a positive integer which we call the order of the basis set. A first order basis set contains four elements, $\{1, \rr\cdot\pp, p^2, r^2\}$ and is equivalent to the linearized scaling ansatz discussed earlier. In general, a basis set of order $M$ has $(M+1)(M+2)(M+3)/6$ elements. Expressions useful for numerical evaluation of the matrix elements of $\mathsf{M}$, $\mathsf{H}_0$, $\mathsf{\Sigma}$ and $\mathsf{I}_c$ in the monopole basis can be found in Appendix~\ref{sec:matrixmon}.

\subsection{Variational basis set for quadrupole oscillations}
A quadrupolar function in two dimensions is a function that changes sign upon a simultaneous $\pi/2$ rotation of both $\rr$ and $\pp$ about the $z$-axis. Such functions belong to the $m_z = \pm 2$ representation of $S_z$ which can be expressed as $f(p,r)\,e^{i M \phi_r}\,e^{i N \phi_p}$, where $M$ and $N$ are two integers such that $M - N = \pm 2$, $\phi_r$ and $\phi_p$ are the angles $\rr$ and $\pp$ make with a fixed axis (we arbitrarily choose the $x$-axis) and $f(p,r)$ is an arbitrary scalar function of $\pp$ and $\rr$. One can identify 12 classes of functions with such symmetry. Apart from an arbitrary scalar function $f(p,r)$, the accompanying multipliers can be:
\begin{center}
$\xi^+_1 \equiv x^2 - y^2,\quad \xi^+_2 \equiv p_x^2 - p_y^2, \quad \xi^+_3 \equiv x p_x - y p_y,$\\
$\eta^+_1 \equiv xy(yp_x - xp_y), \quad \eta^+_2 \equiv p_x p_y(yp_x - xp_y),$\\
$\eta^+_3 \equiv (yp_x + xp_y)(yp_x - xp_y)$,
\end{center}
and:
\begin{center}
$\xi^-_1 \equiv xy,\quad \xi^-_2 \equiv p_x p_y, \quad \xi^-_3 \equiv y p_x + x p_y,$\\
$\eta^-_1 \equiv (y p_x - x p_y)(x^2 - y^2), \quad \eta^-_2 \equiv (y p_x - x p_y)(p_x^2 - p_y^2),$\\
$\eta^-_3 \equiv (y p_x - x p_y)(xp_x - yp_y)$.
\end{center}
The functions with $+$ and $-$ superscript are even and odd eigenfunctions of the reflection operator $R_x$, respectively. Like before, we can drop the odd class. Also, we find the following relations between these pre-factors:
\begin{align}
2\eta_1^+ &= r^2 \xi_3^+ - (\rr \cdot \pp)\,\xi_1^+,\nonumber\\
2\eta_2^+ &= (\rr \cdot \pp)\,\xi_2^+ - p^2\,\xi_3^+,\nonumber\\
2\eta_3^+ &= r^2\,\xi_2^+ - p^2\,\xi_1^+,
\end{align}
using which we can drop the class of functions $f(p,r)\,\eta_i^+$ from the basis set. Since $f(p,r)$ is assumed to be a smooth scalar function of $\pp$ and $\rr$, in can be expanded in the monopole basis. Thus, in summary, we find that any smooth reflection symmetric quadrupolar function can be expanded in terms of $\{\xi_i^+\,\phi_\alpha^+\}$ for $i=1,2,3$ and $\alpha = (m, n, k)$, where $m$, $n$ and $k$ are non-negative integers and $\phi_\alpha^+$ are the previously introduced monopole basis functions. We denote this basis set as the ``extended quadrupole basis''. We also remark that this basis set can still be reduced further in light of the relation $2(\rr\cdot\pp)\,\xi_2^+ = p^2\xi_1^+ + r^2\xi_3^+$, so that the basis functions of the type $\xi_2^+\,r^{2m}p^{2n}(\rr\cdot\pp)^{k+1}$ can be written as a linear combination of $\xi_1^+\,r^{2m}p^{2n+2}(\rr\cdot\pp)^{k}$ and $\xi_3^+\,r^{2m+2}p^{2n}(\rr\cdot\pp)^{k}$. Like before, we drop the $+$ superscript for brevity in the remainder of the paper. An order-$M$ truncation of the quadrupole basis set is the finite set that comprises all quadrupole basis functions satisfying $k + m + n \leq M-1$. The first order basis set contains three elements, $\{x^2 - y^2, p_x^2 - p_y^2, x p_x - y p_y\}$ and is equivalent to the linearized scaling ansatz discussed earlier. In general, a quadrupole basis set of order $M$ contains $M(M+1)(2M+7)/6$ elements. Again, expressions useful for numerical calculation of the matrix elements of $\mathsf{M}$, $\mathsf{H}_0$, $\mathsf{\Sigma}$ and $\mathsf{I}_c$ in the quadrupole basis can be found in Appendix~\ref{sec:matrixquad}.

\subsection{Numerical results}
\subsubsection{Preliminaries}\label{sec:prelim}
In this section, we discuss the numerical results obtained by evaluating the linear responses to monopole and quadrupole perturbations for various parameter. We vary $\lambda_d$ and $T/T_F$ in the range $[0, 2]$ for fixed $N = 2200$. We study the 2D limit $\omega_z = \infty$ as well as a quasi-2D case corresponding to the current experiments with KRb ($\omega_0 = 2\pi \times 36\,\mathrm{Hz}$, $\omega_z = 2\pi \times 23\,\mathrm{kHz}$~\cite{JILA}). This choice of parameters imply $\eta \simeq 0.322$ in the quasi-2D case.

For each configuration, we performed the calculations within a 4th order basis set, comprising 35 and 50 basis functions for the monopole and quadrupole cases respectively, and satisfying all moments of the CBV equation up to 8th order. With the knowledge of the numerically obtained equilibrium solution (see Sec.~\ref{sec:eqb}), the matrix elements of $\mathsf{M}$, $\mathsf{H}_0$ and $\mathsf{\Sigma}$ can be calculated with little computational effort using the expressions provided in Appendices~\ref{sec:matrixmon} and~\ref{sec:matrixquad}.

The most computationally demanding part is the evaluation of the collision matrix elements. Although a considerable number of them vanish due to either conservation laws or symmetries, a 4th order basis set yields 118 (monopole) and 307 (quadrupole) unique matrix elements each of which is a five-dimensional integral that has to be evaluated for each choice of $\lambda_d$, $\eta$ and $T/T_F$. This requires considerably more efforts and computation time compared to the simple scaling ansatz analysis we presented earlier, where only a single collision matrix element had to be dealt with. We calculated the collision matrix elements using the Monte-Carlo integration method with $5 \times 10^8$ integration points. The estimate of the statistical error is $\sim 10^{-3}$ (relative).

We incorporate the self-energy corrections into the collision integral within a local effective mass approximation (see Sec.~\ref{sec:Ic}), which we found to be an excellent approximation in all cases. However, in order to assess the accuracy of this approximation and the consistency of the obtained results, we (1) we performed exact calculation of the collision integrals for a few representative parameter choices (without the effective mass approximation), and (2) checked the satisfaction of conservation laws. We will discuss both of these consistency checks later.\\

For the monopole case, we calculate the dimensionless spectral function $A_{r^2}(\omega)$ defined as:
\begin{equation}
A_{r^2}(\omega) \equiv -(2N)^{-\frac{1}{2}}\mathfrak{Im}[\chi_{r^2}(\omega)],
\end{equation}
for a $\delta$-kick in the potential (see Eq.~\ref{eq:chimon}). This quantity can be found using Eqs.~(\ref{eq:respphi}) and ~(\ref{eq:phidiag}) by choosing the excitation and observation vectors as $\delta U_\alpha(\omega) = O_\alpha = \delta_{m\alpha}$, where $m$ is the index that corresponds to the basis function $\phi = r^2$. For the quadrupole case, we calculate the spectral function $A_{x^2-y^2}(\omega)$ defined as:
\begin{equation}
A_{x^2-y^2}(\omega) \equiv -(2N)^{-\frac{1}{2}} \mathfrak{Im}[\chi_{x^2 - y^2}(\omega)].
\end{equation}
(see Eq.~\ref{eq:chiquad}). Likewise, this quantity can be evaluated by choosing the excitation and observation vectors as $\delta U_\alpha(\omega) = O_\alpha = \delta_{q\alpha}$, where $q$ is the index that corresponds to the basis function $\phi = \xi_1 = x^2 - y^2$. These spectral functions can be directly measured in the experiments in different ways (Ref. to Sec.~\ref{sec:expt}).

Although the evolution matrix has a large number of poles, some of which are clearly isolated and some may belong to branch lines, we found that only a few of them get excited and contribute to the response. Many of such poles
lie inside the null space of $\mathsf{H}_0$, are unphysical and do not get excited (see the discussion at the end of Sec.~\ref{sec:monscl}). In all cases, we found that the spectral functions can be reproduced accurately by a fit function with two simple poles in the lower half plane: 
\begin{equation}\label{eq:fit}
A_\mathrm{fit}(\omega) = \mathfrak{Im}\left[\frac{\mathcal{A}}{\omega-\Omega - i\Gamma} - \frac{\mathcal{A}^*}{\omega+\Omega - i\Gamma} + \frac{i\mathcal{B}}{\omega-i\Gamma'}\right], 
\end{equation}
corresponding to damped oscillations and overdamped components. Such a fit function extracts the most important information from the numerically obtained spectral functions. Moreover, this method allows us to present the results in clear and concise way.

\subsection{Monopole oscillations}
\begin{figure}
\center
\includegraphics[scale=0.7]{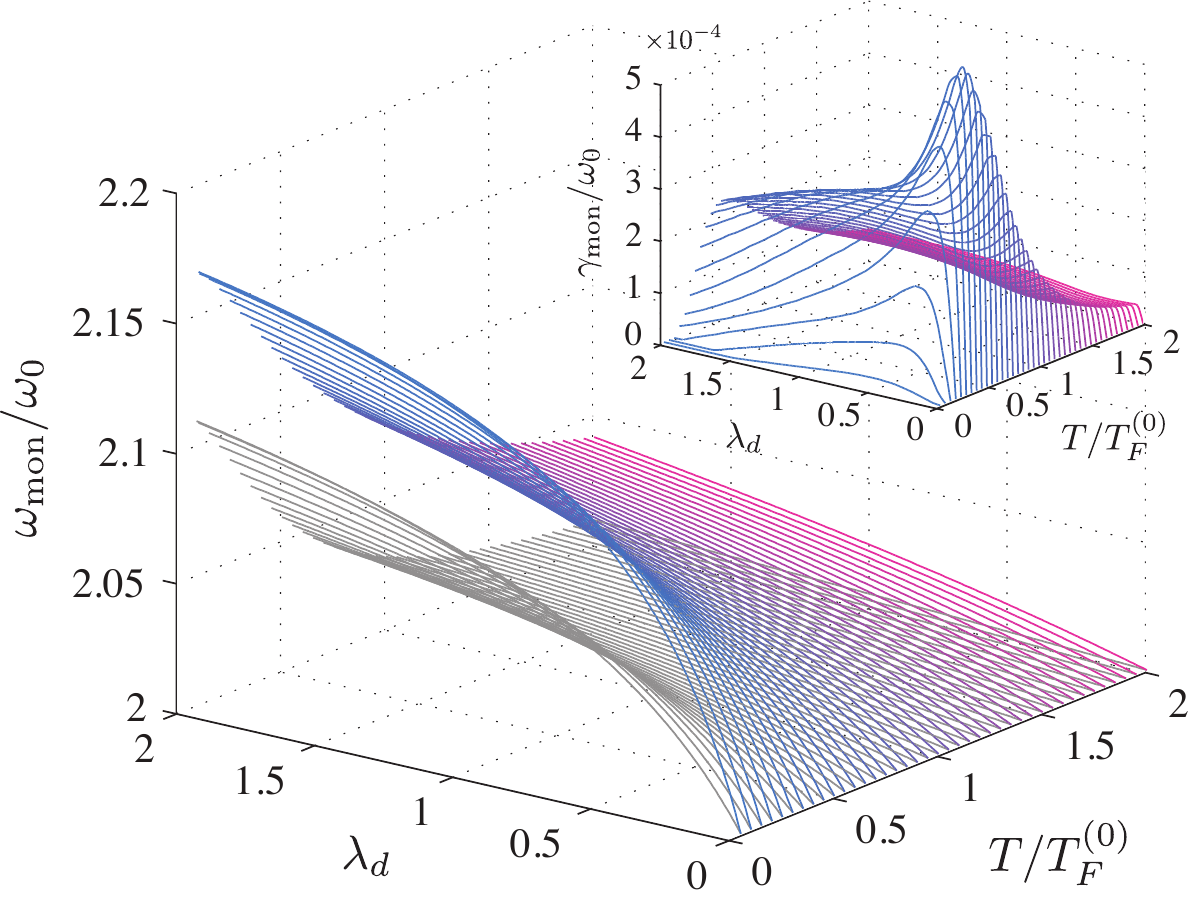}
\caption{(Color online) The oscillation frequency and the damping (inset) of the monopole mode extracted from the numerically obtained spectral functions using a 4th order basis set (including self-energy corrections). The colored and grayscale (upper and lower) graphs correspond to an ideal 2D system ($\eta=0$) and a quasi-2D system ($\eta\simeq 0.322$) respectively. Red and blue graphs correspond to high and low temperatures respectively. In all cases, $N=2200$. The inset plot shows the damping rate in the 2D case ($\eta=0$).}
\label{fig:monOmGam}
\end{figure}
As mentioned earlier in Sec.~\ref{sec:monscl}, in the absence of self-energy corrections, the CBV equation for harmonically trapped gases admits an exact solution corresponding to a scaling velocity field $\mathbf{v} \sim \rr$ which has a fixed oscillation frequency of $2\omega_0$ with no damping, independent of the interaction strength and temperature. This is due to fact that the Boltzmann equation admits a closed set of equations for the moments of $r^2$, $p^2$ and $\rr \cdot \pp$, all of which are immune to collisions due to conservation laws. Taking self-energy corrections into account, the quasiparticle dispersion relations no longer remain quadratic and one finds that this simple chain of moment equations can not be closed anymore. In particular, contributions from higher order moments, many of which are strongly influenced by the collisions, become important. Therefore, we expect the monopole oscillations to be damped to a certain degree.

\begin{figure}
\center
\includegraphics[scale=0.80]{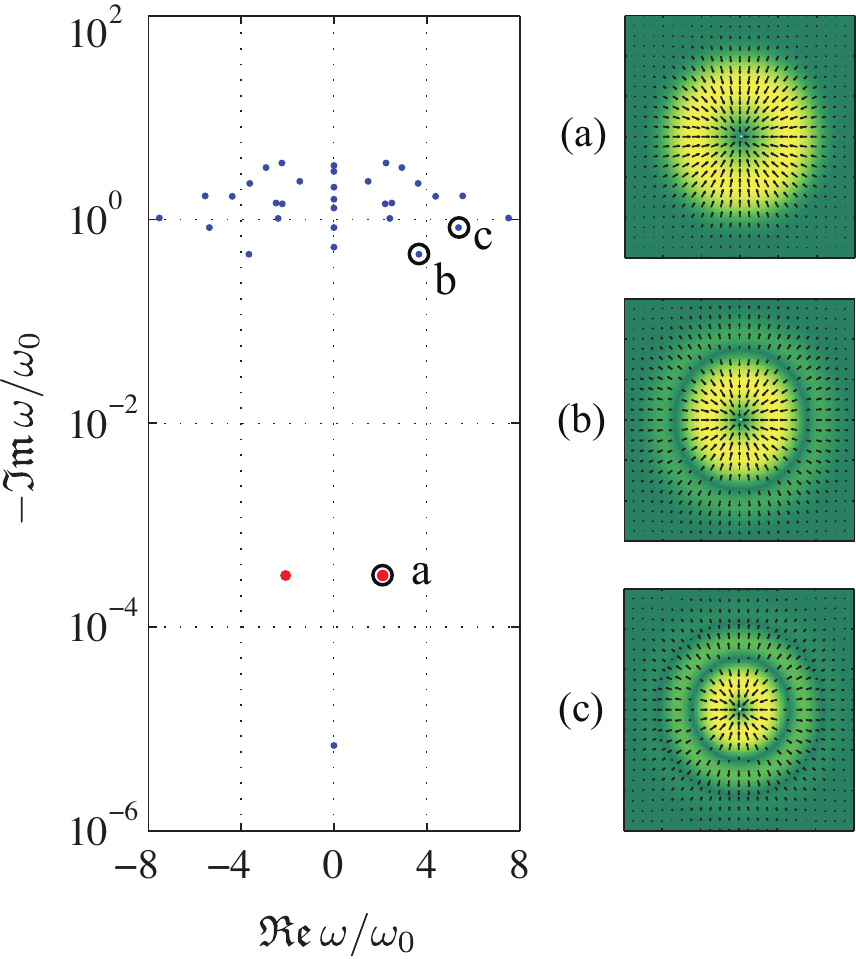}
\caption{(Color online) (left) A typical picture of the poles of the evolution matrix ($T/T_F=0.45$, $\lambda_d = 2$, $N=2200$ and $\eta=0$). (right) the mass currents associated to the indicated poles. The chosen poles correspond to the three modes with lowest energy that survive in the HD regime.}
\label{fig:monpoles}
\end{figure}

\begin{figure*}[ht!]
\center
\includegraphics[scale=0.75]{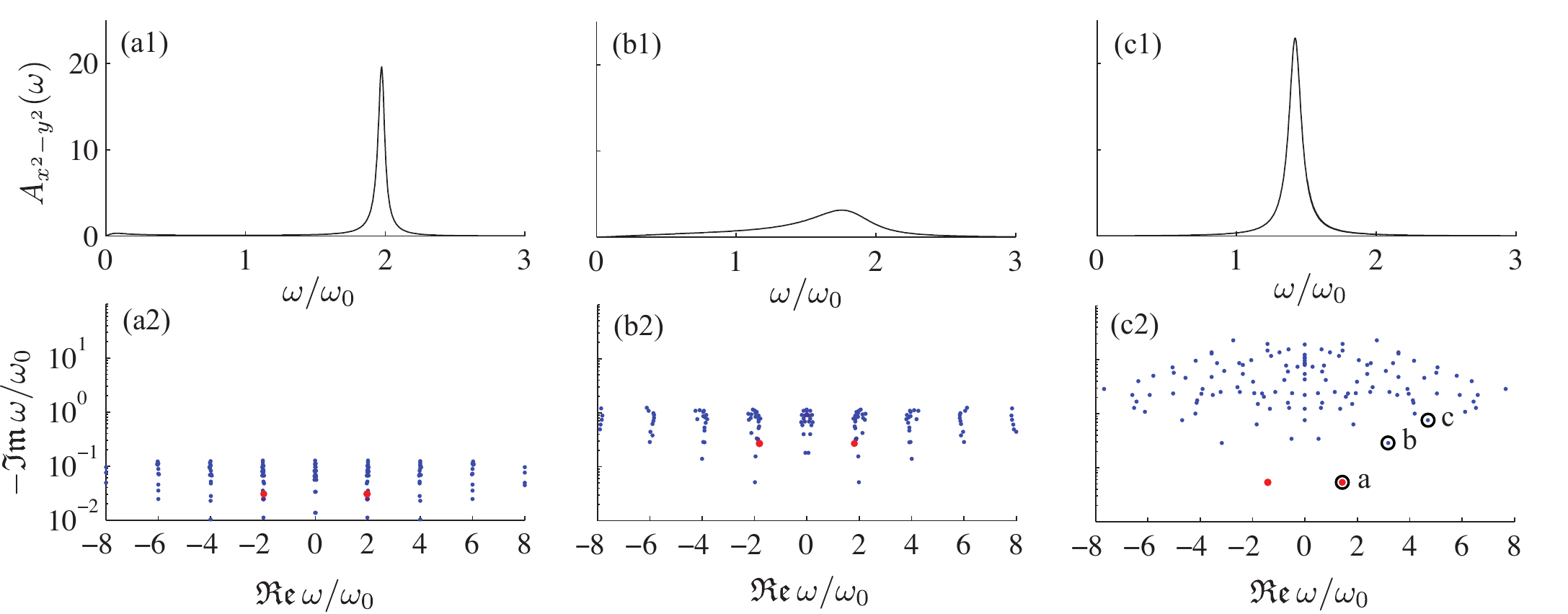}
\caption{(Color online) Evolution of the quadrupole oscillations from collisionless (CL) to hydrodynamic (HD) regime upon increasing the interaction strength (left to right). In all cases, $T/T_F = 0.45$ and $\eta=0$ ($\omega_z = \infty$). The top row shows the spectral function and the bottom row shows the location of the poles of the evolution matrix on the complex plane. The pole shown as red is the pole that makes the dominant contribution to the response. (a1) and (a2): $\lambda_d = 0.1$, (b1) and (b2): $\lambda_d = 0.4$, (c1) and (c2): $\lambda_d = 2$. See Fig.~\ref{fig:quadvel} for a plot of the mass currents associated to the encircled poles. Refer to Sec.~\ref{sec:expt} for a discussion on the experimental methods for measuring the spectral functions.}
\label{fig:quadevol}
\end{figure*}

Fig.~\ref{fig:monOmGam} shows the frequency and damping of the monopole oscillations extracted from the numerically obtained spectral functions. The colored and grayscale (top and bottom) plots show correspond to the 2D limit ($\eta=0$) and the quasi-2D example ($\eta \simeq 0.322$). The repulsive dipole-dipole interaction clearly results in a significant increase in the oscillation frequency. Also, as one expects, deviations from the 2D limit leads to a weaker repulsive effective interaction and thus, a smaller increase in the frequency of collective modes.

We find that the dominant contribution to the response results from a single isolated pole of the evolution matrix, which is the one that has the lowest energy. The relative residues of the other poles were found to be of the order of $\sim 10^{-4}$ in all cases.

The most interesting finding is that this mode exhibits a very small damping, $\gamma_\mathrm{mon} < 10^{-3}\omega_0$, in all of the studied cases (see the inset plot of Fig.~\ref{fig:monOmGam}) despite the presence of significant self-energy corrections. In Sec.~\ref{sec:disc}, we discuss the possibility that the smallness of damping could be a result of the effective mass approximation adopted in evaluating the collision integrals and confirm that even an exact treatment of self-energy corrections does not change this finding appreciably.

We remark that the mode which makes the dominant contribution to the linear response is the one that lies at the bottom of an infinite hierarchy of possible monopole oscillations. The reason that it is the only mode that gets excited is most likely a consequence of harmonic confinement. Inclusion of higher order moments not only yields a more accurate calculation of the frequency of this mode, but also it allows one to investigate higher order modes. Fig.~\ref{fig:monpoles} shows a typical picture of the poles of the evolution matrix, along with plots of the mass current associated to three indicated low-lying modes. In contrast to the scaling mode (shown as $a$ in the figure), the two other modes (b and c) have a significant damping rate. In the experiments, these modes can be excited by non-harmonic perturbations in the trap potential, such as $\sim r^4$.

In the absence of interactions, all of the poles lie on the real frequency axis at discrete locations $2n\omega_0$, $n \in \mathbb{Z}$ (here, up to $|n|=4$ due to the 4th order truncation of the basis set). Each discrete frequency is multiply degenerate (the collisionless Boltzmann equation, i.e. the Liouville equation, has infinitely many degenerate discrete poles in case of harmonic external potential). Upon increasing interactions, the poles spread to the lower half complex frequency plane, signaling the transition to the dissipative CL-HD crossover regime. Upon further increment of the interactions, most of the poles diverge to $-i\infty$ while a few migrate back to the real axis and form the discrete hydrodynamic modes. The absence of damping in this limit, as discussed in Sec.~\ref{sec:quadscl}, is due to the emergence of local equilibrium. The three chosen poles in Fig.~\ref{fig:monpoles} correspond to isolated poles that survive in HD regime.

\subsection{Quadrupole oscillations}
\begin{figure}
\center
\includegraphics[scale=0.48]{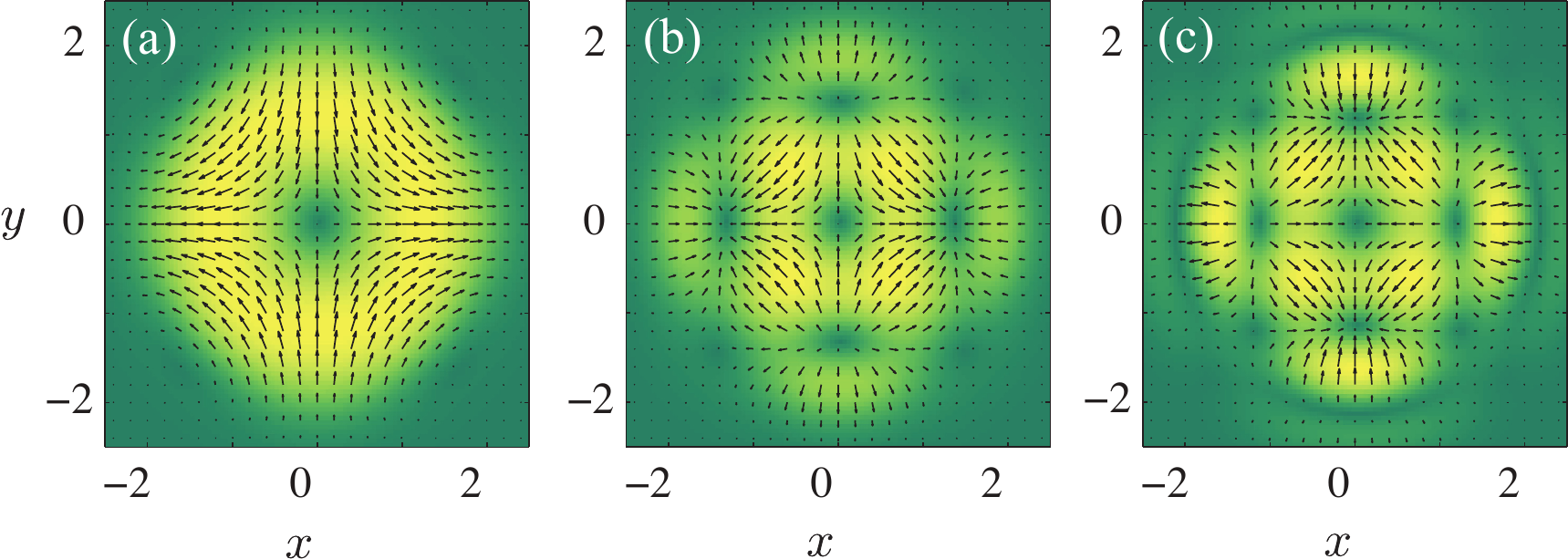}
\caption{(Color online) The mass current associated to the three modes marked in Fig.~\ref{fig:quadevol}c2. (a) is the lowest lying mode, known as the surface mode, characterized by the velocity field $\mathbf{v} \sim x\mathbf{e}_x - y\mathbf{e}_y$. (b) and (c) are the next two modes. The nodal structure of the mass current is clearly noticeable. These modes constitute the three lowest lying HD modes.}
\label{fig:quadvel}
\end{figure}

A typical scenario for quadrupole response is shown in Fig.~\ref{fig:quadevol}. The top and bottom rows show the quadrupole spectral function and the location of the poles on the complex frequency plane respectively. The mode that has the largest residue is marked as red. For small interactions ($\lambda_d \ll 1$, Fig.~\ref{fig:quadevol}a1-2), the spectral function is sharply peak around $2\omega_0$ and the poles of the evolution matrix lie very close to the real axis. Upon increasing the interactions, the poles spread to the lower half complex frequency plane, indicating the entrance to the dissipative CL-HD crossover regime. The spectral function is significantly broadened (see Fig.~\ref{fig:quadevol}b1) in this regime. For stronger interactions, the local equilibrium picture starts to emerge, indicated by a reduction in damping. Fig.~\ref{fig:quadevol}c2 clearly shows a sharply peaked spectral function near $\sqrt{2}\omega_0$in the strongly interacting regime. This is exactly the universal frequency of the hydrodynamic quadrupole surface mode discussed earlier. 

Similar to the monopole case, we find that quadrupole perturbations of the trap potential predominantly excite the lowest lying mode. Here, we find a small contribution from a few overdamped modes as well, specially in the crossover regime. This is in agreement to the result found from the scaling ansatz. Fig.~\ref{fig:quadvel} shows the mass current associated to the three lowest lying modes marked in Fig.~\ref{fig:quadevol}c2. These modes are found to be the ones that survive in the strongly interacting regime and constitute the lowest lying HD modes. Such higher order modes may be excited by applying anharmonic perturbation to the trap potential, such as $\sim r^2(x^2 - y^2)$.

\begin{figure}
\center
\includegraphics[scale=0.65]{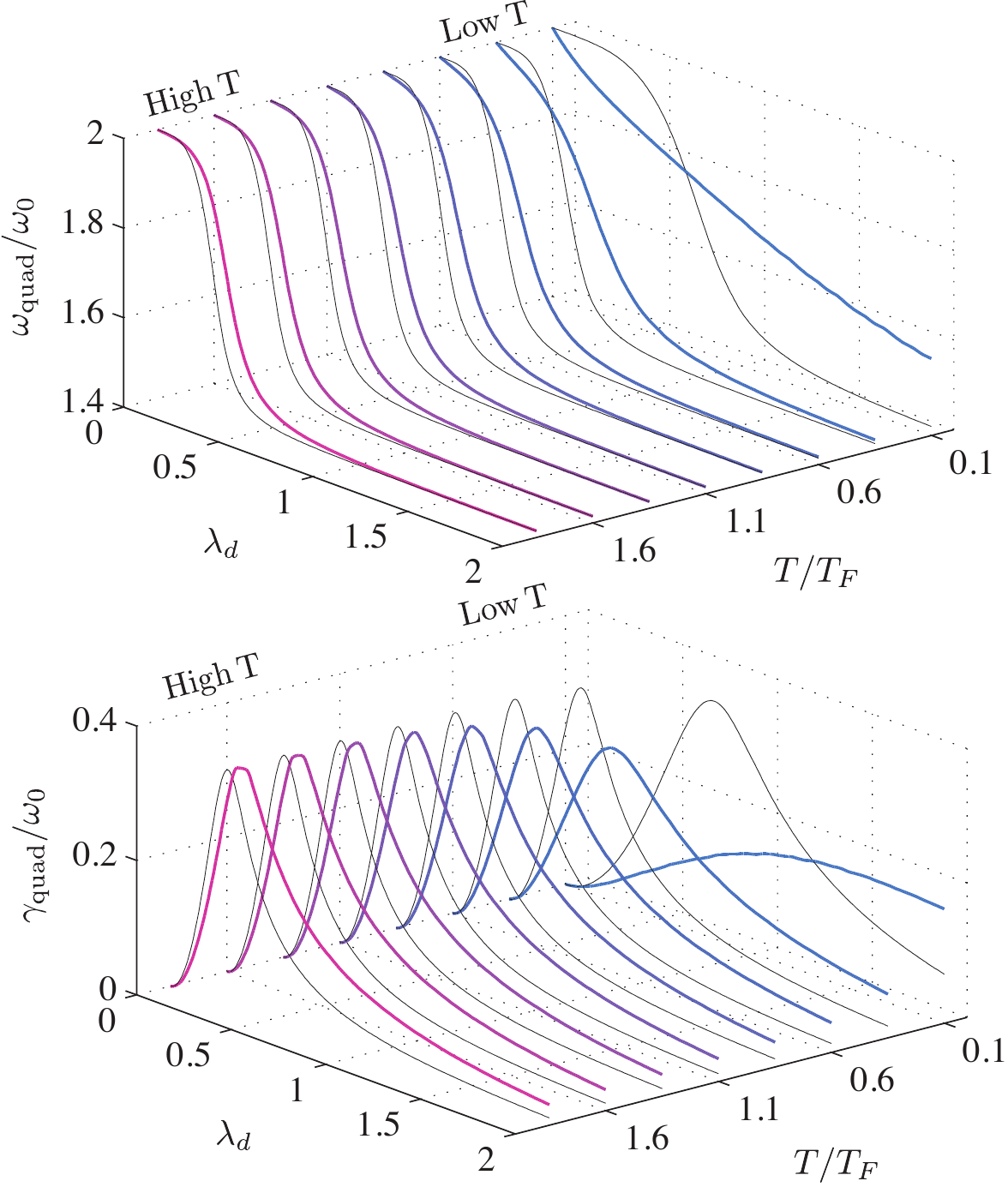}
\caption{(Color online) Frequency and damping (top and bottom graphs respectively) of quadrupole oscillations in a 2D system ($\eta=0$) with $N=2200$ particles. The thick colored lines are the numerical results for a 4th order basis set, including self-energy corrections. The thin black lines correspond to the analytic scaling ansatz analysis presented earlier (Sec.~\ref{sec:quadscl}).}
\label{fig:quad2D}
\end{figure}

\begin{figure}
\center
\includegraphics[scale=0.65]{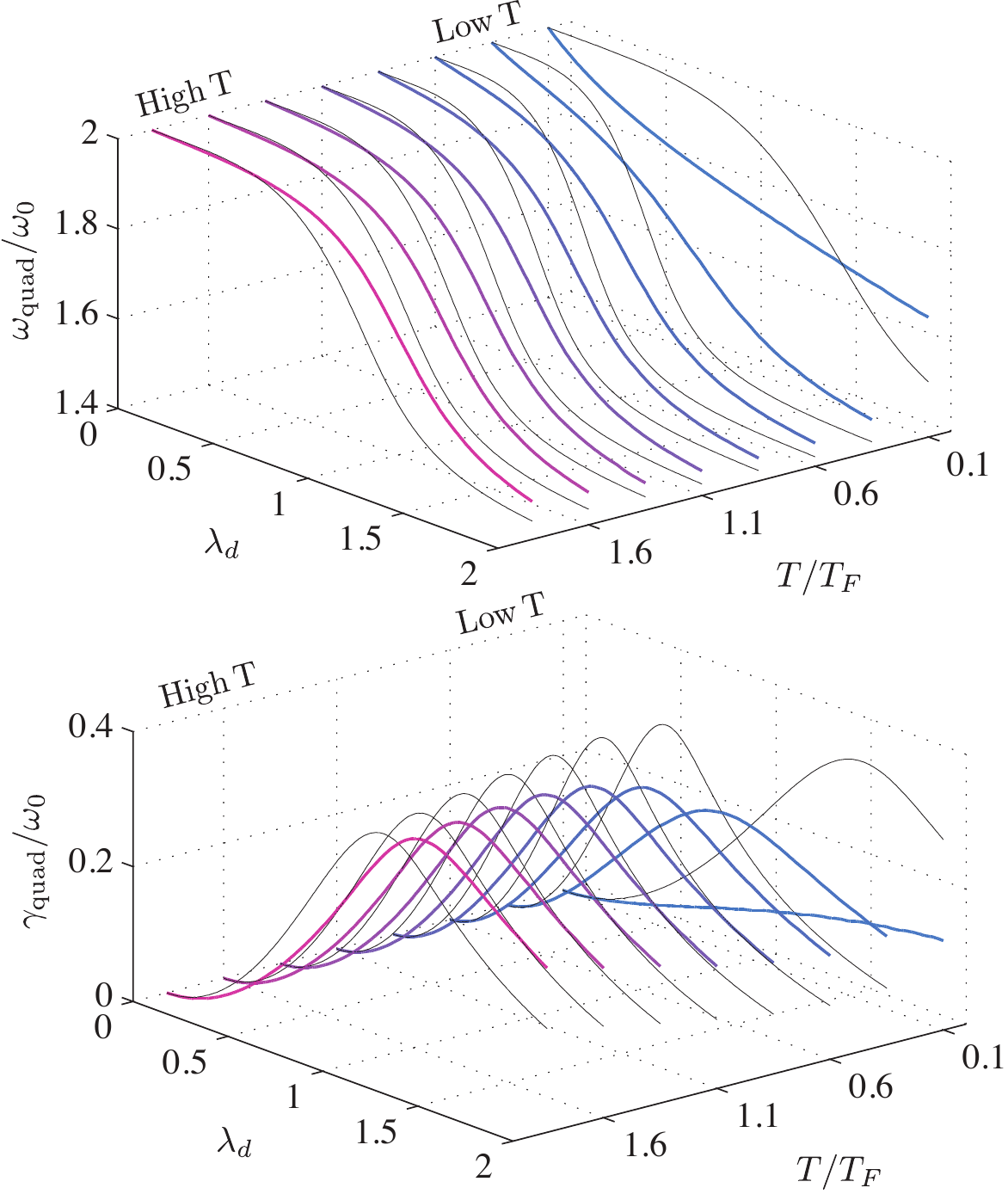}
\caption{(Color online) Frequency and damping of quadrupole oscillations for a quasi-2D system corresponding to $\eta \simeq 0.322$ (refer to the caption of Fig.~\ref{fig:quad2D} for details)}
\label{fig:quadQ2D}
\end{figure}

Figs.~\ref{fig:quad2D} and~\ref{fig:quadQ2D} show the frequency and damping rate of the quadrupole oscillations obtained from the two-mode fit to the quadrupole spectral function. The result obtained from the scaling ansatz analysis presented earlier is also shown as thin black lines for reference.

Clearly, the corrections are significant. In the low temperature regime, the self-energy corrections are dominant, yielding a $\propto \lambda_d$ correction to the frequencies (see the rightmost plot on the top panel of Fig.~\ref{fig:quad2D}). Note that the collisional corrections are only $\propto \lambda_d^4$ in the weakly interacting regime (see Eq.~\ref{eq:approxpoles1} and note that $\nu_c \propto \lambda_d^2$). The corrections resulting from the inclusion of higher order moments can also be seen in the same figure. Generally, scaling ansatz without self-energy corrections overestimates the collision rates and predicts the crossover to the hydrodynamic regime too early (notice that the peak of the damping rate occurs earlier for the scaling ansatz). At lower temperatures, this overestimation is predominantly a density effect (the gas expands due to repulsive interactions and lowers the density, which in turn leads to a lower collision rate). At higher temperatures, the overestimation results from the density inhomogeneity of the trapped gas, which is not accounted for well by the scaling ansatz. We discuss this in more detail in the discussions section.

Finally, we note that the same arguments apply to the quasi-2D case (Fig.~\ref{fig:quadQ2D}) regarding the comparison between the scaling ansatz and the extended analysis. The additional feature of the quasi-2D case, which was also discussed in detail in Sec.~\ref{sec:quadscl}, is the reappearance of the collisionless limit at higher temperatures. A signature of this can be seen in Fig.~\ref{fig:quadQ2D}b by observing the non-monotonic behavior of the location of the peak of the damping rate. At higher temperatures, entrance to the crossover regime is delayed and the CL region expands.

\section{Discussions}\label{sec:disc}
Most of the relevant discussions were already given in the main text. Here, we give a brief summary of the main results and discussions, along with several complementary comments.

In this paper, we started our analysis by investigating the equilibrium state of quasi-two-dimensional dipolar fermions in isotropic traps. In order to study the collective modes of the system, we solved the collisional Boltzmann-Vlasov equation for small perturbations of the trap potential with monopole and quadrupole symmetries. The self-energy corrections to quasiparticle dispersions and collisions were taken into account via the self-consistent Hartree-Fock and Born approximations respectively. The validity of these approximations were assessed at the end of Sec.~\ref{sec:kinetic}. In particular, the usage of Born approximation restricts the validity domain to near-threshold scattering energies (see Eq.~\ref{eq:Tdip}). We confined our attention to the regime where $T_F \ll T_\mathrm{dip}$, so that the scatterings remain in the near-threshold regime up to $T \simeq T_\mathrm{dip} \gg T_F$ and make the thermal regime accessible to the scope of this work.

We emphasize that once the conditions for the validity of CBV equation is met, this formalism is universally applicable to both CL and HD regimes, as well as the intermediate crossover regime.

We carried out the analysis of collective modes in two stages: as a first approximation, we used bare quasiparticles and studied the response functions using the simple picture of scaling ansatz. This analysis implied that the monopole oscillations occur at a fixed frequency of $2\omega_0$, are undamped, and are independent of temperature and dipolar interaction strength. In case of quadrupole oscillations, however, we found a transition from the CL limit to the HD limit, indicated by oscillation frequencies of $2\omega_0$ and $\sqrt{2}\omega_0$ respectively. We investigated the collisional relaxation rate of quadrupole oscillations, $\nu_c$, the single parameter that appears in the characteristic equation of quadrupole oscillations and yields the frequency and damping of the lowest lying quadrupole mode. This quantity was calculated for various temperatures and vertical trap frequencies and was shown to be expressible in terms of a universal function of $T/T_F$ and $\eta$. We found that in the 2D limit ($\eta=0$), $\nu_c$ is a monotonically increasing function of temperature and reaches to a plateau for large $T/T_F$. This plateau persists up to $T \simeq T_\mathrm{dip}$ beyond which the scattering enter the semiclassical regime and the scattering cross section starts to decrease as a function of temperature. The existence of this plateau, which is a novel feature of dipolar interaction implies that (1) the character of trap excitations of a polarized 2D dipolar gas becomes weakly dependent on temperature in the regime $T_F \lesssim T \lesssim T_\mathrm{dip}$, and (2) collisional effects persists despite the fact that gas becomes very dilute. This behavior differentiates 2D dipolar fermionic gases from $s$-wave fermions where rarefaction of the gas at high temperatures takes the system back to the collisionless regime for $T \gtrsim T_F$. Also, the temperature window for collisional behavior is universal for $s$-wave fermions and is not amenable to tuning, whereas for quasi-2D dipolar fermions, one can expand this window by (1) making the vertical confinement stronger to approach the 2D limit, and (2) either increase $T_\mathrm{dip}$ by using weaker dipoles or decrease $T_F$ by using a weaker transverse trap. 

The existence of the plateau in $\nu_c$ is guaranteed as long as the scale separation $T_F \ll T_\mathrm{dip}$ is met. By combining Eqs.~(\ref{eq:Tdip}) and~(\ref{eq:nuc}), one finds the condition for the plateau to lie in the collision dominated (hydrodynamic) regime:
\begin{equation}\label{eq:hydrocond}
N^\frac{1}{4} \ll \frac{a_0}{a_d} \ll N^\frac{1}{2},\qquad\text{(HD plateau)}
\end{equation}
The left and right hand sides of this inequality are equivalent to $T_F \ll T_\mathrm{dip}$ and $N(a_d/a_0)^2 \gg 1$ respectively, where the latter condition implies $\nu_c \gg 1$. The above inequality may be used as a simple experimental guideline to achieve hydrodynamics with dipolar fermions.\\

In the second stage of calculations, we extended the analysis by (1) including self-energy corrections and (2) satisfying all of the moments of the CBV equation up to 8th order. Chiacchiera {\it et al.}~\cite{Chiacchiera2011} and Pantel~{\it et al.}~\cite{Pantel2012} have carried out similar extended moments analysis (up to 4th order moments) for the case $s$-wave fermions and have found the corrections to be significant and improve the matching between the theory and the experiment.

This extended analysis allowed us to (1) investigate higher order modes for both monopole and quadrupole oscillations, and (2) evaluate the reliability of the simple scaling ansatz analysis. We found that despite the fact that satisfaction of higher order moments results in numerous new normal modes, the responses to the monopole and quadrupole perturbations ($\sim r^2$ and $x^2 - y^2$ respectively) are predominantly governed by the lowest lying mode (in the quadrupole cases, we found contributions from a few overdamped modes as well). We remark that the frequency and damping of dominant mode, however, is significantly modified by both self-energy corrections and inclusion of higher order moments.

We argued that self-energy corrections is expected to result in the damping of the lowest lying monopole mode, since the closure of moment equations that ensures the absence of damping relies sensitively on quadratic dispersions. We found that although this expectation is met, the damping remains very small ($< 10^{-3}\omega_0$) even in the strongly interacting regime. The frequency of oscillations, however, is significantly increased from its non-interacting value of $2\omega_0$. This correction was found to be most significant at lower temperatures where self-energy corrections are strongest. In order to rule out the possibility that the smallness of damping could be a result of the local effective mass approximation used in evaluating the collision integrals, we evaluated the collision integrals using the exact quasiparticle dispersion (altbeits, at the costs of a significantly increased computation time; see Appendix.~\ref{sec:exact}) for a few representative cases and found that the damping remains within the same order of magnitude.

By investigating the velocity field of lowest lying monopole mode, we found that it retains its scaling character to an good approximation in all cases (i.e. $\mathbf{v} \sim \mathbf{r}$) with negligible temperature fluctuations. It is known from the hydrodynamic theory of non-ideal fluids that for a true isotropic and isothermal scaling flow, no dissipation results from shear viscosity or thermal conduction and the only source of dissipation is the bulk viscosity~(for instance, see Ref. \cite{LLFluid}, $\S$49). For such flows, one finds $\mathrm{d}S/\mathrm{dt} = \int\mathrm{d}^2\rr\,n_0^{-1} T^{-1}\zeta(\nabla\cdot\mathrm{v})^2$ where $S$ is the total entropy and $\zeta$ is the bulk viscosity. Note that the dissipation rate is small since it is second order in $\mathbf{v}$.

The extended analysis of the lowest lying quadrupole mode has the same qualitative behavior as the scaling ansatz analysis, albeit with significant quantitative corrections. At lower temperatures, self-energy corrections lowers the frequency of oscillations from the non-interacting value of $2\omega_0$ proportionally to $\lambda_d$. Also, we find that much stronger interaction is required to reach the HD regime. This is simply a consequence of the rarefaction of the gas at low temperatures due to repulsive interactions. Generally, we found that satisfying higher order moments results in a delayed entrance to the crossover (and the HD) regime. One may explain this finding by observing that the scaling ansatz overestimates the collision rate since it describes the dynamics simply as a uniform time-dependent rescaling of the equilibrium distribution: in fact, a perturbation like $x^2 - y^2$ is most effective for the gas elements located furthest from the center of the trap while it does not affect the particles sitting at the center of the trap as strongly. Neglecting this fact clearly results in an overestimation of the collisional relaxation, specially since the gas is most dense at the center of the trap.

\begin{figure}
\center
\includegraphics[scale=0.65]{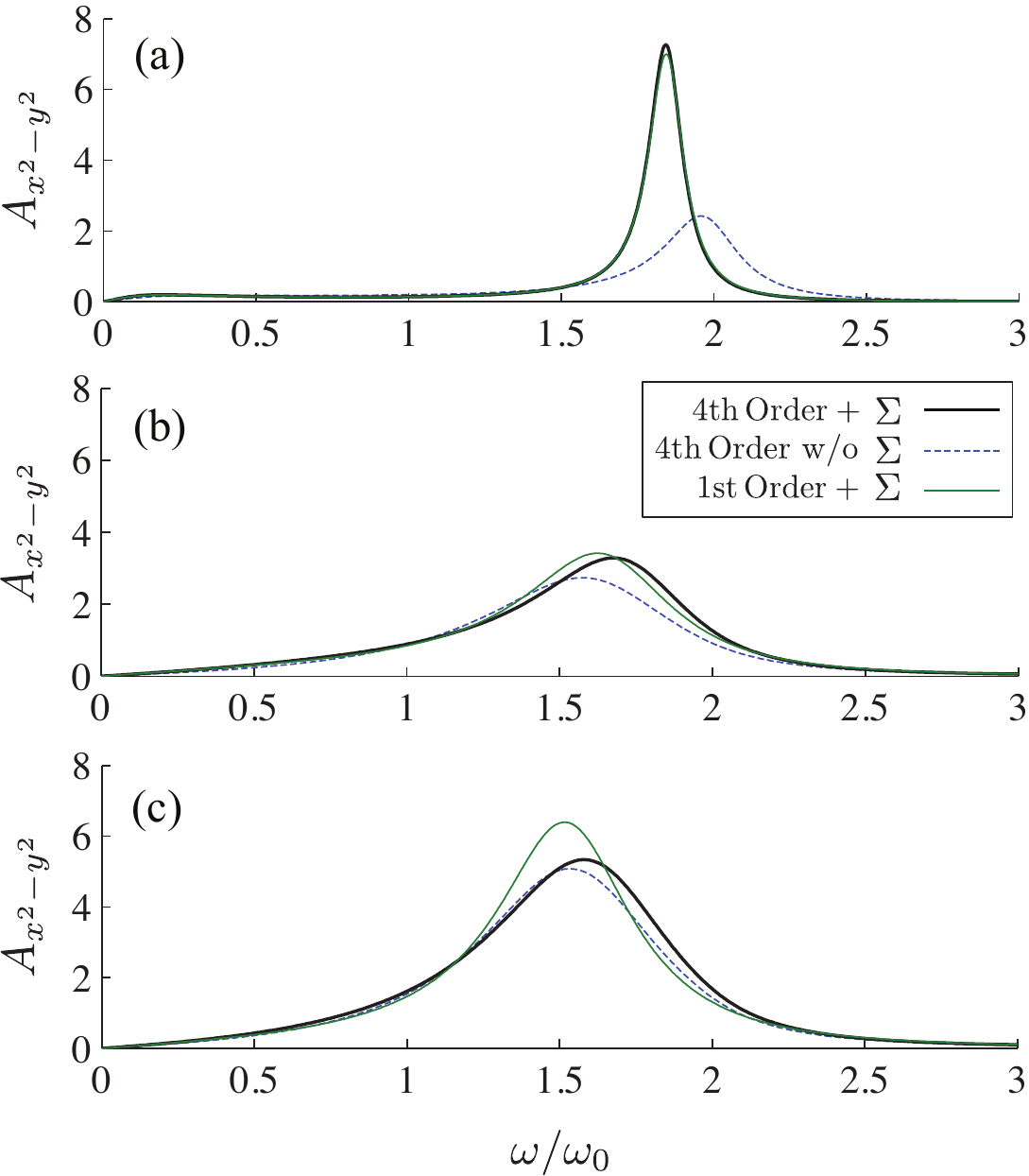}
\caption{(Color online) A comparison of the quadrupole spectral functions obtained using three different approximations. The legend is shown on the plot ($+\Sigma$ and w/o $\Sigma$ means with and without self-energy corrections). $\lambda_d = 0.45$, $\eta=0$ and $N=2200$ in all three plots. (a) $T/T_F = 0.1$, (b) $T/T_F = 0.5$, (c) $T/T_F = 1$.}
\label{fig:mfcomp}
\end{figure}

In order to study the effects self-energy inclusion and higher moments separately, we have shown the quadrupole spectral function for three different temperatures and a fixed interaction strength using three different approximations in Fig.~\ref{fig:mfcomp}: 4th order basis set with and without self-energy, and 1st basis set (scaling ansatz) with self-energy. At low temperatures (panel a), as one would expect, we find a significant correction from the inclusion of self-energy. On the other hand, inclusion of higher order moments yields almost no correction. At higher temperatures, this scenario is reversed. The reason is that self-energy corrections are essentially due to exchange interactions which diminish in the high temperature (classical) regime. On the other hand, the trapped gas assumes a Gaussian density profile at high temperatures (compared to a quadratic profile at $T=0$) and naturally, higher moments are needed to describe the spatial inhomogeneity of the dynamics accurately.

We also found that the most important corrections to the scaling ansatz stems from the 4th order moments, beyond which the corrections become increasingly less significant. In practice, a second order basis set is sufficient to obtain the frequencies of the lowest lying collective modes within $0.1\%$ range of the exact solution. Higher order modes naturally require inclusion of higher order moments.

\begin{figure}
\center
\includegraphics[scale=0.65]{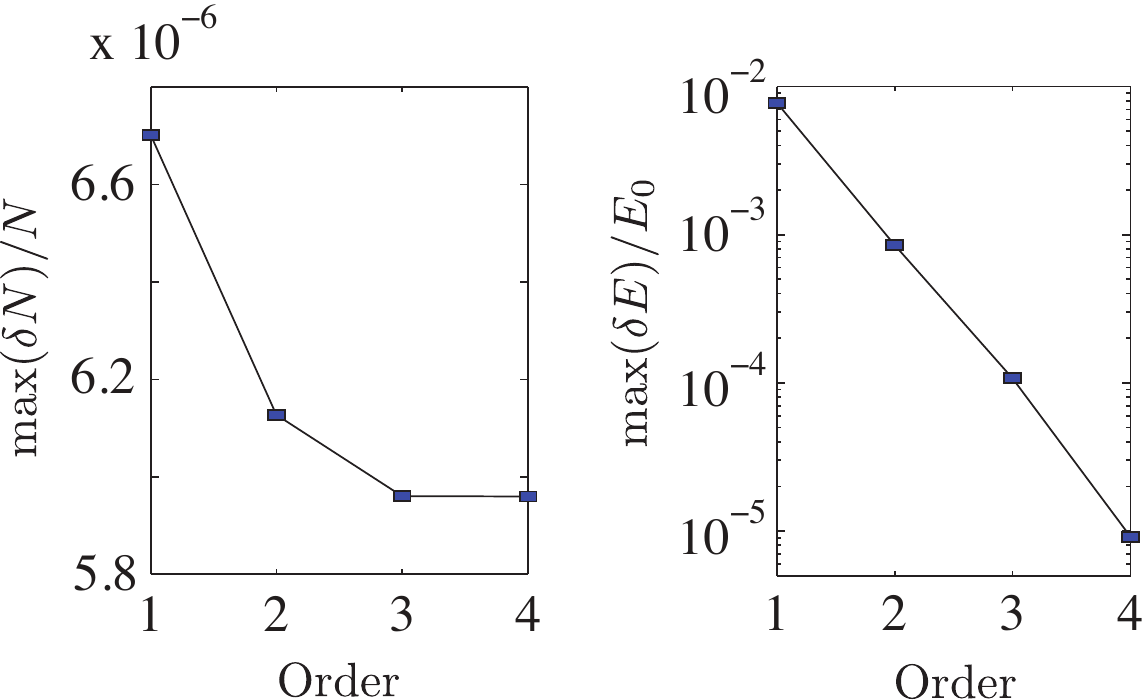}
\caption{Maximum relative deviations of the conserved quantities ($N$ and $E$) for monopole oscillations for a sample configuration ($T/T_F = 0.1$, $\lambda_d = 0.5$, $\eta=0$ and $N=2200$)}
\label{fig:cons}
\end{figure}

As a consistency check for our numerical calculations, we investigated the satisfaction of the conservation laws (see Appendix~\ref{sec:cons}). The CBV equation conserves the particle number, momentum and energy, both in the differential form and the integral form. The quadrupole oscillations trivially satisfy these conservation laws due to the axial symmetry of the equilibrium state. This is not trivial for monopole modes since they have the same symmetry as the equilibrium state. Fig.~\ref{fig:cons} shows the maximum relative deviations of the particle number and energy in monopole oscillations as a function of moment satisfaction order for a sample case. We find that the particle number is conserved within a relative error of $\sim 10^{-6}$ even in a first order basis set (this is because one of the moment equations is in fact the statement of conservation of particle number). On the other hand, we find that conservation of energy improves substantially upon expanding the basis set. For the 4th order basis set, the relative error is $\sim 10^{-5}$.\\

\section{Experimental outlook}\label{sec:expt}
The collective modes can be probed experimentally in various ways. As described earlier, one common method is to perturb the trap potential with a short pulse and monitor the evolution of the cloud using in-situ or absorption imaging techniques (for example, see Ref.~\cite{Altemeyer2007}). The relevant observables are the radius and anisotropy of the cloud in case of isotropic and quadrupolar perturbations respectively. The frequency and damping of the collective modes are found by fitting the measured time evolution of the observable $O_\mathrm{exp}(t)$ to a function of the form $O_\mathrm{fit}(t) = A e^{-\gamma t}\sin(\omega t) + B e^{-\gamma_\mathrm{OD}t}$, where $\omega$ is the frequency of oscillations, and $\gamma$ and $\gamma_\mathrm{OD}$ are damping rate of the oscillatory and overdamped components. If required, the spectral function can be subsequently found by taking a Fourier transform of the measured impulse response signal $O_\mathrm{exp}(t)$.

Another method, which may yield more accurate results, is the measurement of spectral functions via modulation spectroscopy. This is done by introducing a low-amplitude periodic modulation to the trap potential at a fixed frequency $\Omega$ for a long duration $\tau \gg \omega_0^{-1}, \Omega^{-1}$ and measuring the absorbed energy. For a finite trap modulation pulse like $\delta U \sim e^{-|t|/\tau}\cos(\Omega t)\,v(\rr)$, a simple linear response analysis yields~\cite{Babadi2011a}:
\begin{equation}\label{eq:Eabs}
\Delta E_\mathrm{abs} \sim -\tau\,\Omega\,\mathfrak{Im}[\chi_{v(\rr)}(\Omega+i/\tau)],
\end{equation}
where $\Delta E_\mathrm{abs}$ is the absorbed energy, $v(\rr)$ is the shape of the trap perturbation (i.e. $x^2 + y^2$ and $x^2 - y^2$ for probing monopole and quadrupole modes respectively), and $\chi_{v(\rr)}$ is the retarded correlator of $v(\rr)$. Eq.~(\ref{eq:Eabs}) implies that the absorbed energy in a modulation experiment yields a direct measurement of the spectral function. The absorbed energy can be measured in various ways. One method is to let the system re-thermalize after the modulation pulse, followed by mapping it to a non-interacting system by switching off the interactions adiabatically and finally measuring the rise in temperature of the non-interacting gas through a time of flight expansion experiment. The location of the peak in the spectral function and its width yield the frequency and damping of the collective mode.
 
We conclude this section by making predictions for the experiments with KRb. In the recent experiments with a quasi-2D configurations~\cite{JILA}, the vertical and transverse trap frequencies are $\omega_z = (2\pi) \times 23$ kHz and $\omega_0 = (2\pi) \times 36$ Hz respectively. The central layer has $2200$ molecules, the temperature is $T=500$ nK and dipole moment is $D = 0.158$ Debye, using which we find $T/T_F \approx 4.36$, $\eta \approx 0.322$ and $\lambda_d \approx 0.252$. The dipolar temperature is $T_\mathrm{dip} \sim 1.6~\mu$K and $T_F/T_\mathrm{dip} \approx 6.4 \times 10^{-2}$. Therefore, the near-threshold scattering condition can be satisfied for quantum degenerate temperatures. However, $T/T_\mathrm{dip} \approx 2.8$ in the current experiments, implying that the system lies on the margin of semi-classical scatterings. Therefore, the predictions give here may not be very accurate.

The mean-field corrections are small at this temperature and we can use the scaling ansatz analysis of Sec.~\ref{sec:quadscl}. We find $Q(\bar{T} = 4.36,\eta = 0.322) \approx 0.019$ which yields $\nu_c \approx 0.04$. Eq.~(\ref{eq:approxpoles1}) yields the frequency shift and damping rate of the quadrupole mode as $\delta\omega/\omega_0 \approx 1.25 \times 10^{-4}$ and $\gamma \approx 0.01\,\omega_0 \approx 2.3$~Hz. While the frequency shift is too small to be easily experimentally observable, the damping rate is sufficiently fast and can be easily observed. We remark that $Q$ has a strong dependence on $\eta$ and rapidly decreases as $\eta$ is increased (see Fig.~\ref{fig:quadsc}d). Thus, one may increase the collisional relaxation rate significantly simply by making the vertical confinement stronger. In the hypothetical 2D limit, we get $\nu_c^\mathrm{2D} \approx 1.5$ and consequently, $\omega^\mathrm{quad} \approx 1.8\,\omega_0 \approx 415\,$Hz and $\eta^\mathrm{quad} \approx 0.3\,\omega_0 \approx 71\,$Hz. The dramatic enhancement of the collisional effects in 2D compared to the quasi-2D configuration is remarkable. Also, we note that since $T \sim T_\mathrm{dip}$ in the current experiments, the system is expected to remain in the plateau upon further cooling down to degeneracy at $T \sim T_F \approx 115\,$nK.

Finally, we note that recent experimental progress with lanthanide atoms such as ${}^{161}$Dy~\cite{Burdick2012} and ${}^{168}$Er~\cite{Aikawa2012} which have large magnetic dipole moments (10~$\mu_B$ and 7~$\mu_B$ respectively) is another possibility for studying collective many-body excitations of fermionic dipolar gases.

\section{Acknowledgements}
The computations in this paper were run on the Odyssey cluster supported by the FAS Science Division Research Computing Group at Harvard University. The authors acknowledge the support from Harvard-MIT CUA, NSF Grant No. DMR-07-05472, DARPA OLE program, AFOSR Quantum Simulation MURI, AFOSR MURI on Ultracold Molecules, and the ARO-MURI on Atomtronics.
\appendix

\section{Conservation laws of the linearized collisional Boltzmann-Vlasov equation}\label{sec:cons}
The CBV equation admits three important differential conservation laws for of mass density, mass current and energy, which can be established by multiplying the sides of CBV equation by $1$, $\mathbf{p}$ and energy density $\mathcal{E}$ respectively, and integrating over $\mathbf{p}$~\cite{LL}. The collision integrals vanish identically in all three cases due to the existence of the same conservation laws in the level of 2-body scatterings. We state these conservation laws in their integral form here to utilize them later as a consistency check for our numerical calculations. The conservation of mass (or equivalently, particle number) is:
\begin{equation}
\frac{\mathrm{d}}{\mathrm{d}t}\int\mathrm{d}\Gamma\,n(\pp;\rr,t) = 0.
\end{equation}
The linearized equation, with the parametrization given by Eq.~(\ref{eq:fvar}), yields:
\begin{equation}
\frac{\mathrm{d}}{\mathrm{d}t}\int\mathrm{d}\Gamma\,\Delta_0\Phi(\pp;\rr,t) = 0.
\end{equation}
In the same parametrization, the conservation of momentum reads as:
\begin{equation}
\frac{\mathrm{d}}{\mathrm{d}t}\int\mathrm{d}\Gamma\,\mathbf{p}\,\Delta_0\Phi(\pp;\rr,t) = 0.
\end{equation}
Finally, the energy density is given by $\mathcal{E}_\mathrm{HF} = p^2/(2m) + m\omega_0^2/2 + \Sigma^+[n]/2$ in the Hartree-Fock approximation using which we get the following linearized form of the statement of conservation of energy:
\begin{equation}\label{eq:encons}
\frac{\mathrm{d}}{\mathrm{d}t}\int\mathrm{d}\Gamma\left(\delta \mathcal{E}\, n_0 + \mathcal{E}_0\Delta_0\Phi(\pp;\rr,t)\right) = 0,
\end{equation}
where $\mathcal{E}_0 \equiv \mathcal{H}_0$ is the equilibrium energy density and $\delta\mathcal{E} = \Sigma^+[\delta n]]/2 = \Sigma^+[\Delta_0 \Phi]/2$. Using the properties of Hartree-Fock energy density functional, we get $\int\mathrm{d}\Gamma\delta \mathcal{E}\, n_0 = (1/2)\int\mathrm{d}\Gamma\Sigma^+[\Delta_0 \Phi]\,n_0 \equiv (1/2)\int\mathrm{d}\Gamma\Sigma^+[n_0]\,\Delta_0 \Phi$. Using this identity, the two terms in Eq.~(\ref{eq:encons}) can be combined to yield:
\begin{equation}
\frac{\mathrm{d}}{\mathrm{d}t}\int\mathrm{d}\Gamma\,\mathcal{H}_0\Delta_0\Phi(\pp;\rr,t) = 0.
\end{equation}

In case of quadrupole oscillations, these conservation laws as trivially satisfied due to difference between the symmetry of deviations and the equilibrium state. In the monopole case, while the conservation of momentum is still trivially satisfied, the mass and energy conservations may only be fulfilled if $\Phi$ is a legitimate solution of the kinetic equation.

\section{Asymptotic analysis of $Q(\bar{T},\eta=0)$}\label{sec:Qasym}
In the 2D limit ($\eta=0$), the asymptotic behavior of $Q(\bar{T},\eta)$ can be studied analytically. Setting $\eta=0$, the $\mathrm{Erfcx}$ functions appearing in the collision integral (see Eq.~\ref{eq:I33}) evaluate to $1$ and the expression in the brackets in the second line simply becomes $[\chi_1 - \chi_2]^2 = \sin^2\xi\,\sin^2\nu\,[1-|\sin(\phi-\phi')|]$. This will result in significant simplifications.

\subsection{Low temperature expansion}
In the low temperature regime, $\bar{\mu}/\bar{T} \rightarrow \infty$, we may use the following identity:
\begin{multline}
\lim_{\bar{\mu}/\bar{T} \rightarrow \infty}(\bar{\mu}/\bar{T})^{-3}\int_0^\infty \rho^5\,\mathrm{d}\rho\,\bigg[\frac{1}{\cosh(\rho-\bar{\mu}/\bar{T}) + \cosh(b_1 \rho)}\\
\times  \frac{1}{\cosh(\rho-\bar{\mu}/\bar{T}) + \cosh(b_2 \rho)}\bigg] = \frac{4\pi^2}{3}\,\delta(b_1)\,\delta(b_2),
\end{multline}
in order to carry out the $\rho$ integration. This identity can be established by observing that for large $\bar{\beta}\bar{\mu}$ the integrand will be exponentially unless $\rho \sim \bar{\beta}\bar{\mu}$ and $b_1, b_2 \sim (\bar{\beta}\bar{\mu})^{-1}$. In the limit $\bar{\beta}\bar{\mu} \rightarrow \infty$, this implies that the right hand side must be proportional to $\delta(b_1)\delta(b_2)$. The proportionality constant can be found by integrating the left hand side over $b_1$ and $b_2$ which yields the $4\pi^2/3$ prefactor. Identifying $b_1$ and $b_2$ as $\sin^2\xi\,\sin2\nu\,\cos\phi$ and $\sin^2\xi\,\sin2\nu\,\cos\phi'$ respectively, we can carry out the $\xi$ and $\nu$ integrations using the $\delta$-functions and we finally get: 
\begin{equation}\label{eq:lowT}
Q(\bar{T} \rightarrow 0, \eta=0) \approx C\,\frac{(\bar{\mu}/\bar{T})^3}{\Langle \pb^4 \Rangle},
\end{equation}
where $C$ is given by:
\begin{equation}
C=\frac{32}{9}\int_0^{2\pi}\mathrm{d}\phi\int_0^{2\pi}\mathrm{d}\phi'\frac{[1-|\sin(\phi-\phi')|]\,\sin(\phi-\phi')^2}{\cos^2\phi + \cos^2\phi'},
\end{equation}
and is equal to $19.176999$ to six significant digits. $\Langle \pb^4 \Rangle$ can be found analytically with little effort and we get:
\begin{equation}\label{eq:p4}
\Langle \pb^4 \Rangle = -8 \bar{T}^3\,\mathrm{Li}_3(-e^{\bar{\mu}/\bar{T}}).
\end{equation}
Using the asymptotic expansion of $\mathrm{Li}_3(-x)$ for large $x$ and the low temperature expansion of $\bar{\mu}$ mentioned after Eq.~(\ref{eq:mueq}), we the following low temperature expansion:
\begin{equation}
-\mathrm{Li}_3(-e^{\bar{\mu}/\bar{T}}) = 1/(6\bar{T}^3) + \pi^2/(12 \bar{T}) + \mathcal{O}(\bar{T}).
\end{equation}
Combining the last four equations, we finally get:
\begin{equation}
Q(\bar{T} \rightarrow 0, \eta=0) \approx \frac{2}{3}\,C\,\bar{T}^2 \approx 12.784666\,\bar{T}^2, 
\end{equation}
to leading order. This asymptotic limit is shown in Fig.~\ref{fig:quadsc}e as a blue dashed line and agrees with the numerical result.

\subsection{High temperature expansion}
The analysis of the classical limit ($\bar{\beta}\bar{\mu}\rightarrow 0$) is simpler. First, we rewrite the hyperbolic functions in the denominator as $\cosh(\rho-\ln z) \equiv e^{\rho}/(2z) + (z/2)e^{-\rho}$. Here, $z \equiv \exp(\bar{\mu}/\bar{T})$ is the fugacity and goes to zero in the high temperature limit. Thus, $\cosh(\rho-\ln z) \approx e^\rho/(2z)$ to leading order. In fact, the denominator of Eq.~(\ref{eq:I33}) is dominated by the first $\cosh$ term since the second ones are $\mathcal{O}(e^{\rho}) \ll e^{\rho}/(2z)$, so that we can neglect them as well. With this simplification, all of the integrations become elementary and we get:
\begin{equation}
Q(\bar{T} \rightarrow 0, 0) \approx \frac{8(8-3\pi)z^2\bar{T}^5}{\Langle \pb^4 \Rangle}.
\end{equation}
The fugacity in the classical limit can be found from Eq.~(\ref{eq:mueq}), yielding $z = 1/(2\bar{T}^2) + \mathcal{O}(\bar{T}^{-4})$. Using the asymptotic expansion $-\mathrm{Li}_3(-z) = z + \mathcal{O}(z^2)$, we finally get:
\begin{equation}
Q(\bar{T} \rightarrow \infty,0) \approx \frac{1}{2}(3\pi - 8) \approx 0.712389.
\end{equation}
This asymptotic limit is shown in Fig.~\ref{fig:quadsc}e as a red dashed line and is in agreement with the numerical result.

\section{Matrix elements of the evolution matrix in the monopole basis}\label{sec:matrixmon}
The variational linear response analysis of the CBV equation, as described in Sec.~\ref{sec:linresBV} requires the calculation of a large number of matrix elements. This task, however, can be somewhat simplified as the angular integrations in the matrix elements of $\mathsf{M}$, $\mathsf{\Sigma}$ and $\mathsf{H}_0$ can be carried out analytically using the symmetries of the basis functions and the equilibrium state, reducing the problem to the evaluation of a two-dimensional integral over $\bar{p}$ and $\bar{r}$. The latter computation can be done numerically accurately and efficiently.

In this appendix, we provide readily computable formulas for the required matrix elements in the monopole basis. We define the shorthands $R_\alpha \equiv 2m_\alpha + k_\alpha$, $P_\alpha \equiv 2n_\alpha + k_\alpha$ for given basis function $\phi_\alpha$. $R_\alpha$ and $P_\alpha$ count the powers of $r$ and $p$ in $\phi_\alpha$ respectively.

\subsection{Matrix elements of $\mathsf{M}$}
By definition, we have:
\begin{align}
\mathsf{M}_{\alpha\beta} &=\int\mathrm{d}\bar{\Gamma}\,\Delta_\LE(\bar{p},\bar{r})\,\phi_\alpha\phi_\beta\nonumber\\
&=\int(2\pi)\,\rb\,\mathrm{d}\rb\,\frac{1}{(2\pi)^2}\,\pb\,\mathrm{d}\pb\,\Delta_\LE(\pb,\rb)\,\rb^{R_\alpha + R_\beta}\nonumber\\
&\times \pb^{P_\alpha+P_\beta}\,\int_0^{2\pi}(\cos\psi)^{k_\alpha + k_\beta}\,\mathrm{d}\psi\nonumber\\
&=\frac{E(k_\alpha + k_\beta)(k_\alpha + k_\beta)!}{2^{k_\alpha+k_\beta}\left[\left(\frac{k_\alpha + k_\beta}{2}\right)!\right]^2}\Bigg[\int \rb^{R_\alpha + R_\beta + 1}\nonumber\\
&\times \pb^{P_\alpha + P_\beta + 1}\,\Delta_\LE(\pb,\rb)\,\mathrm{d}\rb\,\mathrm{d}\pb\Bigg],
\end{align}
where $E(n) = 1$ for even $n$ and $E(n) = 0$ for odd $n$. For future reference, we define:
\begin{equation}
h(n) = \frac{E(n)\,n!}{2^{n}\left[\left(n/2\right)!\right]^2},
\end{equation}
and:
\begin{equation}
I^{m}_{n}[A(\pb,\rb)] = \int A(\pb,\rb)\,\rb^{m+1}\,\pb^{n+1}\,\mathrm{d}\rb\,\mathrm{d}\pb,
\end{equation}
using which we can write $\mathsf{M}_{\alpha\beta} = h(k_\alpha+k_\beta)\,I^{(R_\alpha+R_\beta)}_{(P_\alpha+P_\beta)}[\Delta_\LE]$.

\subsection{Matrix elements of $\mathsf{H_0}$}
First, we evaluate the Poisson bracket $\{\phi_\beta,\bar{\mathcal{H}}_0\}$:
\begin{align}\label{eq:poiss}
\{\phi_\beta,\bar{\mathcal{H}}_0\} &= \nabla_{\bar{\rr}}\phi_\beta \cdot \nabla_{\bar{\pp}} \bar{\mathcal{H}}_0 - \nabla_{\bar{\pp}}\phi_\beta \cdot \nabla_{\bar{\rr}} \bar{\mathcal{H}}_0\nonumber\\
&=\gamma_p\,(\bar{\pp} \cdot \nabla_{\bar{\rr}}) \phi_\beta - \gamma_r\,(\bar{\rr} \cdot \nabla_{\bar{\pp}}) \phi_\beta\nonumber\\
&=\gamma_p\big[2m_\beta\,\psi_{(m_\beta-1,n_\beta,k_\beta+1)}\nonumber\\
&+ k_\beta\,\psi_{(m_\beta,n_\beta+1,k_\beta-1)}\big]\nonumber\\
&- \gamma_r \big[2n_\beta\,\psi_{(m_\beta,n_\beta-1,k_\beta+1)}\nonumber\\
&- k_\beta\,\psi_{(m_\beta+1,n_\beta,k_\beta-1)}\big],
\end{align}
where:
\begin{align}
\gamma_r &\equiv \bar{r}^{-2}\bar{\rr}\cdot\nabla_{\bar{\rr}}\bar{\mathcal{H}}_0 = 1 + \bar{r}^{-2}\bar{\rr}\cdot\nabla_{\bar{\rr}}\bar{\Sigma}_\LE,\nonumber\\
\gamma_p &\equiv \bar{p}^{-2}\bar{\pp}\cdot\nabla_{\bar{\pp}}\bar{\mathcal{H}}_0 =  1 + p^{-2}\pp\cdot\dpp\bar{\Sigma}_\LE.
\end{align}
Plugging Eq.~(\ref{eq:poiss}) into the definition of $(\mathsf{H}_0)_{\alpha\beta}$, we get:
\begin{align}
(\mathsf{H}_0)_{\alpha\beta} &= \int\mathrm{d}\bar{\Gamma}\,\Delta_\LE\,\phi_\alpha\{\phi_\beta,\mathcal{H}_0\}\nonumber\\
&=\left[2m_\beta\,h(k_\alpha+k_\beta+1)+k_\beta\,h(k_\alpha+k_\beta-1)\right]\nonumber\\
&\times I^{(R_\alpha+R_\beta-1)}_{(P_\alpha+P_\beta+1)}[\gamma_p\Delta_\LE]\nonumber\\
&-\left[2n_\beta\,h(k_\alpha+k_\beta+1)+k_\beta\,h(k_\alpha+k_\beta-1)\right]\nonumber\\
&\times I^{(R_\alpha+R_\beta+1)}_{(P_\alpha+P_\beta-1)}[\gamma_r\Delta_\LE].
\end{align}

\subsection{Matrix elements of $\mathsf{\Sigma}$}
By definition,
\begin{equation}\label{eq:sigF}
\bar{\Sigma}[\Delta_\LE \phi_\beta] = \lambda_d\int\frac{\mathrm{d}^2\bar{\pp}'}{(2\pi)^2}\,u(|\bar{\pp}-\bar{\pp}'|,\eta)\,\Delta_\LE(\bar{p}',\bar{r})\,\phi_\beta(\bar{\pp}',\bar{\rr}).
\end{equation}
It is easy to verify that a simultaneous rotation of $\brr$ and $\bpp$ leaves $\bar{\Sigma}[\Delta_\LE \phi_\beta]$ invariant, so that $\Sigma_F[\Delta_\LE \phi_\beta]$ may only depend on $\bar{r}$, $\bar{p}$ and $\phi$, the angle between $\brr$ and $\bpp$. Let $\cos\psi = (\bpp\cdot\bpp')/(\bar{p} \bar{p}')$ and $\cos\phi = (\brr\cdot\bpp)/(\bar{r}\bar{p})$, so that $\brr\cdot\bpp' = \bar{r} \bar{p}' \cos(\phi+\psi)$. Expanding $u(|\bpp-\bpp'|,\eta)$ in a cosine series,
\begin{equation}
u(|\bpp-\bpp'|,\eta) = \sum_{n=0}^\infty u^{(n)}(\bar{p},\bar{p}';\eta)\cos(n\psi),
\end{equation}
where:
\begin{multline}
u^{(n)}(\pb,\pb') = \frac{1}{\pi(\delta_{n,0}+1)}\\
\times\int_0^{2\pi}\mathrm{d}\psi\,u\left(\sqrt{\pb^2 + \pb'^2 - 2\pb \pb' \cos\psi},\eta\right)\,\cos n\psi,
\end{multline}
and plugging into Eq.~(\ref{eq:sigF}), we get:
\begin{multline}\label{eq:sigF2}
\bar{\Sigma}[\Delta_\LE \phi_\beta](\pb,\rb,\phi) = \lambda_d\int\frac{\pb'\,\mathrm{d}\pb'}{2\pi}\,\Delta_\LE(\pb',\rb)\,\pb'^{P_\beta}\rb^{R_\beta}\\
\times\sum_{n=0}^\infty\,u(\pb,\pb';\eta)\int_0^{2\pi}\frac{\mathrm{d}\psi}{2\pi}\,\cos(n\psi)\cos(\phi+\psi)^{k_\beta}.
\end{multline}
The angular integration can be evaluated using contour integral techniques:
\begin{multline}
\int_0^{2\pi}\frac{\mathrm{d}\psi}{2\pi}\,\cos(n\psi)\cos(\phi+\psi)^{k}\\
= \left[\frac{k!}{2^k}\frac{\theta(k-n)\,E(k+n)}{\left[\left(\frac{k-n}{2}\right)!\right]\left[\left(\frac{k+n}{2}\right)!\right]}\right]\cos(n\phi),
\end{multline}
where $\theta(n)=1$ if $n \ge 0$ and $\theta(n)=0$ otherwise. We denote the numerical prefactor in the brackets of the above equation by $g(n,k)$. Plugging this into Eq.~(\ref{eq:sigF2}), we get:
\begin{equation}
\bar{\Sigma}[\Delta_\LE \phi_\beta](\pb,\rb,\phi) = \lambda_d\sum_{n=0}^{k_\beta}\,Q^{(n)}_\beta(\pb,\rb)\cos(n\phi),
\end{equation}
where:
\begin{multline}
Q^{(n)}_\beta(\pb,\rb) = -g(n,k_\beta)\,\rb^{R_\beta}\int\frac{\mathrm{d}\pb'}{2\pi}\,\Delta_\LE(\pb',\rb)\,\pb'^{(P_\beta+1)}\\
\times u^{(n)}(\pb,\pb';\eta).
\end{multline}
The last integral can be easily evaluated numerically. Also, note that we only need $u^{(n)}$ up to $n = k_\beta$ in order to evaluate $\bar{\Sigma}[\Delta_\LE \phi_\beta]$ exactly. This is due to the fact that $g(n,k_\beta)$ vanishes for $n > k_\beta$. Having evaluated $\bar{\Sigma}[\Delta_\LE \phi_\beta]$, $\mathsf{\Sigma}_{\alpha\beta}$ can be evaluated readily by appealing to its definition:
\begin{align}
\mathsf{\Sigma}_{\alpha\beta} &= \lambda_d\sum_{n=0}^{k_\beta}\Big(\left[2m_\alpha\,g(n,k_\alpha+1) + k_\alpha\,g(n,k_\alpha-1)\right]\nonumber\\
&\times I^{(R_\alpha-1)}_{(P_\alpha+1)}[Q^{(n)}_\beta\Delta_\LE\gamma_p]- [2n_\alpha\,g(n,k_\alpha+1)\\
&+ k_\alpha\,g(n,k_\alpha-1)]\,I^{(R_\alpha+1)}_{(P_\alpha-1)}[Q^{(n)}_\beta\Delta_\LE\gamma_r]\Big).
\end{align}

\subsection{Matrix elements of $\mathsf{I}_c$}\label{sec:Ic}
The evaluation of the matrix elements of the linearized collision integral operator is the most computationally expensive part of the calculation. In particular, the deviation of quasiparticle dispersion from the bare quadratic dispersion makes the calculations even more challenging. To our knowledge, all of the previous works along this line have evaluated the collision integrals for bare particles. This approximation is justified when one is dealing with the the Boltzmann equation where one neglects mean-field corrections altogether. However, since we have included mean-field effects on the dynamics, we must also use the dressed quasiparticles dispersion in order to satisfy conservation of energy. In order to do this in a numerically tractable way, we have found that the quasiparticle dispersions can be approximated well using a local effective mass approximation (within an error of less than 2 percents). To this end, we approximate the dressed quasiparticle energies as:
\begin{equation}
\bar{\mathcal{H}}_0(\pb,\rb) \approx \varepsilon_0(\rb) + \frac{\pb^2}{2m^*(r)} + \frac{\rb^2}{2},
\end{equation}
where:
\begin{align}
\varepsilon_0(\rb) &= \bar{\Sigma}_\LE(\rb;0),\nonumber\\
m^*(\rb) &= \left[1 + \partial_{\pb}^2\,\Sigma_\LE(\rb;\pb)\Big|_{\pb=0}\right]^{-1}.
\end{align}
As we will see shortly, this approximation allows us to put the collision integral into a simple form suitable for numerical treatments. As a first step, we go to the center of mass frame of the colliding particles and define:
\begin{align}
&\bpp = \frac{\bar{\mathbf{P}}}{2} + \bar{\qq}, \,\,\,\,\,\qquad \bar{\pp}_1 = \frac{\bar{\mathbf{P}}}{2} - \bar{\qq},\nonumber\\
&\bar{\pp}' = \frac{\bar{\mathbf{P}}'}{2} + \bar{\qq}', \qquad \bar{\pp}'_1 = \frac{\bar{\mathbf{P}}'}{2} - \bar{\qq}',
\end{align}
using which we get:
\begin{multline}\label{eq:changevar}
\mathrm{d}^2\brr\,\dd{\bpp}\dd{\bpp_1}\dd{\bpp'}\dd{\bpp'_1}\,(2\pi)\delta(\Delta \bar{E})\,(2\pi)^2\delta(\Delta\bar{\mathbf{P}})\\
\rightarrow \frac{m^*(\rb)}{2}\,\rb\,\mathrm{d}\rb\,\mathrm{d}\psi\,\frac{\bar{P}\,\mathrm{d}\bar{P}}{2\pi}\,\frac{\bar{q}\,\mathrm{d}\bar{q}}{2\pi}\,\frac{\mathrm{d\phi}}{2\pi}\,\frac{\mathrm{d\phi'}}{2\pi},
\end{multline}
where $\phi$, $\phi'$ and $\psi$ are defined as $\cos\phi = \bar{\qq} \cdot \bar{\mathbf{P}} /(\bar{q} \bar{P})$, $\cos\phi' = \bar{\qq}' \cdot \bar{\mathbf{P}} /(\bar{q}' \bar{P})$, and $\cos\psi = \brr \cdot \bar{\mathbf{P}}/(\bar{r}\bar{P})$. Note that $\bar{\mathbf{P}} \equiv \bar{\mathbf{P}}'$ and $\bar{q} \equiv \bar{q}'$ in the remainder of the integrand due to conservation of momentum and energy respectively. The scattering amplitude $\bar{\mathcal{M}} = \lambda_d[u(|\bpp-\bpp'|,\eta) - u(|\bpp-\bpp'_1|,\eta)] \rightarrow \lambda_d[u(2\bar{q}|\sin[(\phi-\phi')/2]|,\eta) - u(2\bar{q}|\cos[(\phi-\phi')/2]|,\eta)]$. The product of the equilibrium distribution functions, $n_\LE\,n_{\LE,1}(1-n'_\LE)(1-n'_{\LE,1})$ can be conveniently written as:
\begin{multline}
n_\LE\,n_{\LE,1}(1-n'_\LE)(1-n'_{\LE,1})\nonumber\\
\rightarrow \frac{1}{4}\,\frac{1}{\cosh E + \cosh\gamma}\,\frac{1}{\cosh E + \cosh\gamma'},
\end{multline}
where $E=\bar{\beta}(\bar{P}^2/4 + \bar{q}^2)/[2m^*(\bar{r})] + \bar{\beta}\bar{r}^2/2 - \bar{\beta} \bar{\mu}$, $\gamma = \bar{\beta} \bar{P} \bar{q} \cos\phi / [2m^*(\bar{r})]$, $\gamma' = \bar{\beta} \bar{P} \bar{q} \cos\phi' / [2m^*(\bar{r})]$. The angle $\psi$ is only present in $\mathrm{S}[\phi_\alpha]\mathrm{S}[\phi_\alpha]$. Therefore, the integration over $\psi$ is immediate and elementary. We evaluate the required integrals using Mathematica and define $\mathrm{S}_{\alpha\beta}(\bar{r},\bar{P},\bar{q},\phi,\phi') \equiv \int\mathrm{d}\psi\,\mathrm{S}[\phi_\alpha]\mathrm{S}\,[\phi_\beta]$. The integral can be put in a more useful form using the change of variables $\bar{P} = (8\rho/\bar{\beta})^{1/2}\,\sin\xi\,\cos\nu$, $\bar{q} = (2\rho/\bar{\beta})^{1/2}\,\sin\xi\,\sin\nu$ and $\bar{r} = (2\rho/\beta)^{1/2}\,\cos\xi$, where $\rho \in [0,\infty)$, $\nu \in [0,\pi/2]$ and $\xi \in [0,\pi/2]$. The final expression is:
\begin{widetext}
\begin{align}\label{eq:Iinteg}
\II_{\alpha\beta} =& -\frac{(2N)^\frac{1}{2}\lambda_d^2}{8(2\pi)^2\,\bar{\beta}^{N_\alpha+N_\beta+3}}\int_0^\infty \rho^2\,\mathrm{d}\rho\int_0^{2\pi}\frac{\mathrm{d}\phi}{2\pi}\int_0^{2\pi}\frac{\mathrm{d}\phi'}{2\pi}\int_0^{\frac{\pi}{2}}\mathrm{d}\xi\,\sin^3\xi\,\cos \xi\int_0^{\frac{\pi}{2}}\mathrm{d}\nu\,\sin 2\nu\nonumber\\
&\times\mathrm{S}_{\alpha\beta}(\sqrt{2\rho}\cos\xi,\sqrt{8\rho}\sin\xi\,\cos\nu,\sqrt{2\rho}\sin\xi\,\sin\nu,\phi,\phi')\,m^*(\bar{r})\nonumber\\
&\times\left[\sqrt{\bar{\beta}}\,u\left(2\sqrt{2\rho/\bar{\beta}}\,\sin\xi\,\sin\nu\,|\sin[(\phi-\phi')/2]|,\eta\right)- \sqrt{\bar{\beta}}\,u\left(2\sqrt{2\rho/\bar{\beta}}\,\sin\xi\,\sin\nu\,|\cos[(\phi-\phi')/2]|,\eta\right)\right]^2\nonumber\\
&\times\left\{\left[\cosh\left(\rho\sin^2\xi/m^*(\bar{r}) + \rho\cos^2\xi + \bar{\beta} \varepsilon_0(\bar{r}) -\bar{\beta}\bar{\mu}\right)+\cosh\left(\rho\,\sin^2\xi\sin 2\nu\cos\phi/m^*(\bar{r})\right)\right]\times (\phi \leftrightarrow \phi')\right\}^{-1},
\end{align}
\end{widetext}
where $N_{a(b)} = m_{a(b)} + n_{a(b)} + k_{a(b)}$ and $\bar{r} \equiv \sqrt{2\rho/\bar{\beta}}\,\cos\xi$ in $m^*(\bar{r})$ and $\varepsilon(\bar{r})$. We evaluate the above 5-dimensional integral for all pairwise combination of basis functions using a numerical Monte-Carlo integration with $5 \times 10^8$ points which we found to yield a relative statistical error of less than $10^{-3}$ in all cases.

\section{Matrix elements of the evolution matrix in the quadrupole basis}\label{sec:matrixquad}
In this appendix, we provide readily computable expressions for various matrix elements in the quadrupole basis by carrying out the angular integrations analytically. For a given quadrupole basis function $\xi_i\phi_{\alpha}$, we define a pair of numbers $(\mu_i,\nu_i)$ as the number of powers of $r$ and $p$ in $\xi_i$ respectively, i.e. $(\mu_1,\nu_1) = (2,0)$, $(\mu_2,\nu_2) = (1,1)$, and $(\mu_3,\nu_3) = (0,2)$.

\subsection{Matrix elements of $\mathsf{M}$}
The angular integrations in $\mathsf{M}$ can be easily carried out using the variables $\cos\phi = \hat{\rr} \cdot \hat{x}$ and $\cos\psi = \brr \cdot \bpp / (\rb \pb)$. In this variables, we get $\xi_i = \rb^{\mu_i}\pb^{\nu_i}\cos(2\phi + \nu_j\psi)$. The angular integration are elementary and we find:
\begin{align}
\mathsf{M}_{\alpha\beta}^{ij} &= \int\mathrm{d}\bar{\Gamma}\,\Delta_\LE\,\xi_i\,\xi_j\,\phi_\alpha\phi_\beta\nonumber\\
&=\frac{1}{2}\,g(|\nu_i-\nu_j|,k_\alpha+k_\beta)\,I^{(R_\alpha+R_\beta+\mu_i+\mu_j)}_{(P_\alpha+P_\beta+\nu_i+\nu_j)}[\Delta_\LE].
\end{align}

\subsection{Matrix elements of $\mathsf{H}$}
As a first step, we evaluate the Poisson bracket $\{\xi_j\phi_\beta,\mathcal{H}_0\} = \xi_j\{\phi_\beta,\bar{\mathcal{H}}_0\} + \phi_\beta\{\xi_j,\bar{\mathcal{H}}_0\}$. The expression for $\{\phi_\beta,\mathcal{H}_0\}$ is known from the previous appendix (Eq.~\ref{eq:poiss}). We find $\{\xi_j,\bar{\mathcal{H}}_0\} = X_{jk}(\pb,\rb)\,\xi_k$ (sum over $k$ is implied), where:
\begin{equation}
X_{jk} = \left(\begin{tabular}{ccc}
$0$ & $2\gamma_p$ & $0$\\
$-\gamma_r$ & $0$ & $\gamma_p$\\
$0$ & $-2\gamma_r$ & $0$
\end{tabular}\right).
\end{equation}
Thus, we get:
\begin{align}
\left(\mathsf{H_0}\right)_{\alpha\beta}^{ij} &= \int\mathrm{d}\bar{\Gamma}\,\Delta_\LE\,\xi_i\,\phi_\alpha \{\xi_j \phi_\beta, \bar{\mathcal{H}}_0\}\nonumber\\
&=\underbrace{\int\mathrm{d}\bar{\Gamma}\,\Delta_\LE\,\phi_\alpha\{\phi_\beta,\mathcal{H}_0\}\,\xi_i\xi_j}_{(\mathsf{H}_0)_{\alpha\beta,1}^{ij}}\nonumber\\
&\qquad\qquad\qquad\qquad+ \underbrace{\int\mathrm{d}\bar{\Gamma}\,\Delta_\LE\,\phi_\alpha\phi_\beta\,X_{jk}\,\xi_i\,\xi_k}_{(\mathsf{H}_0)_{\alpha\beta,2}^{ij}}.
\end{align}
The angular integrations in $(\mathsf{H}_0)_{\alpha\beta,1}^{ij}$ can be most easily evaluated using the variables defined earlier, $\cos\phi = \hat{\rr} \cdot \hat{x}$ and $\cos\psi = \brr \cdot \bpp / (\rb\pb)$. The final result is:
\begin{multline}
(\mathsf{H}_0)_{\alpha\beta,1}^{ij} = \frac{1}{2}\Big[2m_\beta\,g(|\nu_i-\nu_j|,k_\alpha+k_\beta+1)\\
+ k_\beta\,g(|\nu_i-\nu_j|,k_\alpha + k_\beta - 1)\Big]\,I^{(R_\alpha+R_\beta+\mu_i + \mu_j-1)}_{(P_\alpha+P_\beta+\nu_i + \nu_j+1)}[\Delta_\LE\,\gamma_p]\nonumber\\
-\frac{1}{2}\Big[2n_\beta\,g(|\nu_i-\nu_j|,k_\alpha+k_\beta+1)\\
+ k_\beta\,g(|\nu_i-\nu_j|,k_\alpha + k_\beta - 1)\Big]\,I^{(R_\alpha+R_\beta+\mu_i + \mu_j+1)}_{(P_\alpha+P_\beta+\nu_i + \nu_j-1)}[\Delta_\LE\,\gamma_r].
\end{multline}
The angular integrations in $(\mathsf{H}_0)_{\alpha\beta,2}^{ij}$ are similar to those in $(\mathsf{M})_{\alpha\beta}^{ij}$. The result is:
\begin{multline}
(\mathsf{H}_0)_{\alpha\beta,2}^{ij} = \frac{1}{2}\,g(|\nu_i-\nu_k|,k_\alpha+k_\beta)\\
\times\,I^{(R_\alpha+R_\beta+\mu_i + \mu_k)}_{(P_\alpha+P_\beta+\nu_i + \nu_k)}[\Delta_\LE\,X_{jk}].
\end{multline}

\subsection{Matrix elements of $\mathsf{\Sigma}$}
Like the monopole case, the first step is evaluating $\bar{\Sigma}[\Delta_\LE\xi_j\phi_\beta]$:
\begin{multline}\label{eq:Sigquad}
\bar{\Sigma}[\Delta_\LE\xi_j\phi_\beta] =\\
\lambda_d\int\frac{\pb'\mathrm{d}\pb'}{2\pi}\sum_{n=0}^{\infty}u^{(n)}(\pb,\pb';\eta)\,\Delta_\LE(\pb',\rb)\,\rb^{R_\beta+\mu_j}\,\pb'^{P_\beta+\nu_j}\\
\times \int \frac{\mathrm{d\psi'}}{2\pi}\cos^{k_\beta}(\psi+\psi')\,\cos[2\phi + \nu_j(\psi+\psi')]\cos(n\psi'),
\end{multline}
where we have expressed $u(|\bpp-\bpp'|,\eta)$ in a cosine series.
The angular integration (the second integral) can be carried out using the contour technique and yields $\tilde{g}(\nu_j,n,k_\beta)\cos(2\phi)\cos(n\psi) - \tilde{h}(\nu_j,n,k_\beta)\sin(2\phi)\sin(n\psi)$, where:
\begin{align}
\tilde{g}(0,n,k) &\equiv g(n,k),\nonumber\\
\tilde{g}(1,n,k) &\equiv g(n,k+1),\nonumber\\
\tilde{g}(2,n,k) &\equiv 2g(n,k+2)-g(n,k),\nonumber\\
\tilde{h}(\nu,n,k) &\equiv \tilde{g}(\nu,n,k) - g(\nu+n,k).
\end{align}
Plugging this back into Eq.~(\ref{eq:Sigquad}), we get:
\begin{multline}
\bar{\Sigma}[\Delta_\LE\xi_j\phi_\beta] = \sum_{n=0}^{k_\beta+2} Q_{\beta,j}^{(n)}(\pb,\rb)\Big[\tilde{g}(\nu_j,n,k_\beta)\cos(2\phi)\\
\times\cos(n\psi)- \tilde{h}(\nu_j,n,k_\beta)\sin(2\phi)\sin(n\psi)\Big],
\end{multline}
where:
\begin{multline}
Q_{\beta,j}^{(n)}(\pb,\rb) = \lambda_d\int\frac{\mathrm{d}\pb'}{2\pi}\,\rb^{R_\beta+\mu_j}\,\pb'^{P_\beta+\nu_j+1}\\
\times u(\pb,\pb';\eta)\,\Delta_\LE(\pb',\rb).
\end{multline}
The last integral can be evaluated easily numerically. The final result can be expressed easily in terms of the last two expressions:
\begin{multline}
\left(\mathsf{\Sigma}^F\right)_{\alpha\beta}^{ij} = \sum_{n=0}^{k_\beta+2}\Bigg[\frac{1}{2}\,G_{(\nu_j,n,k_\beta)}^{(\nu_k,n,k_\alpha)}\,I^{(R_\alpha+\mu_k)}_{(P_\alpha+\nu_k)}[\Delta_\LE Q_{\beta,j}^{(n)}X_{ik}]\\
+\frac{1}{2}\left(2m_\alpha\,G^{(\nu_i,n,k_\alpha+1)}_{(\nu_j,n,k_\beta)}+ k_\alpha\,G^{(\nu_i,n,k_\alpha-1)}_{(\nu_j,n,k_\beta)}\right)\\
\times I^{(R_\alpha+\mu_i-1)}_{(P_\alpha+\nu_i+1)}[\Delta_\LE Q_{\beta,j}^{(n)}\gamma_p]\\
+\frac{1}{2}\left(2n_\alpha\,G^{(\nu_i,n,k_\alpha+1)}_{(\nu_j,n,k_\beta)}+ k_\alpha\,G^{(\nu_i,n,k_\alpha-1)}_{(\nu_j,n,k_\beta)}\right)\\
\times I^{(R_\alpha+\mu_i+1)}_{(P_\alpha+\nu_i-1)}[\Delta_\LE Q_{\beta,j}^{(n)}\gamma_r]\Bigg],
\end{multline}
where we have defined the shorthand notation $G^{(\nu_1,n_1,k_1)}_{(\nu_2,n_2,k_2)} = \tilde{g}(\nu_1,n_1,k_1)\,\tilde{g}(\nu_2,n_2,k_2) + \tilde{h}(\nu_1,n_1,k_1)\,\tilde{h}(\nu_2,n_2,k_2)$.

\subsection{Matrix elements of $\mathsf{I}_c$}\label{sec:Icquad}
The matrix elements of the collision integral in the quadrupole basis is identical in form to those of the monopole basis (Eq.~\ref{eq:Iinteg}). The only differences are (1): $\mathrm{S}_{\alpha\beta}$ must be replaced with:
\begin{equation}
\mathrm{S}^{ sij}_{\alpha\beta}(\rb,\bar{P},\bar{q},\phi,\phi') \equiv \int \frac{\mathrm{d}\theta}{2\pi}\mathrm{d}\psi\,\mathrm{S}[\xi_i \phi_\alpha]\,\mathrm{S}[\xi_j \phi_\beta],
\end{equation}
where we introduced an extra angle $\cos\theta = \mathbf{e}_x \cdot \bar{\mathbf{P}}/\bar{P}$. (2): The prefactor $\bar{\beta}^{N_\alpha + N_\beta + 3} \rightarrow \bar{\beta}^{N_\alpha + N_\beta + 5}$ in the denominator due to the extra powers of $\bar{\beta}^{-1}$ $\xi_i$ and $\xi_j$ introduce. The definition of $N_{\alpha(\beta)}$ is the same as before.

\section{Calculation of the collision integrals with exact Hartree-Fock quasiparticle dispersions}\label{sec:exact}
\begin{figure}[!ht]
\center
\includegraphics[scale=0.6]{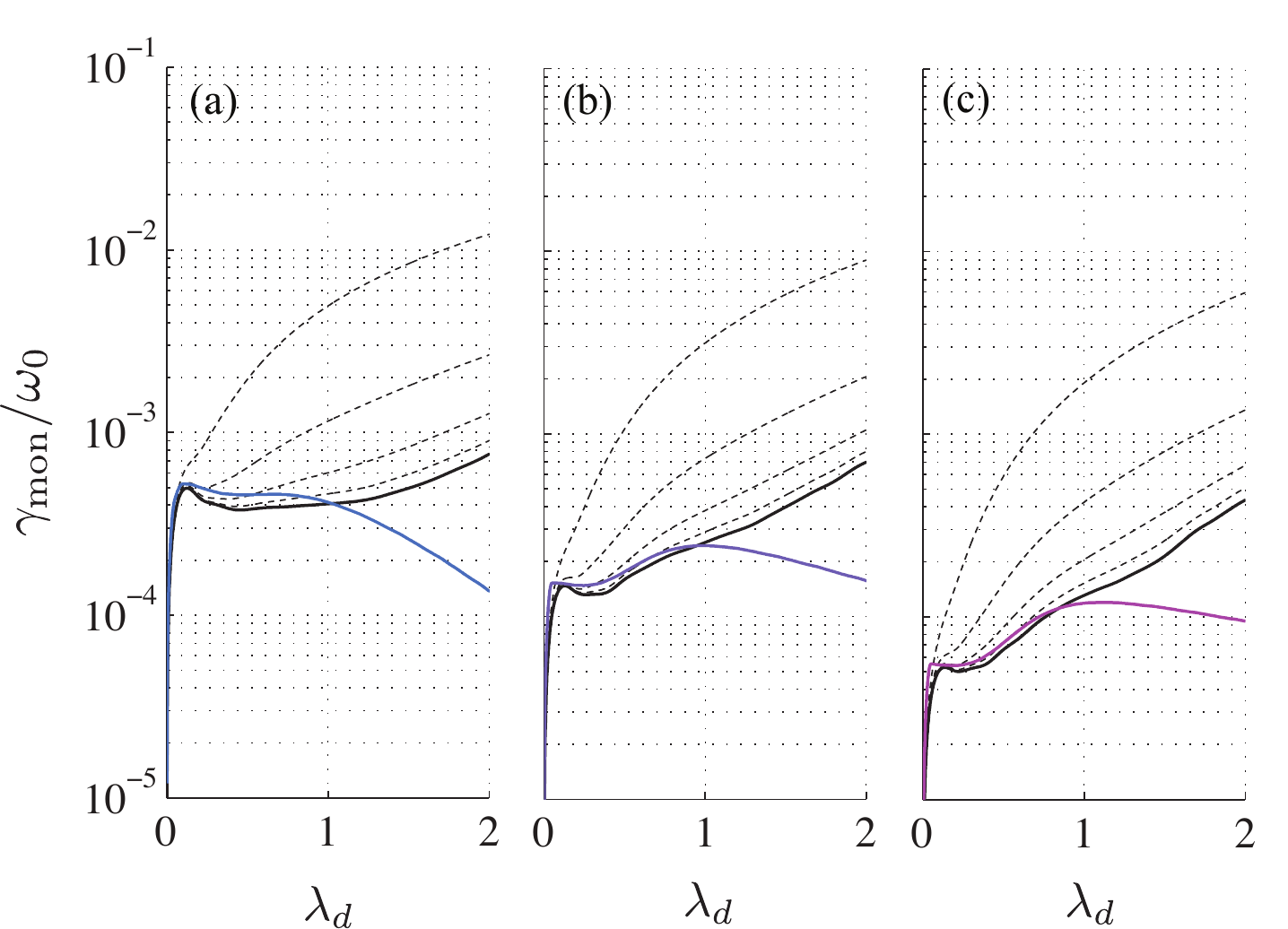}
\caption{(Color online) The damping rate of the monopole oscillations in 2D and with $N=2200$ particles. (a) $T/T_F = 0.5$, (b) $T/T_F = 1.0$ and (c) $T/T_F = 1.5$. The thick colored lines are the previously discussed result obtained using the local effective mass approximation of the the dispersions in the collision integral. The dashed lines denote approximate solutions obtained by relaxing the conservation of energy (from top to bottom, $\sigma=0.05$, $0.02$, $0.01$ and $0.005$). The thick black line is extrapolation to $\sigma=0$ (the exact solution).}
\label{fig:exact}
\end{figure}

In Sec.~\ref{sec:Ic}, we simplified the expression for the collision integral matrix elements using the local effective mass approximation (LEMA) for the quasiparticle dispersions. Although we found this scheme to be a decent approximation in the weakly interacting regime (the approximate dispersions lie within a few percents of the exact Hartree-Fock dispersions), one may argue that an exact treatment is necessary for stronger interactions. This objection is more serious when one is looking at the effects that crucially depend on self-energy corrections, such as the damping of the monopole mode. In this section, we address this issue and present numerical justification for the reliability of LEMA.

The major simplification resulting from LEMA is the possibility of an analytic treatment of the $\delta$-function in the collision integral associated to the conservation of energy (see Eq.~\ref{eq:changevar}). In that case, one simply finds $q=q'$, where $q$ and $q'$ are the magnitude of the momenta of the initial and final scattering pairs in the center of mass frame. Without a spatially local quadratic dispersion, this result does not hold anymore and in general, there is no easy way of treating the $\delta$-function analytically. Here, we adopt a simple numerical approach to overcome this difficulty. Using a limiting process to to define the delta functions,
\begin{equation}
\delta(\Delta \bar{E}) = \lim_{\sigma \rightarrow 0} \frac{1}{\sqrt{2\pi}\sigma}\,e^{-\Delta\bar{E}^2/(2\sigma^2)},
\end{equation}
we may replace the $\delta$-function with Gaussians and calculate the collision integrals for various values of $\sigma$. The $\sigma \rightarrow 0$ limit may be found by extrapolating the obtained results. This approach is considerably more computationally demanding than LEMA, however, it yields an accurate calculation of the collision integral matrix elements. The integrals are six dimensional in this case (over the variables $\bar{r}$, $\bar{P}$, $\bar{q}$, $\bar{q}'$, $\phi$ and $\phi'$) since $q$ and $q'$ may have different values now.\\

We implemented the above method for the case of monopole oscillations for a 2nd order basis set (satisfying all of the 2nd and 4th order moments of the CBV equation). The extrapolation is carried out using a polynomial fit. Fig.~\ref{fig:exact} shows the damping of monopole oscillations obtained using several choices of $\sigma$, the extrapolated result, and the result obtained from the effective mass approximation (shown earlier in Fig.~\ref{fig:monOmGam}). The matching between the effective mass approximation and the exact result is excellent up to $\lambda_d \sim 1$. The LEMA result, however, deviates from the exact result for $\lambda_d \gtrsim 1$. In any case, we find $\gamma^\mathrm{exact}_\mathrm{mon} < 10^{-3}\omega_0$ and our conclusion about the smallness of the damping of the lowest lying monopole mode is still valid. Finally, we note that the improvement in the prediction for the frequency of oscillations is much smaller (a relative correction of about $10^{-6}$) even in the strongly interacting cases. This is due to the fact that the frequency shift arises essentially from the self-energy corrections on the dynamical side of the CBV equation, which is already treated exactly.

\end{document}